\documentclass[journal,10pt]{IEEEtran}
\usepackage{graphicx}
\usepackage{amsmath}
\usepackage{amssymb}
\usepackage{bbm}
\usepackage{subfigure}
\usepackage{bm}
\usepackage{bbold}
\usepackage[usenames,dvipsnames]{xcolor}
\usepackage{booktabs}
\usepackage{cite}
\allowdisplaybreaks[4]
\usepackage{multirow}
\usepackage{amsthm}
\usepackage{algorithm}
\usepackage{algorithmic}
\usepackage{stmaryrd}

\usepackage{xr}
\makeatletter
\newcommand*{\addFileDependency}[1]{% argument=file name and extension
  \typeout{(#1)}
  \@addtofilelist{#1}
  \IfFileExists{#1}{}{\typeout{No file #1.}}
}
\makeatother

\begin{document}
\title{Experimental Validation of Cooperative RSS-based Localization with Unknown Transmit Power, Path Loss Exponent, and Precise Anchor Location}

% \title{Experimental Validation of Cooperative Localization through RSS Measurements with Unknown Transmit Power and Path Loss Exponent}

%\title{Cooperative localization using RSS measurements with unknown transmit power, path loss exponent with Experiment Validation}

%\title{Cooperative localization using RSS measurements with unknown model parameters}
\author{Yingquan Li,~\IEEEmembership{Graduate Student Member,~IEEE}, Bodhibrata Mukhopadhyay,~\IEEEmembership{Member,~IEEE}, Jiajie Xu,~\IEEEmembership{Student Member,~IEEE}, and Mohamed-Slim Alouini,~\IEEEmembership{Fellow,~IEEE}
\thanks{Y. Li, J. Xu, and M. S. Alouini are with Computer, Electrical, and Mathematical Sciences and Engineering Division (CEMSE), King Abdullah University of Science and Technology (KAUST), Thuwal, 23955-6900, Kingdom of Saudi Arabia. Bodhibrata Mukhopadhyay was with CEMSE, KAUST, Thuwal, 23955-6900, Kingdom of Saudi Arabia. He is now with the department of Electronics and Communication, Indian Institute of Technology Roorkee, Uttarakhand, 247667, India (email: yingquan.li@kaust.edu.sa, bodhibrata@ece.iitr.ac.in, jiajie.xu.1@kaust.edu.sa, slim.alouini@kaust.edu.sa).}}

\maketitle

\begin{abstract}
Received signal strength (RSS)--based cooperative localization has gained significant attention due to its straightforward system architectures and cost-effectiveness. In this paper, we propose Cooperative Localization Techniques (with Unknown Parameters), referred to as CTUP(s), which consider uncertainty in anchor nodes' locations and assume the transmit power and \textcolor{blue}{path loss exponent (PLE)} to be unknown. Unlike prior studies, CTUP(s) address unknowns by estimating these parameters, along with the location of target nodes. The non-convex and non-linear nature of the maximum likelihood (ML) estimator of the problem is addressed through relaxation techniques, employing Taylor series expansion, semidefinite relaxation (SDR), and the epigraph method. The resulting problem is solved using semidefinite second-order cone programming (SDP-SOCP), leveraging the precision of SDP and the simplicity of SOCP. We deployed an extensive network comprising 50 BLE nodes covering an area of 640~m $\times$ 180~m to gather RSS data. The precise location of the nodes is obtained using real-time kinematics global positioning system (RTK-GPS), which is treated as the ground truth. Furthermore, to replicate real-world scenarios, we recorded the positions of the anchor nodes using a standard GPS, thereby introducing uncertainty into the anchor node locations. Extensive simulation and hardware experimentation demonstrate the superior performance of CTUP compared to existing techniques.
\end{abstract}           

%While existing studies assume complete knowledge of nodes' transmit power, path loss exponent (PLE), and anchor nodes' locations (nodes whose locations are assumed to be known), this paper addresses the uncertainties associated with them. In this paper, we propose Cooperative Localization Techniques (with Unknown Parameters), referred to as CTUP(s), which consider uncertainty in anchor nodes' locations and assume the transmit power and PLE to be unknown. CTUP estimates these parameters along with the location of the target nodes (nodes whose locations are unknown). 
%
%Extensive simulation and hardware experimentation demonstrate the superior performance of CTUP in terms of location, transmit power, and PLE estimation accuracy, as well as computational complexity, compared to existing techniques.
\begin{IEEEkeywords}
Wireless sensor networks, cooperative localization, received signal strength (RSS), transmit power, path loss exponent (PLE).
\end{IEEEkeywords}

\IEEEpeerreviewmaketitle

\section{Introduction}\label{sec:intro}

\IEEEPARstart{W}{ireless} sensor networks (WSNs) have found extensive applications in diverse fields such as environmental monitoring, industrial management, smart cities, and healthcare~\cite{Intro_Loc_Environ, Intro_Loc_TOA, SDP_ToA_ma, Intro_Loc_IoT}. Without the geographical location of the nodes, the sensing data lacks significance or context. \textcolor{blue}{In~\cite{MoeWin2}, the authors propose a framework for designing network localization and navigation (NLN) for the Internet of Things (IoT).} However, equipping the nodes with a Global Positioning System (GPS) device is often impractical and cost-ineffective~\cite{Intro_Gps}. \textcolor{blue}{Additionally, GPS devices not only significantly impact nodes' battery life due to high power consumption but also fail in challenging operational environments, such as urban canyons and buildings~\cite{MoeWin4}.} Consequently, considerable research efforts are directed towards the development of efficient and precise localization techniques. \textcolor{blue}{Liu~\textit{et al.}~\cite{MoeWin5} presents a localization system leveraging inertial measurements and spatial cooperation, featuring a graphical model for single-user and multi-user scenarios.} In WSN, a subset of nodes, specifically a few, are presumed to have known locations and are identified as anchors. Leveraging these anchors, the locations of the remaining nodes, referred to as target nodes, are estimated or determined.

Localization techniques employ diverse measurements such as angle-of-arrival (AOA)~\cite{Intro_AOA}, time-of-arrival (TOA)~\cite{Intro_TOA}, time-difference-of-arrival (TDOA)~\cite{Intro_TDOA}, and received signal strength (RSS)~\cite{DLOC_bodhi} to estimate the location of target nodes. TOA and TDoA techniques stand out for their high accuracy in determining node positions. \textcolor{blue}{However, these techniques depend on precise clocks, which can pose significant challenges during real-world deployment, thereby presenting limited scalability in practical implementations.} Furthermore, their dependence on sophisticated hardware and intricate signal processing algorithms adds complexity and cost to the system, potentially limiting their practicality in resource-constrained environments or large-scale deployments~\cite{SignalStamp}. RSS-based localization determines node positions by analyzing the strength of the received signals from surrounding nodes~\cite{Vaghefi_cooperative}.  This approach doesn't necessitate precise time synchronization or specialized hardware, making it simpler to implement and more cost-effective. Therefore, RSS-based localization proves advantageous in scenarios where accuracy requirements are moderate and when hardware or synchronization constraints exist, making it suitable for localization in large networks~\cite{Intro_RWLS}. 

\begin{table*}[!t]
\centering
\caption{Summary of the RSS-based localization techniques. (NC: non-cooperative, C: cooperative).}
\resizebox{\textwidth}{!}{\begin{tabular}{|c|c|ccc|c|c|c|c|c|}
\hline
\multirow{2}{*}{Algorithm} & \multirow{2}{*}{Unknown parameters}                                                                 & \multicolumn{3}{c|}{Estimated parameters}                                                                                                         & \multirow{2}{*}{Manner}     & \multirow{2}{*}{Method}            & \multirow{2}{*}{Accuracy}      & \multirow{2}{*}{Complexity}        & \multirow{2}{*}{Year}          \\ \cline{3-5}
                           &                                                                                                     & \multicolumn{1}{c|}{Location}                               & \multicolumn{1}{c|}{Transmit power}                         & PLE                   &                             &                                    &                                &                                    &                                \\ \hline
LSRE-Shi~\cite{SDPShi}                   & Transmit power                                                                                      & \multicolumn{1}{c|}{$\checkmark$}                           & \multicolumn{1}{c|}{$\checkmark$}                           & $\times$              & NC                          & SDP                                & Moderate                       & High                               & 2020                           \\ \hline
RWLS-AE~\cite{RWLSAE}                    & Transmit power \& PLE                                                                               & \multicolumn{1}{c|}{$\checkmark$}                           & \multicolumn{1}{c|}{$\checkmark$}                           & $\checkmark$          & NC                          & SDP                                & High                           & High                               & 2021                           \\ \hline
SDP-Zou~\cite{SDP_Zou}                    & \begin{tabular}[c]{@{}c@{}}Transmit power \& PLE\\ Anchors' accurate location\end{tabular}          & \multicolumn{1}{c|}{$\checkmark$}                           & \multicolumn{1}{c|}{$\times$}                               & $\times$              & NC                          & SDP                                & Moderate                       & High                               & 2021                           \\ \hline
RLBM~\cite{Lohrasbipeydeh_Gulliver_2021}                       & Transmit power                                                                                      & \multicolumn{1}{c|}{$\checkmark$}                           & \multicolumn{1}{c|}{$\times$}                               & $\times$              & NC                          & SDP                                & Moderate                       & High                               & 2021                           \\ \hline
MSL~\cite{MSL}                        & PLE                                                                                                 & \multicolumn{1}{c|}{$\checkmark$}                           & \multicolumn{1}{c|}{$\times$}                               & $\checkmark$          & NC                          & LLS                                & Low                            & Low                                & 2022                           \\ \hline
IRGDL~\cite{Invex}                      & None                                                                                                & \multicolumn{1}{c|}{$\checkmark$}                           & \multicolumn{1}{c|}{$\times$}                               & $\times$              & C                           & Invex Relaxation                   & High                           & Low                                & 2022                           \\ \hline
SDP-l2~\cite{SDPl2}                     & None                                                                                                & \multicolumn{1}{c|}{$\checkmark$}                           & \multicolumn{1}{c|}{$\times$}                               & $\times$              & C                           & SDP                                & High                           & High                               & 2022                           \\ \hline
FCUP~\cite{FCUP}                       & Transmit power                                                                                      & \multicolumn{1}{c|}{$\checkmark$}                           & \multicolumn{1}{c|}{$\checkmark$}                           & $\times$              & C                           & SDP-SOCP                           & High                           & Moderate                           & 2023                           \\ \hline
\textbf{CTUP-1}            & \textbf{Anchors' accurate location}                                                                 & \multicolumn{1}{c|}{\multirow{4}{*}{\textbf{$\checkmark$}}} & \multicolumn{1}{c|}{\textbf{$\times$}}                      & \textbf{$\times$}     & \multirow{4}{*}{\textbf{C}} & \multirow{4}{*}{\textbf{SDP-SOCP}} & \multirow{4}{*}{\textbf{High}} & \multirow{4}{*}{\textbf{Moderate}} & \multirow{4}{*}{\textbf{2023}} \\ \cline{1-2} \cline{4-5}
\textbf{CTUP-2}            & \textbf{\begin{tabular}[c]{@{}c@{}}PLE\\ Anchors' accurate location\end{tabular}}                   & \multicolumn{1}{c|}{}                                       & \multicolumn{1}{c|}{\textbf{$\times$}}                      & \textbf{$\checkmark$} &                             &                                    &                                &                                    &                                \\ \cline{1-2} \cline{4-5}
\textbf{CTUP-3}            & \textbf{\begin{tabular}[c]{@{}c@{}}Transmit power\\ Anchors' accurate location\end{tabular}}        & \multicolumn{1}{c|}{}                                       & \multicolumn{1}{c|}{\multirow{2}{*}{\textbf{$\checkmark$}}} & \textbf{$\times$}     &                             &                                    &                                &                                    &                                \\ \cline{1-2} \cline{5-5}
\textbf{CTUP-4}            & \textbf{\begin{tabular}[c]{@{}c@{}}Transmit power \& PLE\\ Anchors' accurate location\end{tabular}} & \multicolumn{1}{c|}{}                                       & \multicolumn{1}{c|}{}                                       & \textbf{$\checkmark$} &                             &                                    &                                &                                    &                                \\ \hline
\end{tabular}}
\label{Tb:ExistingTech}
\end{table*}

%how to solve the RSS: 1) LS and 2) ML 
RSS-based localization techniques determine the positions of target nodes using RSS values between anchor-target links and/or target-target links. The localization problem can broadly be solved using methods such as least squares (LS)~\cite{LS_ref} and maximum likelihood (ML)~\cite{UTSDSOCP}. LS-based techniques are used for scenarios where the distribution of noise in RSS measurement is unknown. However, ML takes into account the noise statistics to achieve asymptotically optimal performance. Nevertheless, the ML is non-convex, non-linear, and has non-removable discontinuities, making it unsolvable using off-the-shelf techniques.  Therefore, researchers have reformulated the ML problem into a tractable problem using various relaxation techniques~\cite{CRS, ADMMSF, SOCPU}.  

In~\cite{TLLS}, the authors proposed a two-step linear least squares (LLS) estimator to obtain the location of the target nodes for a scenario where transmit power is unknown. In the first step, the authors obtained the ratio-of-distance estimates from the RSS measurement.  Subsequently, they obtained the location by solving a set of linear equations. However, it has been observed that LLS demonstrates inadequate performance, particularly in scenarios with high levels of noise. To overcome the drawbacks of LLS, we explore non-linear least squares (NLS)-based methods.

In~\cite{ADMMSF}, the authors reformulated the ML optimization into a convex distributed problem, utilizing the alternating direction method of multipliers (ADMM). Then, they forced the solution towards the local minimum of the non-relaxed problem using soft transition. In~\cite{Invex}, the authors relaxed the ML objective function into an invex optimization problem and solved it using gradient descent and coordinate descent. Researchers extensively used semidefinite programming (SDP) to solve the ML due to its guaranteed convergence and high accuracy~\cite{Intro_SDPLSREWang,SDPl2}. In~\cite{Intro_SDPLSREWang}, authors proposed a cooperative localization technique that first converts the log-normal shadowing RSS measurement model to a multiplicative model. Subsequently, the ML problem is reformulated into a non-convex estimator using relative error and finally solved using the semidefinite relaxation (SDR) technique. Wang~\textit{et al.}~\cite{SDPl2} addressed the localization problem for scenarios where RSS measurements are biased. They proposed two estimators using SDP with $\ell_1$-norm and $\ell_2$-norm, respectively. The above mentioned techniques require complete knowledge of transmit power, path loss exponent (PLE), and precise locations for anchor nodes.

The key model parameters of RSS-based localization techniques are the transmit power of the nodes, PLE, and anchor location uncertainty~\cite{RSSTOA}\footnote{In this study, the transmit power of a node is defined as the power received from the node at a reference distance.}. However, many studies assume extensive knowledge of these parameters, overlooking their susceptibility to various factors such as the battery level of the nodes~\cite{Battery_level}, channel state~\cite{Channel_state}, and antenna orientation~\cite{AntennaOrientation}. The commonly used Log-normal model implies that the signal strength attenuation is caused by obstacles and diverse environmental conditions~\cite{simon2001digital}, posing challenges in accurately measuring transmit power. Additionally, PLE varies due to changes in transmitter frequency~\cite{PLE_Radiofrequency}, weather patterns~\cite{PLE_weather}, and temperature fluctuations~\cite{PLE_temperature}. Studies have indicated that PLE is dynamic, typically ranging between 2 and 4~\cite{Beta0Ref, PLE_AI, PLE_2}.
Generally, researchers use GPS to obtain the location of the anchor nodes; however, their measurement accuracy depends on several factors like satellite visibility, atmospheric conditions, and quality of the GPS module~\cite{Anchor_GPS}. Moreover, in both maritime and aerial wireless networks, obtaining precise anchor locations (buoys in water and drones in the air) is hindered by factors like water currents and wind, respectively~\cite{Anchor_vary}. This proves the significance of incorporating uncertainty in anchor location as a model parameter. 
Researchers have designed and developed several non-cooperative localization (only using anchor-target links as measurement) techniques by assuming transmit power and (or) PLE to be unknown and (or) considering anchor location uncertainty. Shi~\textit{et al.}~\cite{SDPShi} utilized the least squares relative error (LSRE) to develop an estimator in the presence of unknown transmit power. The log-normal model was reformulated into a multiplicative form to facilitate semidefinite relaxation (SDR). In~\cite{MSL}, the authors applied a weighted LLS framework to transform the ML estimator for unknown PLE scenarios. They proposed a bisection-based method to estimate the location of the target nodes. Zou~\textit{et al.}~\cite{SDP_Zou} used Taylor expansion and penalty factors to introduce a SDP-based estimator without knowing the transmit power and PLE. Then, the proposed estimator was extended for the scenarios with anchor location uncertainty. Sun~\textit{et al.}~\cite{RWLSAE}  introduced a method that alternately estimates unknown transmit power and PLE. Based on this, the authors presented two SDP-based estimators, each addressing one unknown parameter, and improved their efficacy using an iterative approach. Researchers deal with the unknown transmit power scenarios using RSS difference (RSSD)-based techniques. In~\cite{Lohrasbipeydeh_Gulliver_2021}, the authors proposed an iterative SDP-based technique by considering a mean squared error (MSE) minimization problem and finally transforming it into a convex optimization problem using SDR. However, the aforementioned methods are exclusively applicable in non-cooperative localization.  In Non-cooperative localization, each target node is required to solve one optimization problem, which leads to increased complexity, limited scalability, and increased power consumption. Unlike cooperative localization techniques~\cite{CRS},  non-cooperative approaches face challenges in achieving high precision due to their inability to leverage target-target links. 

There exist few cooperative localization techniques that incorporate both anchor-target links and target-target links, assuming certain model parameters to be unknown. In~\cite{FCUP}, the authors proposed a cooperative localization technique that jointly estimates the location and transmit power of target nodes. However, they assumed the PLE to be known and did not consider the anchor location uncertain. To address the above-mentioned issues, in this study, we propose SDP-SOCP-based cooperative localization techniques known as CTUP-1, CTUP-2, CTUP-3, and CTUP-4\footnote{CTUP: Cooperative Technique with Unknown model Parameters}. These techniques aim to estimate the locations of target nodes across four distinct scenarios: i) transmit power and PLE are known, ii) PLE is unknown, iii) transmit power is unknown, and iv) both transmit power and PLE are unknown, respectively. We investigate the anchor-anchor wireless links to get an initial estimate of PLE as reported in~\cite{A_Alink1, A_Alink2, A_Alink3}. In Table~\ref{Tb:ExistingTech}, we
provide a brief summary of the existing RSS-based localization techniques. Unlike our approach, none of the existing studies addressing anchor location uncertainty have conducted performance tests on real experimental data. The main contributions of this paper are summarized as follows:
\begin{table}
%\captionsetup{justification=centering, labelsep=newline}
\centering
\caption{Table of notations}
\label{Tb:Notations}
\resizebox{.48\textwidth}{!}{\begin{tabular}{|c|p{5.2cm}|}
\hline
\multicolumn{1}{|c|}{\textbf{Notations}} & \multicolumn{1}{c|}{\textbf{Description}}\\ 
\hline
$N_a$ & The number of anchor nodes \\\hline
$N_t$ & The number of target nodes \\\hline
$\mathbf{t}_j$ & The location of the $j^{\text{th}}$ target node \\\hline
$\breve{\mathbf{s}}_i$ & The location of the $i^{\text{th}}$ anchor node \\\hline
$\mathbf{s}_i$ & The erroneous location of the $i^{\text{th}}$ anchor node \\\hline
$\mathcal{T}$ & The index set for target nodes \\\hline
$\mathcal{A}$ & The index set for anchor nodes \\\hline
$\mathcal{T}_j$ & The index set for the neighboring target nodes to $\mathbf{t}_j$ \\\hline
$\mathcal{A}_j$ & The index set for the neighboring anchor nodes to $\mathbf{t}_j$ \\\hline
$P_{ij}$ & The RSS measurement at $\breve{\mathbf{s}}_i$ (or $\mathbf{t}_i$) when $\mathbf{t}_j$ is the transmitter \\\hline
$P_j$ & The transmit power of $\mathbf{t}_j$ \\\hline
$\beta$ & PLE \\\hline
$d_{ij}$ & The Euclidean distance between $\mathbf{t}_j$ and $\breve{\mathbf{s}}_i$ (or $\mathbf{t}_i$) \\\hline
$d_0$ & The reference distance \\\hline
$n_{ij}$ & The RSS measurement noise of $P_{ij}$ \\\hline
$\eta_i$ & The anchor location uncertainty of $\mathbf{s}_i$ \\\hline
$\sigma_{ij}$ & The standard deviation of $n_{ij}$ \\\hline
$\delta_i$ & The standard deviation of $\eta_i$ \\\hline
$\mathbf{t}$ & The location of target nodes \\\hline
$\breve{\mathbf{s}}$ & The location of anchor nodes \\\hline
$\mathbf{p}$ & The transmit power of target nodes \\\hline
$\mathbf{e}_j$ & The $j^{\text{th}}$ column of $\mathbf{I}_{2(N_t+N_a)}$ \\\hline
$\breve{P}_{ij}$ & The RSS measurement at $\breve{\mathbf{s}}_i$ when $\breve{\mathbf{s}}_j$ is the transmitter  \\\hline
$\breve{P}_j$ & The transmit power of $\breve{\mathbf{s}}_j$ \\\hline
$\breve{d}_{ij}$ & The Euclidean distance between $\mathbf{s}_j$ and $\mathbf{s}_i$ \\\hline
$\breve{\sigma}_{ij}$ & The noise standard deviation of $\breve{P}_{ij}$ \\\hline
$\beta_0$ & The initial estimate of PLE \\\hline
$\mathcal{S}_j$ & The index set for the neighboring anchor nodes to $\mathbf{s}_j$ \\\hline
$\lvert\mathcal{S}\rvert$ & The number of anchor-anchor links \\\hline
$\lvert\mathcal{H}\rvert$ & The number of target-anchor links and target-target links \\\hline
\end{tabular}}
\end{table}
\begin{itemize} 
    %\item We reformulate the non-convex, non-linear, and discontinuous ML objective function into a standard SDP by using SDR. Subsequently, we relax the SDP into a mixed SDP-SOCP problem by employing the epigraph method, thereby decreasing the computational complexity without affecting the performance. 
    \item We proposed four cooperative techniques (CTUP-X) addressing scenarios dependent on the knowledge of transmit power and PLE, while also accounting for anchor location uncertainty. The proposed techniques can jointly estimate the location of target nodes, transmit power, and  PLE depending on the scenarios.
    \item  Cramer-Rao lower bound (CRLB) for RSS-based cooperative localization with unknown transmit power, PLE, and anchor location uncertainty. 
    \item We created an indigenous dataset comprising RSS measurements between 50 nodes covering an area of 640~m $\times$ 180~m.  The dataset also includes measurements of anchor uncertainty acquired through real-time kinematics (RTK) GPS and a standard GPS. To our knowledge, this is the most extensive and densely populated network utilized for localization. 
     \item Extensive numerical simulations and real field experiments demonstrate the superior performance of CTUPs in terms of accuracy and computational complexity as compared to state-of-the-art techniques. 
\end{itemize}

The rest of this paper is organized as follows. Section~\ref{sec:SystemModel} presents the system model for RSS-based cooperative localization with anchor location uncertainty. Section~\ref{sec:Algorithm} discusses the proposed SDP-SOCP-based estimators. The CRLB in the presence of unknown model parameters is derived in Section~\ref{sec:CRLB}. Numerical results and experimental performance are given in Section~\ref{sec:Numerical}. Finally, we conclude the paper in Section~\ref{sec:conclusions}.

$Notation$: $\lvert\mathcal{X}\rvert$ denotes the cardinality of set $\mathcal{X}$. Vectors and matrices are represented by bold lowercase and bold uppercase letters, respectively. $\ell_2$-norm of a vector is given by $\lVert\cdot\rVert$.  $\mathbf{I}_N$ and  $\mathbf{1}_N$ denote $N\times N$ identity matrix, and all-ones vectors with $N$ rows respectively. $\mathbf{0}_{M,N}$ and $\mathbf{1}_{M,N}$ represent the all-zeros and all-ones matrices with $M$ rows and $N$ columns, respectively. $\mathbf{x}\left(m:n\right)$ represents the elements of $\mathbf{x}$ from the $m^{\text{th}}$ row to the $n^{\text{th}}$ row. $\mathbf{X}_{m,n}$ represents the element of $\mathbf{X}$ located at the intersection of the $m^{\text{th}}$ row and the $n^{\text{th}}$ column. $\mathbf{X}_{m:n,m:n}$ denotes a submatrix composed of rows $m$ to $n$ and columns $m$ to $n$ of $\mathbf{X}$. $\text{diag}\left(\mathbf{x}\right)$ denotes the diagonal matrix with the elements of $\mathbf{x}$ on the main diagonal. For a symmetric matrix $\mathbf{X}$, $\mathbf{X}\succcurlyeq\mathbf{0}$ implies that $\mathbf{X}$ is positive semidefinite. $\text{tr}\left(\cdot\right)$ represents the trace of a matrix. $\text{Var}\left(\cdot\right)$ denotes the variance of a random variable. A summary of notations is presented in Table~\ref{Tb:Notations}.

\section{System Model}\label{sec:SystemModel}
%\subsection{System Model}
We consider a two-dimensional network consisting of $N_a$ anchor nodes and $N_t$ target nodes
%Assume $N_t$ target nodes and $N_a$ anchor nodes are deployed in a two-dimensional wireless network. 
The location of the $j^{\text{th}}$ target node and the $i^{\text{th}}$ anchor node are represented by $\mathbf{t}_j$ and $\breve{\mathbf{s}}_i$, respectively. We denote the set of indices of the target nodes and anchor nodes as $\mathcal{T}$ and $\mathcal{A}$, respectively. A level of uncertainty in the positioning of the anchor nodes is taken into account. Based on the Log-normal signal propagation model~\cite{SDP_Zou}, the received power and the erroneous anchor location are expressed using:
\begin{IEEEeqnarray}{lCl}
    \IEEEyesnumber\IEEEyessubnumber*
    P_{ij} &=& P_j - 10\beta \log_{10}\frac{d_{ij}}{d_0} + n_{ij},\ j \in \mathcal{T},\ i \in \mathcal{A}_j \cup \mathcal{T}_j,\label{LogNormal}\IEEEeqnarraynumspace\\
    \mathbf{s}_i &=& \mathbf{\breve{s}}_i + \eta_{i}\mathbf{1}_2,\ i\in \mathcal{A},\label{AnchorLoc}
\end{IEEEeqnarray}
where $P_{ij}$ is the received power at $\mathbf{\breve{s}}_i$ (or $\mathbf{t}_i$) when $\mathbf{t}_j$ is transmitting. The Euclidian distance between two nodes are given by $d_{ij} = \|\mathbf{t}_j-\breve{\mathbf{s}}_i\| \left(\text{or}\ \|\mathbf{t}_j-\mathbf{t}_i\|\right)$. The reference distance $d_0$ is considered to be 1~m. $P_j$ is the received power at $d_0$ and will be referred to as the transmit power of $\mathbf{t}_j$, and $\beta$ is PLE. The neighboring target nodes and anchor nodes of $\mathbf{t}_j$ are indexed by $\mathcal{T}_j$ and $\mathcal{A}_j$, respectively. The RSS measurement noise $\left(n_{ij}\right)$ and anchor location uncertainly $\left(\eta_i\right)$ are assumed to follow Gaussian distribution and are represented as $n_{ij}\sim\mathcal{N}\left(0,\sigma_{ij}^2\right)$ and $\eta_{i}\sim\mathcal{N}\left(0,{\delta_i}^2\right)$, respectively. We assume the components of $\mathbf{s}_i$ are independent random variables. Let $\mathbf{t} = \left[\mathbf{t}_1^T,\dots,\mathbf{t}_{N_t}^T\right]^T$  and $\mathbf{p} = \left[P_1,\dots,P_{N_t}\right]^T$ be the location and transmit power of the target nodes, respectively. We define the precise location of anchor nodes as $\breve{\mathbf{s}} = \left[\breve{\mathbf{s}}_1^T,\dots,\breve{\mathbf{s}}_{N_a}^T\right]^T$. Let $\bm{\theta} = \left[\mathbf{t}^T,\breve{\mathbf{s}}^T,\mathbf{p}^T,\beta\right]^T$ be the unknown parameter vector and $\mathbf{m} = \left[\dots,P_{ij},\dots,\mathbf{s}_1^T,\dots,\mathbf{s}_{Na}^T\right]^T$ be the collection of RSS and anchor location measurements. The probability density function (PDF) of $\mathbf{m}$ given $\bm{\theta}$ is  expressed as
$p(\mathbf{m};\bm{\theta}) =$
\begin{IEEEeqnarray}{lr}
    \label{PDF_theta}
    &\prod_{j\in\mathcal{T} \atop i\in \mathcal{A}_j\cup\mathcal{T}_j}\frac{1}{\sqrt{2\pi \sigma_{ij}^2}}\exp\left[\frac{\left(P_{ij}-P_j+10\beta\log_{10}d_{ij}\right)^2}{-2\sigma_{ij}^2}\right]\IEEEeqnarraynumspace\nonumber\\
    &\times\prod_{i\in\mathcal{A}}\frac{1}{\sqrt{2\pi {\delta_i}^2}}\exp\left[\frac{\|\mathbf{s}_i-\breve{\mathbf{s}}_i\|^2}{-2{\delta_i}^2}\right]. \IEEEeqnarraynumspace
\end{IEEEeqnarray}
The ML estimator for $\bm{\theta}$ is given by using~\eqref{PDF_theta}~\cite{kay1993fundamentals}
\begin{IEEEeqnarray}{lr}
    \label{MLEstimator}
    \mathop{\rm{min}}_{\bm{\theta}} &\sum_{j\in\mathcal{T} \atop i\in \mathcal{A}_j\cup\mathcal{T}_j}\sigma_{ij}^{-2}\left(P_{ij}-P_j+10\beta\log_{10}d_{ij}\right)^2 \nonumber\\
    &+\sum_{i\in \mathcal{A}}\delta_i^{-2}\|\mathbf{s}_i - \breve{\mathbf{s}}_i\|^2. \IEEEeqnarraynumspace
\end{IEEEeqnarray}

\section{Cooperative localization algorithms}\label{sec:Algorithm}

\subsection{Scenario $\uppercase\expandafter{\romannumeral1}$: Transmit power and PLE are known }\label{S1}
In the first scenario, we assume that both the transmit power of target nodes and PLE are known. Through the utilization of Taylor expansion for $n_{ij}$,~\eqref{LogNormal} can be reformulated as
\begin{IEEEeqnarray}{lCl}
    \label{S1_multiplicative}
    d^2_{ij} &=& 10^{\frac{P_{j}-P_{ij}}{5\beta}}10^{\frac{n_{ij}}{5\beta}}\IEEEeqnarraynumspace\nonumber\\
    &\approx& 10^{\frac{P_{j}-P_{ij}}{5\beta}}\left(1+\frac{\ln10}{5\beta}n_{ij}\right)
    = 10^{\frac{P_{j}-P_{ij}}{5\beta}}+\zeta_{ij},\IEEEeqnarraynumspace
\end{IEEEeqnarray}
where $\zeta_{ij}\sim\mathcal{N}\left(0,\left(10^{\frac{P_{j}-P_{ij}}{5\beta}}\frac{\ln10}{5\beta}\sigma_{ij}\right)^2\right)$. Now, let $\bm{\mu} = \left[\mathbf{t}^T,\breve{\mathbf{s}}^T\right]^T$ be a vector representing the location of the target nodes and anchor nodes. The PDF of $\mathbf{m}$ can be expressed as $p(\mathbf{m};\bm{\mu})=$
\begin{IEEEeqnarray}{lr}
    \label{PDF_zeta}
    &\prod_{j\in\mathcal{T} \atop i\in \mathcal{A}_j\cup\mathcal{T}_j}\frac{1}{\sqrt{2\pi \Tilde{\zeta}^2_{ij}}}\exp\left[\frac{\left(d^2_{ij} - 10^{\frac{P_{j}-P_{ij}}{5\beta}}\right)^2}{-2\Tilde{\zeta}^2_{ij}}\right]\IEEEeqnarraynumspace\nonumber\\
    &\times\prod_{i\in\mathcal{A}}\frac{1}{\sqrt{2\pi {\delta_i}^2}}\exp\left[\frac{\|\mathbf{s}_i-\breve{\mathbf{s}}_i\|^2}{-2{\delta_i}^2}\right], \IEEEeqnarraynumspace
\end{IEEEeqnarray}
where $\Tilde{\zeta}_{ij}=10^{\frac{P_{j}-P_{ij}}{5\beta}}\frac{\ln10}{5\beta}\sigma_{ij}$. The ML estimator for $\mathbf{t}$ is given by%
\begin{IEEEeqnarray}{cll}
    \label{S1_ML}
    \IEEEyesnumber\IEEEyessubnumber*
    \mathop{\rm{min}}_{\bm{\mu}} & \sum_{j\in\mathcal{T} \atop i\in \mathcal{A}_j\cup\mathcal{T}_j}\left(\frac{d_{ij}^2-10^{\frac{P_j-P_{ij}}{5\beta}}}{\Tilde{\zeta}_{ij}}\right)^2 + \sum_{i\in \mathcal{A}}\frac{\|\mathbf{s}_i - \breve{\mathbf{s}}_i\|^2}{\delta_i^{2}} \IEEEeqnarraynumspace\\
{\text{s.t.}} 
    & d_{ij} = \|\mathbf{t}_j-\breve{\mathbf{s}}_i\|,\ j\in\mathcal{T},\ i\in\mathcal{A}_j, \label{dij1}\\
    & d_{ij} = \|\mathbf{t}_j-\mathbf{t}_i\|,\ j\in\mathcal{T},\ i\in\mathcal{T}_j. \label{dij2}
\end{IEEEeqnarray}
The optimization problem in~\eqref{S1_ML} is non-convex and can not be solved using standard techniques. \textcolor{blue}{To mitigate the non-convexity resulting from the norm constraint, we firstly square both sides of~\eqref{dij1} and~\eqref{dij2} to reformulate~\eqref{S1_ML} as
\begin{IEEEeqnarray}{cll}
    \label{S1_ML_revise}
    \IEEEyesnumber\IEEEyessubnumber*
    \mathop{\rm{min}}_{\bm{\mu}} & \sum_{j\in\mathcal{T} \atop i\in \mathcal{A}_j\cup\mathcal{T}_j}\left(\frac{d_{ij}^2-10^{\frac{P_j-P_{ij}}{5\beta}}}{\Tilde{\zeta}_{ij}}\right)^2 + \sum_{i\in \mathcal{A}}\frac{\|\mathbf{s}_i - \breve{\mathbf{s}}_i\|^2}{\delta_i^{2}} \IEEEeqnarraynumspace\\
{\text{s.t.}} 
    & d_{ij}^2 = \|\mathbf{t}_j-\breve{\mathbf{s}}_i\|^2,\ j\in\mathcal{T},\ i\in\mathcal{A}_j, \\
    & d_{ij}^2 = \|\mathbf{t}_j-\mathbf{t}_i\|^2,\ j\in\mathcal{T},\ i\in\mathcal{T}_j.
\end{IEEEeqnarray}
Subsequently, we introduce an auxiliary variable $\mathbf{K} = \bm{\mu}\bm{\mu}^T$ to represent $d_{ij}^2$ in a manner satisfying the requirements of SDP. Likewise, $\|\mathbf{s}_i - \breve{\mathbf{s}}_i\|^2$ can also be rewritten in a similar format. Consequently, we can rewrite~\eqref{S1_ML_revise} as}
\begin{IEEEeqnarray}{cll} 
\label{S1_SDP_first}
\IEEEyesnumber\IEEEyessubnumber*
\mathop {{\rm{min}}}\limits_{u_{ij},\lambda_i,\bm{\mu},\mathbf{K}} {\rm{ }} &  \sum_{j\in\mathcal{T} \atop i\in \mathcal{A}_j\cup\mathcal{T}_j}\left(\frac{u_{ij}-10^{\frac{P_j-P_{ij}}{5\beta}}}{\Tilde{\zeta}_{ij}}\right)^2 + \sum_{i\in\mathcal{A}}\delta_i^{-2}\lambda_i\\ 
{\text{s.t.}} 
& u_{ij} = \text{tr}\left(\bm{\Xi}_j\mathbf{K}\bm{\Xi}_j^T-2\bm{\Xi}_j\mathbf{K}\bm{\Psi}_i^T + \bm{\Psi}_i\mathbf{K}\bm{\Psi}_i^T\right),\nonumber\IEEEeqnarraynumspace\\
&\hspace{3.5cm} j \in \mathcal{T},\ i\in \mathcal{A}_j, \label{General_constraint1} \\
& u_{ij} = \text{tr}\left(\bm{\Xi}_j\mathbf{K}\bm{\Xi}_j^T-2\bm{\Xi}_j\mathbf{K}\bm{\Xi}_i^T + \bm{\Xi}_i\mathbf{K}\bm{\Xi}_i^T\right),\nonumber\IEEEeqnarraynumspace\\
&\hspace{3.5cm} j \in \mathcal{T},\ i\in \mathcal{T}_j, \label{General_constraint2} \\
& \lambda_{i} = \text{tr}\left(\bm{\Psi}_i\mathbf{K}\bm{\Psi}_i^T\right) -2\mathbf{s}_i^{T} \bm{\Psi}_i\bm{\mu} + \|\mathbf{s}_i\|^2,\ i \in \mathcal{A}, \IEEEeqnarraynumspace\label{General_constraint3} \\
& \mathbf{K} = \bm{\mu}\bm{\mu}^T, \label{S1_SDP_first_constraint4} \\
& \text{rank}(\mathbf{K}) = 1,\label{S1_SDP_first_constraint5}
\end{IEEEeqnarray}
where $\bm{\Xi}_i = \left[\mathbf{e}_{2i-1},\mathbf{e}_{2i}\right]^T$ and $\bm{\Psi}_i = \left[\mathbf{e}_{2i+2N_t-1},\mathbf{e}_{2i+2N_t}\right]^T$, $\mathbf{e}_j$ is the $j^{\text{th}}$ column of $\mathbf{I}_{2\left(N_t+N_a\right)}$. \textcolor{blue}{In~\eqref{S1_SDP_first}, $u_{ij}$ and $\lambda_{i}$ are auxiliary variables representing $d_{ij}^2$ and $\|\mathbf{s}_i - \breve{\mathbf{s}}_i\|^2$, respectively.} By dropping off the rank constraint and relaxing~\eqref{S1_SDP_first_constraint4}, an estimator based on SDP can be obtained
\begin{IEEEeqnarray}{cll} 
\label{S1_SDP_second}
\IEEEyesnumber\IEEEyessubnumber*
\mathop {{\rm{min}}}\limits_{u_{ij},\lambda_i,\bm{\mu},\mathbf{K}} {\rm{ }} &  \sum_{j\in\mathcal{T} \atop i\in \mathcal{A}_j\cup\mathcal{T}_j}\left(\frac{u_{ij}-10^{\frac{P_j-P_{ij}}{5\beta}}}{\Tilde{\zeta}_{ij}}\right)^2 + \sum_{i\in\mathcal{A}}\delta_i^{-2}\lambda_i\IEEEeqnarraynumspace\\ 
{\text{s.t.}} 
& \eqref{General_constraint1},\ \eqref{General_constraint2},\ \eqref{General_constraint3}, \nonumber\\
& \left[\begin{array}{ll} 1 & \bm{\mu}^T \\ \bm{\mu} & \mathbf{K}\end{array}  \right] \succcurlyeq \mathbf{0}. \label{General_constraint4}
\end{IEEEeqnarray}
The location of target nodes can be estimated by using interior point methods to solve~\eqref{S1_SDP_second}~\cite{vandenberghe1996semidefinite,cvx}. However, SDP-based estimator has higher computational complexity than those based on SOCP. Exploiting the low complexity of SOCP, we rewrite~\eqref{S1_SDP_second} as an epigraph form using the auxiliary variable $\omega$
\begin{IEEEeqnarray}{cll} 
\label{S1_SDP_third}
\IEEEyesnumber\IEEEyessubnumber*
\mathop {{\rm{min}}}\limits_{u_{ij},\lambda_i,\bm{\mu},\atop \mathbf{K},\omega} {\rm{ }} &  \, \, \omega + \sum_{i\in\mathcal{A}}\delta_i^{-2}\lambda_{i}\\ 
{\text{s.t.}} 
& \sum_{j\in\mathcal{T} \atop i\in \mathcal{A}_j\cup\mathcal{T}_j}\Tilde{\zeta}_{ij}^{-2}\left(u_{ij}-10^{\frac{P_j-P_{ij}}{5\beta}}\right)^2 \leq \omega, \label{S1_epigraph_constraint1} \\ 
& \eqref{General_constraint1},\ \eqref{General_constraint2},\ \eqref{General_constraint3},\ \eqref{General_constraint4}. \nonumber
\end{IEEEeqnarray}
The optimization problem~\eqref{S1_SDP_third} is not a SOCP as the constraint~\eqref{S1_epigraph_constraint1} is not a linear matrix inequality form. To address this issue, we use an auxiliary variable $\bm{\xi} = \left[\dots,\xi_{ij},\dots\right]$ and reformulate (9) into a  mixed SDP-SOCP problem. 
\begin{IEEEeqnarray}{cll} 
\label{S1_SDP_SOCP}
\IEEEyesnumber\IEEEyessubnumber*
\mathop {{\rm{min}}}\limits_{u_{ij},\lambda_i,\bm{\mu},\atop \mathbf{K},\omega,\bm{\xi}} {\rm{ }} &  \, \, \omega + \sum_{i\in\mathcal{A}}\delta_i^{-2}\lambda_{i}\\ 
\label{S1_SDP_SOCP_oj}
{\text{s.t.}} 
& \left\lVert\left[2\bm{\xi}^T,\omega-1\right]\right\rVert \leq \omega + 1, \label{S1_SOCP1} \\ 
& \text{diag}\big(\big[\dots, \xi_{ij}-\Tilde{\zeta}_{ij}^{-1}\left(u_{ij}-10^{\frac{P_j-P_{ij}}{5\beta}}\right),\dots,  \nonumber \\
& \hspace{.5cm} \dots,-\xi_{ij}+\Tilde{\zeta}_{ij}^{-1}\left(u_{ij}-10^{\frac{P_j-P_{ij}}{5\beta}}\right),\dots \big]\big) \succcurlyeq\mathbf{0},\nonumber\\
&\hspace{3.5cm} j \in \mathcal{T},\ i\in \mathcal{A}_j \cup \mathcal{T}_j, \label{S1_SOCP2} \\ 
& \eqref{General_constraint1},\ \eqref{General_constraint2},\ \eqref{General_constraint3},\ \eqref{General_constraint4}. \nonumber
\end{IEEEeqnarray}
The proposed estimator in \eqref{S1_SDP_SOCP} benefits from the high accuracy of SDP and low complexity of SOCP, which is referred to as CTUP-1. The estimated location of the target nodes is obtained using $\hat{\mathbf{t}}_j = \hat{\bm{\mu}}\left(2j-1:2j\right)^T$. 

\subsection{Scenario $\uppercase\expandafter{\romannumeral2}$: PLE is unknown}\label{S2}
In this scenario, we consider $\beta$ to be unknown. The anchor nodes communicating with $\breve{\mathbf{s}}_j ( j\in \mathcal{A})$ are indexed by $\mathcal{S}_j$. The total number of available links in a network is given by $\lvert\mathcal{S}\rvert = \sum_{j}\lvert\mathcal{S}_j\rvert$. The RSS measurement and the Euclidean distance between the anchor nodes $\breve{\mathbf{s}}_j ( j\in \mathcal{A})$ and $\breve{\mathbf{s}}_i ( i\in \mathcal{S}_j)$ are represented by $\breve{P}_{ij}$ and $\breve{d}_{ij}$, respectively. The transmit power of $\breve{\mathbf{s}}_j$, represented by $\breve{P}_j$, is assumed to be known. The RSS measurements are affected by additive zero-mean Gaussian noise, which is characterized by $\mathcal{N}\left(0,\breve{\sigma}_{ij}^2\right)$. By letting $\breve{\mathbf{m}} = \left[\dots,\breve{P}_{ij},\dots\right]^T$ be the collection of RSS measurements, the conditional joint PDF of $\breve{\mathbf{m}}$ given PLE is expressed as $p(\breve{\mathbf{m}};\beta)=$
\begin{IEEEeqnarray}{lCr}
    \label{PDF_AA}
     \prod_{j\in\mathcal{A} \atop i\in \mathcal{S}_j}\frac{1}{\sqrt{2\pi \breve{\sigma}_{ij}^2}}\exp\left[\frac{\left(\breve{P}_{ij}-\breve{P}_j+10\beta\log_{10}\breve{d}_{ij}\right)^2}{-2\breve{\sigma}_{ij}^2}\right]. \IEEEeqnarraynumspace
\end{IEEEeqnarray}
Thus, the ML estimator of $\beta$ obtained using~\eqref{PDF_AA} is given by
\begin{IEEEeqnarray}{lCr}
    \label{MLBeta}
     \mathop{\rm{min}}_{\beta} \sum_{j\in\mathcal{A} \atop i\in \mathcal{S}_j}\breve{\sigma}_{ij}^{-2}\left(\breve{P}_{ij}-\breve{P}_j+10\beta\log_{10}\breve{d}_{ij}\right)^2. \IEEEeqnarraynumspace
\end{IEEEeqnarray}
The objective function is convex with respect to $\beta$ and achieves the global minimum $(\beta_0)$ at its stationary point. An initial estimate of $\beta$ is expressed as
\begin{IEEEeqnarray}{lCr}
    \label{PreBeta}
    \beta_0 = \frac{\mathbf{1}_{\lvert\mathcal{S}\rvert}^T\breve{\mathbf{Q}}\breve{\mathbf{q}}}{\mathbf{1}_{\lvert\mathcal{S}\rvert}^T\breve{\mathbf{Q}}\breve{\bm{\phi}}}, \IEEEeqnarraynumspace
\end{IEEEeqnarray}
where $\breve{\mathbf{Q}} = \text{diag}\left(\left[\dots,\breve{\sigma}_{ij}^{-2},\dots\right]\right)$, $\breve{\mathbf{q}} = \left[\dots,\breve{P}_j-\breve{P}_{ij},\dots\right]^T$ and $\breve{\bm{\phi}} = \left[\dots,10\log_{10}\breve{d}_{ij},\dots\right]^T$, $j\in\mathcal{A}$, $i\in\mathcal{S}_j$. 

We express $\beta = \beta_0\left(1+\epsilon\right)$ where $\epsilon = \frac{\beta-\beta_0}{\beta_0}$. When $\beta_0$ is sufficiently close to $\beta$,~\eqref{LogNormal} can be reformulated into 
\begin{IEEEeqnarray}{lCl}
    \label{S2_multiplicative_1}
    d^2_{ij} = 10^{\frac{P_{j}-P_{ij}}{5\beta_0\left(1+\epsilon\right)}}10^{\frac{n_{ij}}{5\beta}}
    \approx 10^{\frac{P_{j}-P_{ij}}{5\beta_0}\left(1-\epsilon\right)}\left(1+\frac{\ln10}{5\beta_0}n_{ij}\right).\IEEEeqnarraynumspace
\end{IEEEeqnarray}
Applying Taylor expansion into~\eqref{S2_multiplicative_1} yields
\begin{IEEEeqnarray}{lCl}
    \label{S2_multiplicative}
    d^2_{ij} \approx 10^{\frac{P_{j}-P_{ij}}{5\beta_0}}-10^{\frac{P_{j}-P_{ij}}{5\beta_0}}\frac{\left(P_j-P_{ij}\right)\epsilon\ln10}{5\beta_0}+\tau_{ij},\IEEEeqnarraynumspace
\end{IEEEeqnarray}
where $\tau_{ij} \sim \mathcal{N} \left(0,\Tilde{\tau}_{ij}^2\right)$ and $\Tilde{\tau}_{ij}$ is expressed as
\begin{IEEEeqnarray}{lCl}
    \label{tau_ij}
    \Tilde{\tau}_{ij} &=& 10^{\frac{P_{j}-P_{ij}}{5\beta_0}}\left(1-\frac{\left(P_j-P_{ij}\right)\epsilon\ln10}{5\beta_0}\right)\frac{\ln10}{5\beta_0}\sigma_{ij}\nonumber\\
    &\approx& 10^{\frac{P_{j}-P_{ij}}{5\beta_0}}\frac{\ln10}{5\beta_0}\sigma_{ij}.\IEEEeqnarraynumspace
\end{IEEEeqnarray}
By letting $\chi_{ij} = 10^{\frac{P_j-P_{ij}}{5\beta_0}}\frac{\left(P_j-P_{ij}\right)\ln10}{5\beta_0}$, the ML estimator for $\mathbf{t}$ and $\beta$ is formulated using~\eqref{S2_multiplicative} as 
\begin{IEEEeqnarray}{cll}
    \label{S2_ML}
    \IEEEyesnumber\IEEEyessubnumber*
    \mathop{\rm{min}}_{\bm{\mu},\epsilon} &\sum_{j\in\mathcal{T} \atop i\in \mathcal{A}_j\cup\mathcal{T}_j}\Tilde{\tau}_{ij}^{-2}\left(d_{ij}^2-10^{\frac{P_j-P_{ij}}{5\beta_0}}+\chi_{ij}\epsilon\right)^2 \nonumber\\ &\hspace{3.5cm} +\sum_{i\in \mathcal{A}}\delta_i^{-2}\|\mathbf{s}_i - \breve{\mathbf{s}}_i\|^2 \IEEEeqnarraynumspace\\
{\text{s.t.}} 
    & d_{ij} = \|\mathbf{t}_j-\breve{\mathbf{s}}_i\|,\ j\in\mathcal{T},\ i\in\mathcal{A}_j, \\
    & d_{ij} = \|\mathbf{t}_j-\mathbf{t}_i\|,\ j\in\mathcal{T},\ i\in\mathcal{T}_j.
\end{IEEEeqnarray}
Following the approach used in~\eqref{S1_SDP_second}, the minimization problem can be transformed through SDR
\begin{IEEEeqnarray}{cll} 
\label{S2_SDP_first}
\IEEEyesnumber\IEEEyessubnumber*
\mathop {{\rm{min}}}\limits_{u_{ij},\lambda_i,\bm{\mu},\atop \mathbf{K},\epsilon} {\rm{ }} &  \underbrace{\sum_{j\in\mathcal{T} \atop i\in \mathcal{A}_j\cup\mathcal{T}_j}\Tilde{\tau}_{ij}^{-2}\left(u_{ij}-10^{\frac{P_j-P_{ij}}{5\beta_0}}+\chi_{ij}\epsilon\right)^2}_{z} \nonumber\\
&\hspace{3.5cm} +\sum_{i\in\mathcal{A}}\delta_i^{-2}\lambda_i\\ 
{\text{s.t.}} 
& \eqref{General_constraint1},\ \eqref{General_constraint2},\ \eqref{General_constraint3},\ \eqref{General_constraint4}.\nonumber
\end{IEEEeqnarray}
The objective function of~\eqref{S2_SDP_first} can be converted into a linear from by considering the epigraph of $z$ using an
an auxiliary variable $\omega$. Also, with the assistance of another auxiliary variable $\bm{\xi} = \left[\dots,\xi_{ij},\dots\right]$, we convert the epigraph constraint $(z \leq \omega)$ to an LMI constraint and finally represent ~\eqref{S2_SDP_first} into an SDP-SOCP formulation
\begin{IEEEeqnarray}{cll} 
\label{S2_SDP_SOCP}
\IEEEyesnumber\IEEEyessubnumber*
\mathop {{\rm{min}}}\limits_{u_{ij},\lambda_i,\bm{\mu},\atop \mathbf{K},\epsilon,\omega,\bm{\xi}} {\rm{ }} &  \, \, \omega + \sum_{i\in\mathcal{A}}\delta_i^{-2}\lambda_{i}\\ 
\label{S2_SDP_SOCP_oj}
{\text{s.t.}} 
& \left\lVert\left[2\bm{\xi}^T,\omega-1\right]\right\rVert \leq \omega + 1, \label{S2_SOCP1} \\ 
& \text{diag}\big(\big[\dots, \xi_{ij}-\left(\frac{u_{ij}-10^{\frac{P_j-P_{ij}}{5\beta_0}}+\chi_{ij}\epsilon}{\Tilde{\tau}_{ij}}\right),\dots,  \nonumber \\
&\hspace{.2cm} \dots,-\xi_{ij}+\left(\frac{u_{ij}-10^{\frac{P_j-P_{ij}}{5\beta_0}}+\chi_{ij}\epsilon}{\Tilde{\tau}_{ij}}\right),\dots \big]\big) \succcurlyeq\mathbf{0},\nonumber\\
&\hspace{3.5cm} j \in \mathcal{T},\ i\in \mathcal{A}_j \cup \mathcal{T}_j, \label{S2_SOCP2} \\ 
& \eqref{General_constraint1},\ \eqref{General_constraint2},\ \eqref{General_constraint3},\ \eqref{General_constraint4}. \nonumber
\end{IEEEeqnarray}
Solving~\eqref{S2_SDP_SOCP}, we get an estimate of $\mathbf{t}$ and $\beta$ via $\hat{\mathbf{t}}_j = \hat{\bm{\mu}}\left(2j-1:2j\right)^T$ and $\hat{\beta} = \beta_0\left(1+\hat{\epsilon}\right)$, respectively, and we refer to it as CTUP-2.

\subsection{Scenario $\uppercase\expandafter{\romannumeral3}$: Transmit power is unknown}\label{S3}
In this section, the transmit power is unknown and is jointly estimated along with the location of target nodes.
%The derivation is slightly different when the transmit power of target nodes is unknown. 
By rearranging the terms of~\eqref{LogNormal}, we have
\begin{IEEEeqnarray}{lCl}
    \label{S3_multiplicative}
    d^2_{ij}10^{\frac{P_{ij}}{5\beta}} &=& 10^{\frac{P_{j}}{5\beta}}10^{\frac{n_{ij}}{5\beta}}\nonumber\\
    &\approx& 10^{\frac{P_{j}}{5\beta}}\left(1+\frac{\ln10}{5\beta}n_{ij}\right)=10^{\frac{P_{j}}{5\beta}}+\rho_{ij},\IEEEeqnarraynumspace
\end{IEEEeqnarray}
where $\rho_{ij}\sim\mathcal{N}\left(0,\left(10^{\frac{P_{j}}{5\beta}}\frac{\ln10}{5\beta}=\Tilde{\rho}_{ij}\right)^2\right)$. Unlike in Section~\ref{S1} and Section~\ref{S2}, it is not possible to estimate $\mathbf{t}$ and $\mathbf{p}$ by solving the ML estimator obtained from~\eqref{S3_multiplicative} as $\Tilde{\rho}_{ij}$ is unknown.  Thus, in this scenario, we employ the NLS estimator, a commonly used technique known for its verified effectiveness, especially in cases involving unknown noise statistics.~\cite{Vaghefi_cooperative, kay1993fundamentals}. Therefore, we reformulate~\eqref{S3_multiplicative} as an NLS optimization problem aiming to minimize the sum of squared errors
\begin{IEEEeqnarray}{cll}
    \label{S3_NLS}
    \IEEEyesnumber\IEEEyessubnumber*
    \mathop{\rm{min}}_{\bm{\mu},\mathbf{p}} & \sum_{j\in\mathcal{T} \atop i\in \mathcal{A}_j\cup\mathcal{T}_j}\left(d_{ij}^2 10^{\frac{P_{ij}}{5\beta}}-10^{\frac{P_j}{5\beta}}\right)^2 + \sum_{i\in \mathcal{A}}\|\mathbf{s}_i - \breve{\mathbf{s}}_i\|^2 \IEEEeqnarraynumspace \label{S3_NLS_obj} \\
{\text{s.t.}} 
    & d_{ij} = \|\mathbf{t}_j-\breve{\mathbf{s}}_i\|,\ j\in\mathcal{T},\ i\in\mathcal{A}_j, \\
    & d_{ij} = \|\mathbf{t}_j-\mathbf{t}_i\|,\ j\in\mathcal{T},\ i\in\mathcal{T}_j.
\end{IEEEeqnarray}
The non-convex constraints in~\eqref{S3_NLS} are convexified through the utilization of auxiliary variables $\mathbf{K} = \bm{\mu}\bm{\mu}^T$ and $\lambda_i = \|\mathbf{s}_i - \breve{\mathbf{s}}_i\|^2$. By considering $g_j = 10^{\frac{P_j}{5\beta}}$ and $u_{ij}=d_{ij}^2$ we can express~\eqref{S3_NLS} as
\begin{IEEEeqnarray}{cll} 
\label{S3_SDP_first}
\IEEEyesnumber\IEEEyessubnumber*
\mathop {{\rm{min}}}\limits_{u_{ij},g_j,\lambda_i,\atop \bm{\mu},\mathbf{K}} {\rm{ }} &  \sum_{j\in\mathcal{T} \atop i\in \mathcal{A}_j\cup\mathcal{T}_j}\left(u_{ij}10^{\frac{P_{ij}}{5\beta}}-g_j\right)^2 + \sum_{i\in\mathcal{A}}\lambda_i \label{S3_SDP_obj}\\ 
{\text{s.t.}} 
& \eqref{General_constraint1},\ \eqref{General_constraint2},\ \eqref{General_constraint3},\ \eqref{General_constraint4}.\nonumber
\end{IEEEeqnarray}
We introduce an epigraph variable $\omega$ for the first term of~\eqref{S3_SDP_obj} and an auxiliary variables $\bm{\xi}$ to obtain the LMI constraint. Thus, we convert the SDP in~\eqref{S3_SDP_first} to an SDP-SOCP problem. 
\begin{IEEEeqnarray}{cll} 
\label{S3_SDP_SOCP}
\IEEEyesnumber\IEEEyessubnumber*
\mathop {{\rm{min}}}\limits_{u_{ij},g_j,\lambda_i,\atop \bm{\mu},\mathbf{K},\omega,\bm{\xi}} {\rm{ }} &  \, \, \omega + \sum_{i\in\mathcal{A}}\lambda_{i}\\ 
\label{S3_SDP_SOCP_oj}
{\text{s.t.}} 
& \left\lVert\left[2\bm{\xi}^T,\omega-1\right]\right\rVert \leq \omega + 1, \label{S3_SOCP1} \\ 
& \text{diag}\big(\big[\dots, \xi_{ij}-u_{ij}10^{\frac{P_{ij}}{5\beta}}+g_j,\dots,  \nonumber \\
& \hspace{.5cm} \dots,-\xi_{ij}+u_{ij}10^{\frac{P_{ij}}{5\beta}}-g_j,\dots \big]\big) \succcurlyeq\mathbf{0},\nonumber\\
& \hspace{3.5cm} j \in \mathcal{T},\ i\in \mathcal{A}_j \cup \mathcal{T}_j,\IEEEeqnarraynumspace \label{S3_SOCP2} \\ 
& \eqref{General_constraint1},\ \eqref{General_constraint2},\ \eqref{General_constraint3},\ \eqref{General_constraint4}. \nonumber
\end{IEEEeqnarray}
By solving~\eqref{S3_SDP_SOCP}, we can obtain the location and transmit power of target nodes using $\hat{\mathbf{t}}_j = \hat{\bm{\mu}}\left(2j-1:2j\right)^T$ and $\hat{P}_j = 5\beta\log_{10}\hat{g}_j$. We refer to the technique as CTUP-3.

\subsection{Scenario $\uppercase\expandafter{\romannumeral4}$: Transmit power and PLE are unknown}\label{S4}
As the transmit power and PLE are unknown, we rewrite~\eqref{LogNormal} as a multiplicative form and square both side
\begin{IEEEeqnarray}{lCr}
    \label{S4_multiplicative}
    d^2_{ij}10^{\frac{P_{ij}}{5\beta}} = 10^{\frac{P_{j}}{5\beta_0\left(1+\epsilon\right)}}10^{\frac{n_{ij}}{5\beta}} \approx 10^{\frac{P_{j}}{5\beta_0}\left(1-\epsilon\right)}10^{\frac{n_{ij}}{5\beta}},
\end{IEEEeqnarray}
where $\epsilon$ is defined in Section~\ref{S2}. By performing Taylor expansion at $\epsilon = 0$ with sufficiently small noise,~\eqref{S4_multiplicative} is reformulated into 
\begin{IEEEeqnarray}{lCl}
    \label{S4_taylor}
    d^2_{ij}10^{\frac{P_{ij}}{5\beta_0}} &\approx& 10^{\frac{P_{j}}{5\beta_0}}\left(1-\frac{P_j\epsilon\ln10}{5\beta_0}\right)\left(1+\frac{\ln10}{5\beta}n_{ij}\right)\nonumber\\
    &=& 10^{\frac{P_{j}}{5\beta_0}} - 10^{\frac{P_{j}}{5\beta_0}}\frac{P_j\epsilon\ln10}{5\beta_0} + \nu_{ij},
\end{IEEEeqnarray}
where $\nu_{ij}$ is a zero-mean Gaussian random variable with its unknown standard deviation given by
\begin{IEEEeqnarray}{lCl}
    \label{nu_ij}
    \Tilde{\nu}_{ij} = 10^{\frac{P_{j}}{5\beta_0}}\left(1-\frac{P_j\epsilon\ln10}{5\beta_0}\right)\frac{\ln10}{5\beta}\sigma_{ij}.
\end{IEEEeqnarray}
Similar to the method in Section~\ref{S3}, the NLS estimator is given by
\begin{IEEEeqnarray}{cl}
    \label{S4_NLS}
    \IEEEyesnumber  \IEEEyessubnumber*
    \mathop{\rm{min}}_{\bm{\mu},\mathbf{p},\beta} & \sum_{j\in\mathcal{T} \atop i\in \mathcal{A}_j\cup\mathcal{T}_j}\left(d^2_{ij}10^{\frac{P_{ij}}{5\beta_0}} - 10^{\frac{P_{j}}{5\beta_0}} + 10^{\frac{P_{j}}{5\beta_0}}\frac{P_j\epsilon\ln10}{5\beta_0}\right)^2 \nonumber\\
    &\hspace{5cm} +\sum_{i\in \mathcal{A}}\|\mathbf{s}_i - \breve{\mathbf{s}}_i\|^2 \IEEEeqnarraynumspace\\
{\text{s.t.}} 
    & d_{ij} = \|\mathbf{t}_j-\breve{\mathbf{s}}_i\|,\ j\in\mathcal{T},\ i\in\mathcal{A}_j, \\
    & d_{ij} = \|\mathbf{t}_j-\mathbf{t}_i\|,\ j\in\mathcal{T},\ i\in\mathcal{T}_j.
\end{IEEEeqnarray}
The auxiliary variables $\mathbf{K}$ and $\lambda_i$ are used to tackle the non-convexity of~\eqref{S4_NLS}. By defining $g_j = 10^{\frac{P_{j}}{5\beta_0}}$ and $r_j = 10^{\frac{P_{j}}{5\beta_0}}\frac{P_j\epsilon\ln10}{5\beta_0}$, we obtain the following optimization problem
\begin{IEEEeqnarray}{cll} 
\label{S4_SDP_first}
\IEEEyesnumber\IEEEyessubnumber*
\mathop {{\rm{min}}}\limits_{u_{ij},g_j,r_j,\atop \lambda_i,\bm{\mu},\mathbf{K}} {\rm{ }} &  \sum_{j\in\mathcal{T} \atop i\in \mathcal{A}_j\cup\mathcal{T}_j}\left(10^{\frac{P_{ij}}{5\beta_0}}u_{ij} - g_j + r_j\right)^2 + \sum_{i\in\mathcal{A}}\lambda_{i}\IEEEeqnarraynumspace\\ 
{\text{s.t.}} 
& \eqref{General_constraint1},\ \eqref{General_constraint2},\ \eqref{General_constraint3},\ \eqref{General_constraint4}.\nonumber
\end{IEEEeqnarray}
The optimization problem~\eqref{S4_SDP_first} can be solved by standard SDP solvers, which have high computational complexity. To address this issue, the objective function is converted to the epigraph form by applying an auxiliary variable $\omega$
\begin{IEEEeqnarray}{cll} 
\label{S4_SDP_epigraph}
\IEEEyesnumber\IEEEyessubnumber*
\mathop {{\rm{min}}}\limits_{u_{ij},g_j,r_j,\atop \lambda_i,\bm{\mu},\mathbf{K},\omega} {\rm{ }} &  \, \, \omega + \sum_{i\in\mathcal{A}}\lambda_{i}\\ 
{\text{s.t.}} 
& \sum_{j\in\mathcal{T} \atop i\in \mathcal{A}_j\cup\mathcal{T}_j}\left(10^{\frac{P_{ij}}{5\beta_0}}u_{ij} - g_j + r_j\right)^2 \leq \omega, \label{S4_SDP_epigraph_constraint1} \\ 
& \eqref{General_constraint1},\ \eqref{General_constraint2},\ \eqref{General_constraint3},\ \eqref{General_constraint4}.\nonumber
\end{IEEEeqnarray}
To convert~\eqref{S4_SDP_epigraph} into an SDP-SOCP problem, we introduce $\bm{\xi}$ which stacks all values $\xi_{ij} = u_{ij}10^{\frac{P_{ij}}{5\beta_0}} - g_j + r_j, \ j \in\mathcal{T},\ i\in \mathcal{A}_j \cup \mathcal{T}_j$ to obtain an LMI 
\begin{IEEEeqnarray}{cll} 
\label{SDP_SOCP}
\IEEEyesnumber\IEEEyessubnumber*
\mathop {{\rm{min}}}\limits_{u_{ij},g_j,r_j,\atop \lambda_i,\bm{\mu},\mathbf{K},\omega,\bm{\xi}} {\rm{ }} &  \, \, \omega + \sum_{i\in\mathcal{A}}\lambda_{i}\\ 
\label{SDP_SOCP_oj}
{\text{s.t.}} 
& \left\lVert\left[2\bm{\xi}^T,\omega-1\right]\right\rVert \leq \omega + 1, \label{SDP_SOCP_constraint1} \\ 
& \text{diag}\big(\big[\dots, \xi_{ij}-10^{\frac{P_{ij}}{5\beta_0}}u_{ij}+g_j-r_j,\dots,  \nonumber \\
& \hspace{.5cm} \dots,-\xi_{ij}+10^{\frac{P_{ij}}{5\beta_0}}u_{ij}-g_j+r_j,\dots \big]\big) \succcurlyeq\mathbf{0},\nonumber\\
& \hspace{3.5cm} j \in \mathcal{T},\ i\in \mathcal{A}_j \cup \mathcal{T}_j, \IEEEeqnarraynumspace\label{SDP_SOCP_constraint2} \\ 
& \eqref{General_constraint1},\ \eqref{General_constraint2},\ \eqref{General_constraint3},\ \eqref{General_constraint4}.\nonumber
\end{IEEEeqnarray}
The proposed estimator in~\eqref{SDP_SOCP} can obtain the location of target nodes $(\mathbf{t})$, their transmit powers $(\mathbf{p})$, and $\beta$. However, the estimated accuracy of $\mathbf{p}$ and $\beta$ is poor. We obtain a better estimate of $\mathbf{p}$ and $\beta$ by exploiting the fact that~\eqref{MLEstimator} is a convex function with respect to $\mathbf{p}$ and $\beta$, respectively
\begin{IEEEeqnarray}{cll}
    \label{MLPtBeta}
    \IEEEyesnumber\IEEEyessubnumber*
    &\hat{P}_j = \frac{\mathbf{1}_{\lvert\mathcal{A}_j\cup\mathcal{T}_j\rvert}^T\mathbf{Q}\mathbf{h}_j}{\mathbf{1}_{\lvert\mathcal{A}_j\cup\mathcal{T}_j\rvert}^T\mathbf{Q}\mathbf{1}_{\lvert\mathcal{A}_j\cup\mathcal{T}_j\rvert}},\label{ML_Pt}\\
    &\hat{\beta} = \frac{\mathbf{1}_{\lvert\mathcal{H}\rvert}^T\mathbf{Q}\mathbf{q}}{\mathbf{1}_{\lvert\mathcal{H}\rvert}^T\mathbf{Q}\bm{\phi}}, \IEEEeqnarraynumspace \label{ML_Beta}
\end{IEEEeqnarray}
where $\mathbf{Q}$ is the covariance matrix of $n_{ij}$, $\mathbf{h}_j$ stacks $P_{ij}+10\beta\log_{10}d_{ij}$, $\mathbf{q}$ stacks $P_j-P_{ij}$, and $\bm{\phi}$ stacks $10\log_{10}d_{ij}$, $j\in\mathcal{T}$, $i\in\mathcal{A}_j\cup\mathcal{T}_j$. 
The total number of anchor-target links and target-target links are expressed as
$\lvert\mathcal{H}\rvert = \sum_{j}\lvert\mathcal{A}_j\cup\mathcal{T}_j\rvert$. In Algorithm~\ref{Alg:CTUP}, we summarize The proposed technique dealing with the scenario for unknown transmit power and PLE. We refer to it as CTUP-4.

\begin{figure*}
\begin{IEEEeqnarray}{c}
    \label{FIM1}
    \mathbf{F}(\bm{\theta}) = \begin{bmatrix}
        -\mathbbm{E}\left[\frac{\partial^2 \ln p(\mathbf{m};\bm{\theta})}{\partial \bm{\theta}^2_1}\right] & \cdots & -\mathbbm{E}\left[\frac{\partial^2 \ln p(\mathbf{m};\bm{\theta})}{\partial \bm{\theta}_1\partial \bm{\theta}_{M_1}}\right]\\
        \vdots & \ddots & \vdots \\
        -\mathbbm{E}\left[\frac{\partial^2 \ln p(\mathbf{m};\bm{\theta})}{\partial \bm{\theta}_{M_1} \partial \bm{\theta}_{1}}\right]& \cdots & -\mathbbm{E}\left[\frac{\partial^2 \ln p(\mathbf{m};\bm{\theta})}{\partial \bm{\theta}^2_{M_1}}\right] 
    \end{bmatrix} 
    = \begin{bmatrix}
        \mathbbm{E}\left[\left(\frac{\partial \ln p(\mathbf{m};\bm{\theta})}{\partial \bm{\theta}_1}\right)^2\right] & \cdots & \mathbbm{E}\left[\frac{\partial \ln p(\mathbf{m};\bm{\theta})}{\partial \bm{\theta}_1}\cdot\frac{\partial \ln p(\mathbf{m};\bm{\theta})}{\partial \bm{\theta}_{M_1}} \right]\\
        \vdots & \ddots & \vdots \\
        \mathbbm{E}\left[\frac{\partial \ln p(\mathbf{m};\bm{\theta})}{\partial \bm{\theta}_1}\cdot\frac{\partial \ln p(\mathbf{m};\bm{\theta})}{\partial \bm{\theta}_{M_1}} \right] & \cdots & \mathbbm{E}\left[\left(\frac{\partial \ln p(\mathbf{m};\bm{\theta})}{\partial \bm{\theta}_{M_1}}\right)^2\right]
    \end{bmatrix}\IEEEeqnarraynumspace
\end{IEEEeqnarray}
\hrulefill
\end{figure*}

% %
% \begin{algorithm}[!t]
% \caption{CTUP-4: The cooperative technique with unknown transmit power and PLE} 
% \label{Alg:CTUP}
% 	\begin{algorithmic}[1]
%         \State $\beta_0$ is obtained from~\eqref{PreBeta}.
%         \State Solve~\eqref{SDP_SOCP} to get $\hat{\bm{\mu}}$ and then compute $\hat{\mathbf{t}}_j = \hat{\bm{\mu}}\left(2j-1:2j\right)^T$.
%         \State Insert $\beta_0$ into~\eqref{ML_Pt} to obtain $\hat{\mathbf{p}}$ as the estimate of $\mathbf{p}$.
% 	\State Apply $\hat{\mathbf{p}}$ into~\eqref{ML_Beta} to get $\hat{\beta}$ to be the estimate of PLE.
% 	\end{algorithmic} 
% \end{algorithm}
% %
\renewcommand{\algorithmicrequire}{\textbf{Input:}}
\renewcommand{\algorithmicensure}{\textbf{Output:}}
\begin{algorithm}[!t]
\caption{CTUP-4: The cooperative technique with unknown transmit power and PLE} 
\label{Alg:CTUP}
    \begin{algorithmic}[1]
        \REQUIRE{$N_a,\ N_t,\ \mathcal{A}_j,\ \mathcal{T}_j,\ \mathcal{S}_j,\ \mathbf{m},\ \breve{\mathbf{Q}},\ \breve{\mathbf{q}},\ \breve{\bm{\phi}},\ \mathbf{Q}.$}
        \STATE Compute $\beta_0$ using~\eqref{PreBeta}.
        \STATE Solve the SDP-SOCP problem in~\eqref{SDP_SOCP} to obtain $\hat{\mathbf{t}}_j = \hat{\bm{\mu}}\left(2j-1:2j\right)^T$ and $\hat{d}_{ij} = \sqrt{\hat{u}_{ij}}$, $j\in\mathcal{T}$, $i\in\mathcal{A}_j\cup\mathcal{T}_j$.
        \STATE Compute $\mathbf{h}_j = P_{ij}+10\beta_0\log_{10}\hat{d}_{ij}$, $j\in\mathcal{T},\ i\in\mathcal{A}_j\cup\mathcal{T}_j$.
        \STATE Compute $\hat{P}_j$ using~\eqref{ML_Pt}, $j\in\mathcal{T}$.
        \STATE Compute $q_{ij} = \hat{P}_j-P_{ij}$ and $\phi_{ij} = 10\log_{10}\hat{d}_{ij}$, $j\in\mathcal{T}$, $i\in\mathcal{A}_j\cup\mathcal{T}_j$. $\mathbf{q}$ comprises all $q_{ij}$, and $\bm{\phi}$ comprises all $\phi_{ij}$.
        \STATE Compute $\hat{\beta}$ using~\eqref{ML_Beta}.
        \ENSURE{$\hat{\mathbf{t}}_j,\ \hat{P}_j,\ \hat{\beta},\ j\in\mathcal{T}$.}
    \end{algorithmic}
\end{algorithm}
\section{Cramer-Rao lower bounds (CRLB)}\label{sec:CRLB}
% CRLB provides a lower bound for the variance of unbiased estimators~\cite{kay1993fundamentals} and is widely used as a benchmark to evaluate the performance of localization techniques. 
\textcolor{blue}{The CRLB provides a theoretical lower bound for unbiased estimators, which serves as an efficient performance benchmark for researchers to identify optimal performance~\cite{MoeWin3, kay1993fundamentals}.} Generally, authors~\cite{SDP_Tomic, MSL, SDP_Zou} have derived the CRLB for location estimates and specific model parameters, but they have not addressed scenarios where all model parameters are jointly considered. In this section, we provide the CRLB for the Root Mean Square Error (RMSE) of cooperative RSS-based localization. We encompass scenarios involving unknown transmit power, unknown PLE, and anchor location uncertainty. The fisher information matrix (FIM) of $\bm{\theta}$ is given in~\eqref{FIM1}, where $\bm{\theta}_{j}$ denotes the $j^{\text{th}}$ element of $\bm{\theta}$, and $M_1 = 3N_t+2N_a+1$. The first and second moments of the RSS and anchor location measurements are given by
\begin{IEEEeqnarray}{rll}
    \label{Moments}
    \IEEEyesnumber\IEEEyessubnumber*
    \mathbbm{E}&\left[P_{ij}\right] = P_j - 10\beta\log_{10}d_{ij},\quad &\text{Var}\left(P_{ij}\right) = \sigma_{ij}^2,\IEEEeqnarraynumspace\\
    \mathbbm{E}&\left[\mathbf{s}_i\right] = \breve{\mathbf{s}}_i,\quad &\text{Var}\left(\mathbf{s}\right) = \delta_{i}^2,\IEEEeqnarraynumspace
\end{IEEEeqnarray}
Thus, substituting~\eqref{PDF_theta} and~\eqref{Moments} into~\eqref{FIM1} yields
\begin{IEEEeqnarray}{ccl}
    \label{FIM2}
    \mathbf{F}(\bm{\theta}) = \mathbf{C}^T\mathbf{Q}\mathbf{C} + \sum_{i\in\mathcal{A}} \delta_i^{-2}\mathbf{R}_i^T\mathbf{R}_i,\IEEEeqnarraynumspace
\end{IEEEeqnarray}
where
\begin{IEEEeqnarray*}{cll}
    \label{FIM3}
    \mathbf{C} &= \begin{bmatrix}
                    \vdots & \vdots & \vdots & \vdots \\
                    \mathbf{c}^{t}_{ij} & \mathbf{c}^{\breve{{s}}}_{ij} & \mathbf{c}^{p}_{ij} & \mathbf{c}^\beta_{ij} \\
                    \vdots & \vdots & \vdots & \vdots
                    \end{bmatrix},\ j\in\mathcal{T},\\
    \mathbf{c}^{t}_{ij} &= \left[\mathbf{0}_{1,2\left(j-1\right)},\frac{10\beta}{\ln10}\frac{\left(\breve{\mathbf{s}}_{i}-\mathbf{t}_j\right)^T}{\|\breve{\mathbf{s}}_{i}-\mathbf{t}_j\|^2},\mathbf{0}_{1,2\left(N_t-j\right)}\right],\ i\in\mathcal{A}_j,\\
    \mathbf{c}^{t}_{ij} &= \left[\mathbf{0}_{1,2\left(j-1\right)},\mathbf{b}_{ij}^T,\mathbf{0}_{1,2\left(i-j-1\right)},-\mathbf{b}_{ij}^T,\mathbf{0}_{1,2\left(N_t-i\right)}\right],\nonumber\\
    &\hspace{5.65cm} i>j,\ i\in\mathcal{T}_j, \\
    \mathbf{c}^{t}_{ij} &= \left[\mathbf{0}_{1,2\left(i-1\right)},\mathbf{b}_{ij}^T,\mathbf{0}_{1,2\left(j-i-1\right)},-\mathbf{b}_{ij}^T,\mathbf{0}_{1,2\left(N_t-j\right)}\right],\nonumber\\
    &\hspace{5.65cm} i<j,\ i\in\mathcal{T}_j, \\
    \mathbf{c}^{\breve{{s}}}_{ij} &= \left[\mathbf{0}_{1,2\left(i-1\right)},\frac{10\beta}{\ln10}\frac{\left(\mathbf{t}_j-\breve{\mathbf{s}}_{i}\right)^T}{\|\breve{\mathbf{s}}_{i}-\mathbf{t}_j\|^2},\mathbf{0}_{1,2\left(N_a-i\right)}\right],\ i\in\mathcal{A}_j,\\
    \mathbf{c}^{\breve{{s}}}_{ij} &= \mathbf{0}_{1,2N_a},\ i\in\mathcal{T}_j,\\
    \mathbf{c}^{p}_{ij} &= \left[\mathbf{0}_{1,j-1},1,\mathbf{0}_{1,N_t-j}\right],\ i\in\mathcal{A}_j\cup\mathcal{T}_j,\\
    \mathbf{c}^\beta_{ij} &= -10\log_{10}\|\mathbf{t}_j-\breve{\mathbf{s}}_{i}\|,\ i\in\mathcal{A}_j,\\
    \mathbf{c}^\beta_{ij} &= -10\log_{10}\|\mathbf{t}_j-\mathbf{t}_{i}\|,\ i\in\mathcal{T}_j,\\
    \mathbf{b}_{ij} &= \frac{10\beta}{\ln10}\frac{\mathbf{t}_{i}-\mathbf{t}_j}{\|\mathbf{t}_{i}-\mathbf{t}_j\|^2},\ i\in\mathcal{T}_j,\\
    \mathbf{R} &= \left[\mathbf{0}_{2,2N_t+2\left(i-1\right)},\mathbf{I}_2,\mathbf{0}_{2,2\left(Na-i\right)+N_t+1}\right],\ i\in\mathcal{A}.
\end{IEEEeqnarray*}
Define $M_2 = 2N_t+2N_a$. Now, the lower bounds of the variance of the estimated parameters are shown as follows:
\begin{IEEEeqnarray}{rcl}
    \label{VarLB}
    \IEEEyesnumber\IEEEyessubnumber*
    \text{Var}\left(\hat{\mathbf{t}}_j\right) &\geq& \ \text{tr}\left(\left[\mathbf{F}^{-1}\left(\bm{\theta}\right)\right]_{2j-1:2j,2j-1:2j}\right),\ j\in\mathcal{T},\IEEEeqnarraynumspace\\
    \text{Var}\left(\hat{P}_j\right) &\geq& \left[\mathbf{F}^{-1}\left(\bm{\theta}\right)\right]_{M_2+j,M_2+j},\ j\in\mathcal{T},\IEEEeqnarraynumspace\\
    \text{Var}\left(\hat{\beta}\right) &\geq& \left[\mathbf{F}^{-1}\left(\bm{\theta}\right)\right]_{M_1,M_1}.
\end{IEEEeqnarray}
It is important to obtain the RMSE of the location and transmit power for all target nodes. In numerous practical scenarios, understanding the system's overall performance holds greater significance than concentrating solely on individual nodes.
Let the RMSE of estimated parameters be
\begin{IEEEeqnarray}{rcl}
    \label{RMSE_def}
    \IEEEyesnumber\IEEEyessubnumber*
    \text{RMSE}\left(\hat{\mathbf{t}}\right) &=& \sqrt{\frac{1}{N_t}\sum_{j\in\mathcal{T}}\|\hat{\mathbf{t}}_j-\mathbf{t}_j\|^2},\\
    \text{RMSE}\left(\hat{\mathbf{p}}\right) &=& \sqrt{\frac{1}{N_t}\sum_{j\in\mathcal{T}}\|\hat{P}_j-P_j\|^2},\\
    \text{RMSE}\left(\hat{\beta}\right) &=& \sqrt{\|\hat{\beta}-\beta\|^2}.
\end{IEEEeqnarray}
Using~\eqref{VarLB}, we can express the CRLB of the estimated parameters as 
\begin{IEEEeqnarray}{rcl}
    \label{CRLB_final}
    \IEEEyesnumber\IEEEyessubnumber*
    \text{RMSE}\left(\hat{\mathbf{t}}\right) &\geq& \sqrt{\frac{1}{N_t}\text{tr}\left(\left[\mathbf{F}^{-1}\left(\bm{\theta}\right)\right]_{1:2N_t,1:2N_t}\right)}\triangleq \text{CRLB}_{{t}},\IEEEeqnarraynumspace\\
    \text{RMSE}\left(\hat{\mathbf{p}}\right) &\geq& \sqrt{\frac{1}{N_t}\text{tr}\left(\left[\mathbf{F}^{-1}\left(\bm{\theta}\right)\right]_{M_2+1:M_2+N_t,M_2+1:M_2+N_t}\right)}\nonumber\\
    &\triangleq & \ \text{CRLB}_{{p}},\\
    \text{RMSE}\left(\hat{\beta}\right) &\geq& \sqrt{\left[\mathbf{F}^{-1}\left(\bm{\theta}\right)\right]_{M_1,M_1}}\triangleq \text{CRLB}_{\beta}.
\end{IEEEeqnarray}
The CRLB for Section~\ref{S1}, Section~\ref{S2}, and Section~\ref{S3} can be similarly calculated by considering $\bm{\theta} = \left[\mathbf{t}^T,\breve{\mathbf{s}}^T\right]^T$, $\bm{\theta} = \left[\mathbf{t}^T,\breve{\mathbf{s}}^T,\beta\right]^T$, and $\bm{\theta} = \left[\mathbf{t}^T,\breve{\mathbf{s}}^T,\mathbf{p}^T\right]^T$, respectively.

\section{Performance analysis}\label{sec:Numerical}
In this section, the performance of CTUP is 
compared with current techniques (refer Table~\ref{Tb:ExistingTech}) with respect to accuracy in estimating location, transmit power, and PLE, alongside computational complexity. The estimation accuracy is verified using both numerical simulations and real-field experimentation. The convergence criterion for iterative techniques (RWLS-AE, RLBM, and IRGDL) is $\left|\mathbf{J}^k-\mathbf{J}^{k-1}\right| \leq 10^{-3}$, with $\mathbf{J}^k$ denoting the values of the corresponding objective functions. IRGDL undergoes random initialization, and CTUP-X, along with other SDP-based methods, is implemented using the CVX toolbox, leveraging SeDuMi with the best precision. We choose the normalized root mean squared error (NRMSE) as a performance benchmark
\begin{IEEEeqnarray}{ccl}
    \label{NRMSE}
    \text{NRMSE}_{t} &=& \sqrt{\frac{1}{N_tM_c}\sum_{k=1}^{M_c}\sum_{j\in\mathcal{T}}\|\hat{\mathbf{t}}_j^k-\mathbf{t}_j\|^2},\nonumber\\
    \text{NRMSE}_{p} &=& \sqrt{\frac{1}{N_tM_c}\sum_{k=1}^{M_c}\sum_{j\in\mathcal{T}}\|\hat{P}_j^k-P_j\|^2},\nonumber\\
    \text{NRMSE}_{\beta} &=& \sqrt{\frac{1}{M_c}\sum_{k=1}^{M_c}\|\hat{\beta}^k-\beta\|^2},\nonumber
\end{IEEEeqnarray}
where $M_c$ is the number of Monte-Carlo simulations. All the simulation results presented in the paper are obtained using 3000 Monte-Carlo simulations.

\subsection{Numerical simulations}\label{Simulations}
The performance of localization techniques is evaluated for four scenarios discussed in Section~\ref{sec:Algorithm}. We employ two networks, denoted as NW-1 and NW-2, to study the impact of noise and anchor location uncertainty, respectively. NW-1 consists of 5 anchor nodes and 10 target nodes, and NW-2 consists of 20 anchor nodes and 10 target nodes. In both NW-1 and NW-2, the target nodes and anchor nodes are randomly deployed over an area of $100\times100$~$m^2$, with fixed locations throughout the simulations. NW-1 and NW-2 are assumed to be fully connected in the simulations, enabling all anchor nodes and target nodes to communicate interchangeably. The transmit power of target nodes is randomly selected from a uniform distribution between [-10 10]~dBm, and the path loss exponent $(\beta)$ in~\eqref{LogNormal} is set to 3. Without loss of generality, we assume that $\sigma_{ij} = \sigma$ and $\delta_i = \delta$~\cite{Vaghefi_cooperative, DeltaSame}. All the results presented in Section~\ref{Sim:S1}, Section~\ref{Sim:S2}, Section~\ref{Sim:S3}, and Section~\ref{Sim:S4} are obtained using NW-1 and the simulation results shown in Section~\ref{Sim:AU}  are based on NW-2.

\begin{figure}
    \centering
    \includegraphics[width=\linewidth]{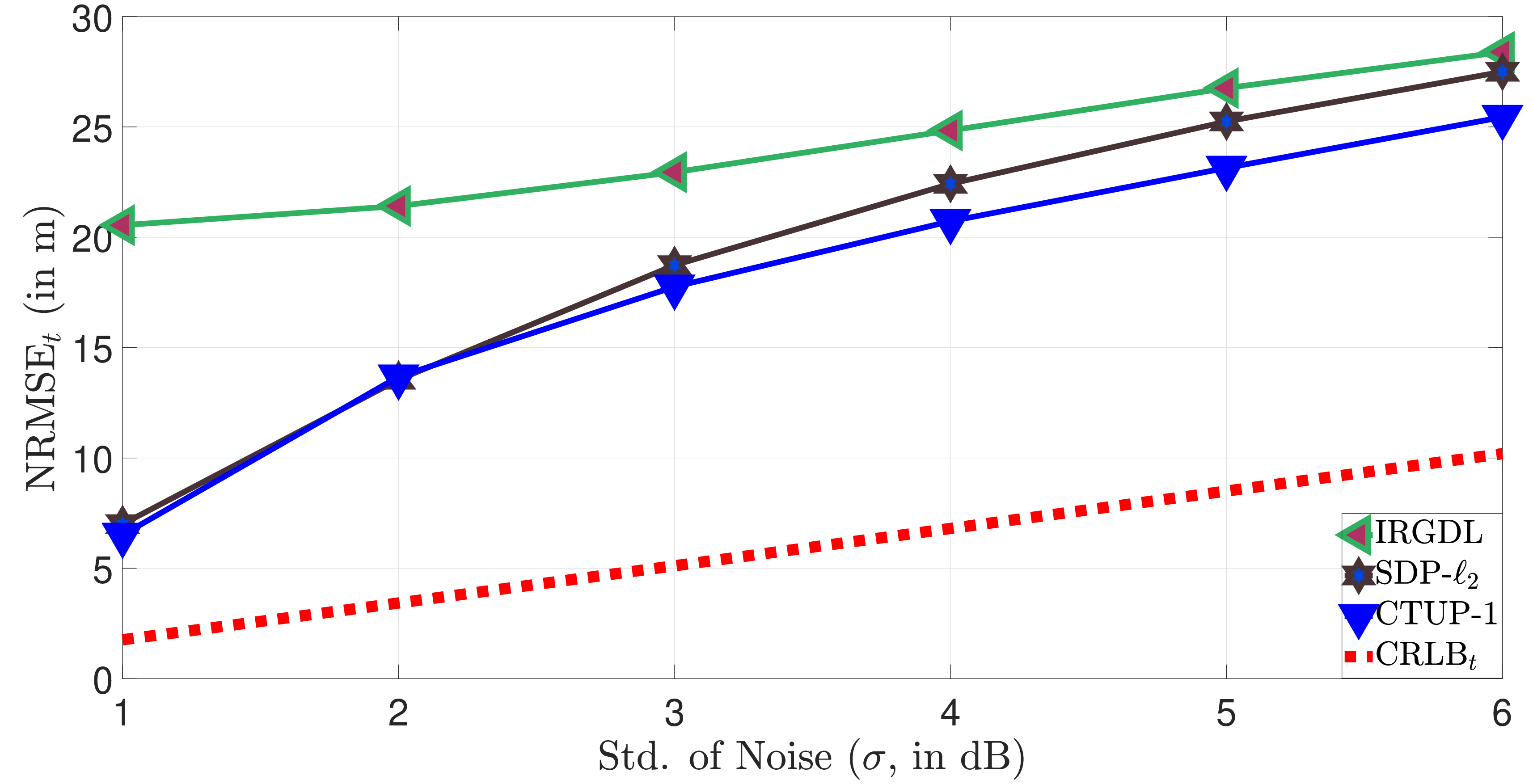}
    \caption{Performance of the localization techniques for scenarios where transmit power and PLE are known ($\delta = 3$~m).}
    \label{Fig:S1_Loc}
\end{figure}
\begin{figure*}[!t]
    \centering
    \subfigure[NRMSE of location estimate as a function of $\sigma$.]{
    \begin{minipage}[t]{0.47\linewidth}
    \centering
    \includegraphics[width=\linewidth]{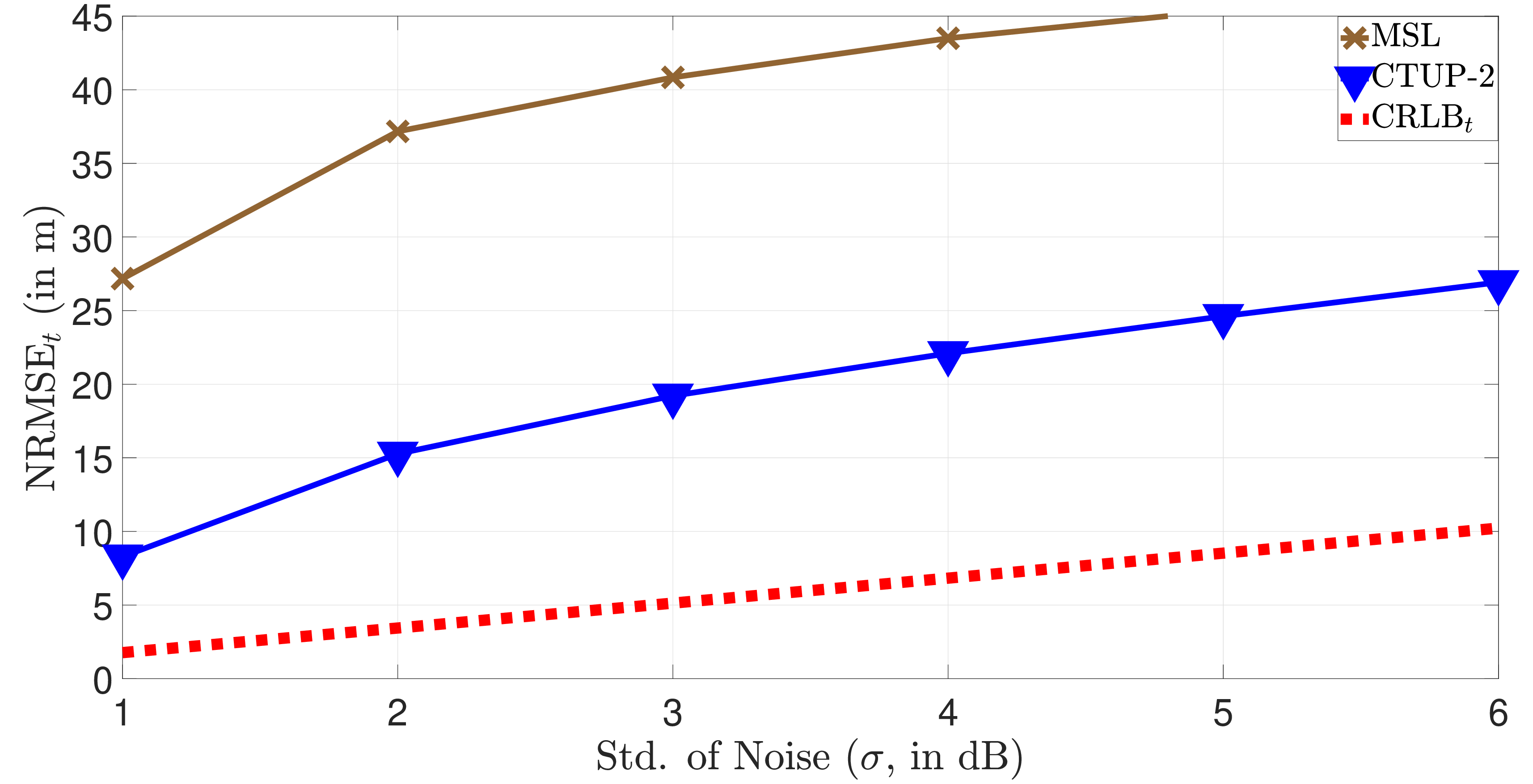}
    \label{Fig:S2_Loc_noise}
    \end{minipage}
    }
    \subfigure[NRMSE of PLE estimate as a function of $\sigma$.]{
    \begin{minipage}[t]{0.47\linewidth}
    \centering
    \includegraphics[width=\linewidth]{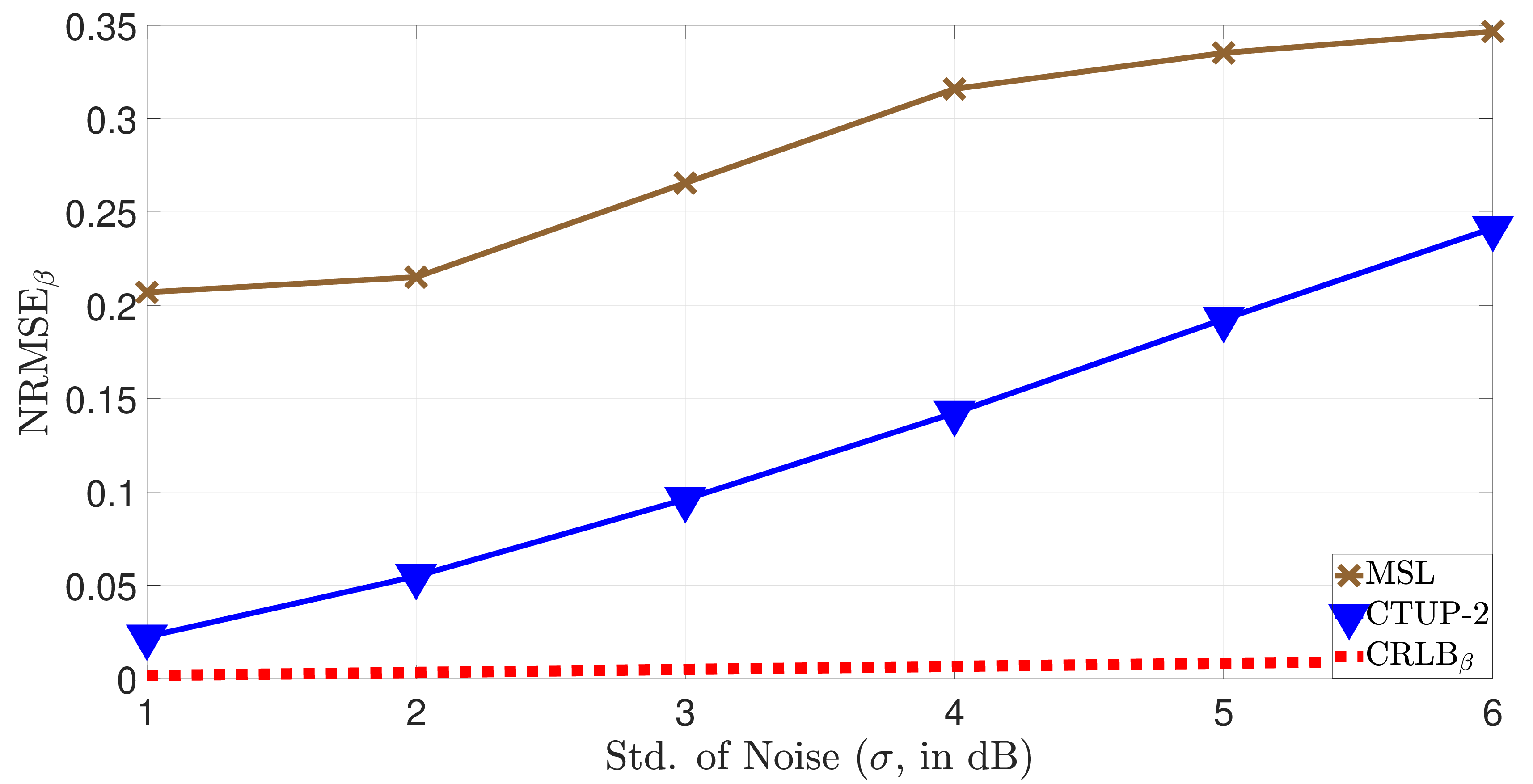}
    \label{Fig:S2_PLE_noise}
    \end{minipage}
    }
    \caption{Performance of the localization techniques for scenarios where PLE is unknown ($\delta = 3$~m).}
    \label{Fig:S2}
\end{figure*}
\begin{figure*}[!t]
    \centering
    \subfigure[NRMSE of location estimate as a function of $\sigma$.]{
    \begin{minipage}[t]{0.47\linewidth}
    \centering
    \includegraphics[width=\linewidth]{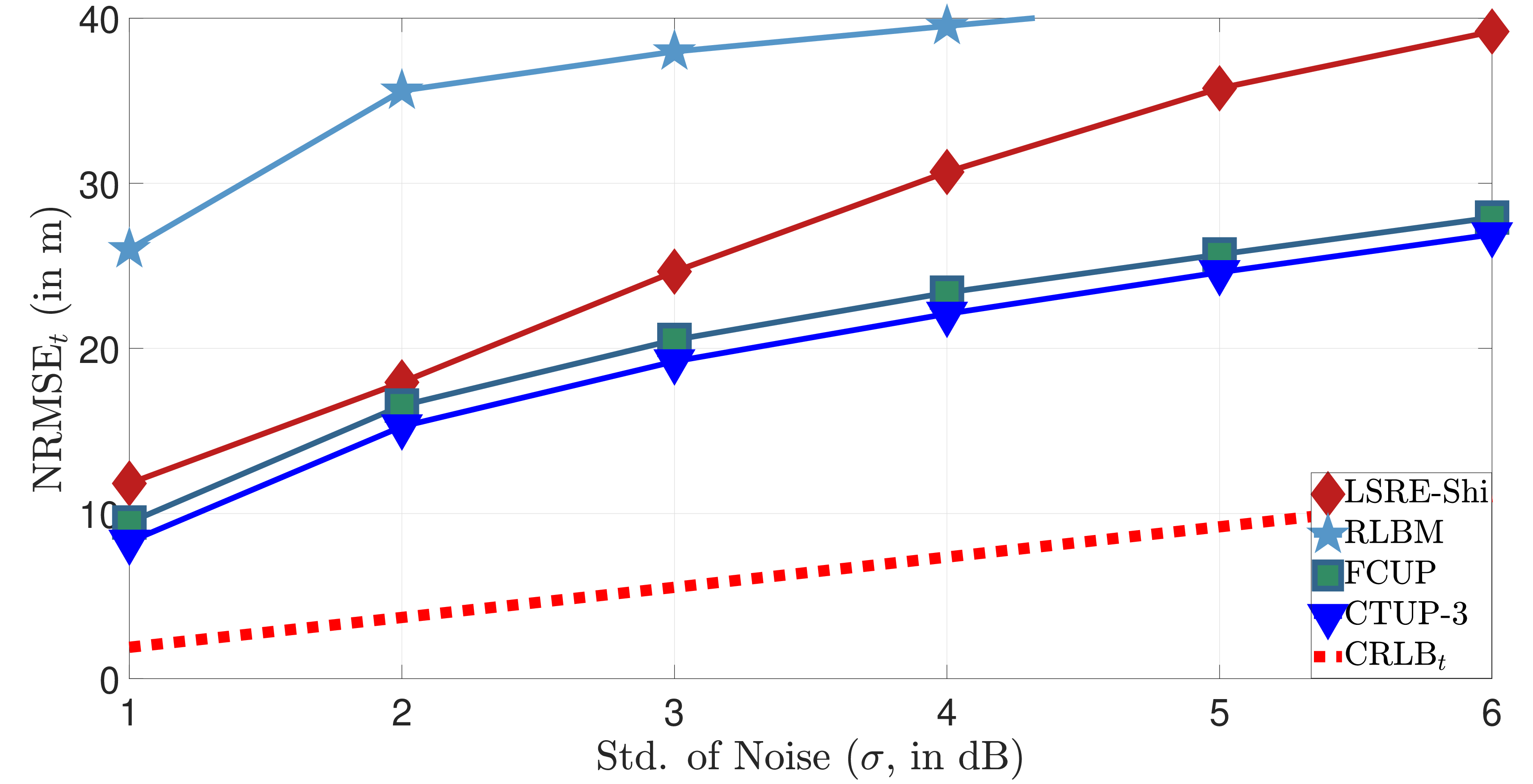}
    \label{Fig:S3_Loc_noise}
    \end{minipage}
    }
    \subfigure[NRMSE of transmit power estimate as a function of $\sigma$.]{
    \begin{minipage}[t]{0.47\linewidth}
    \centering
    \includegraphics[width=\linewidth]{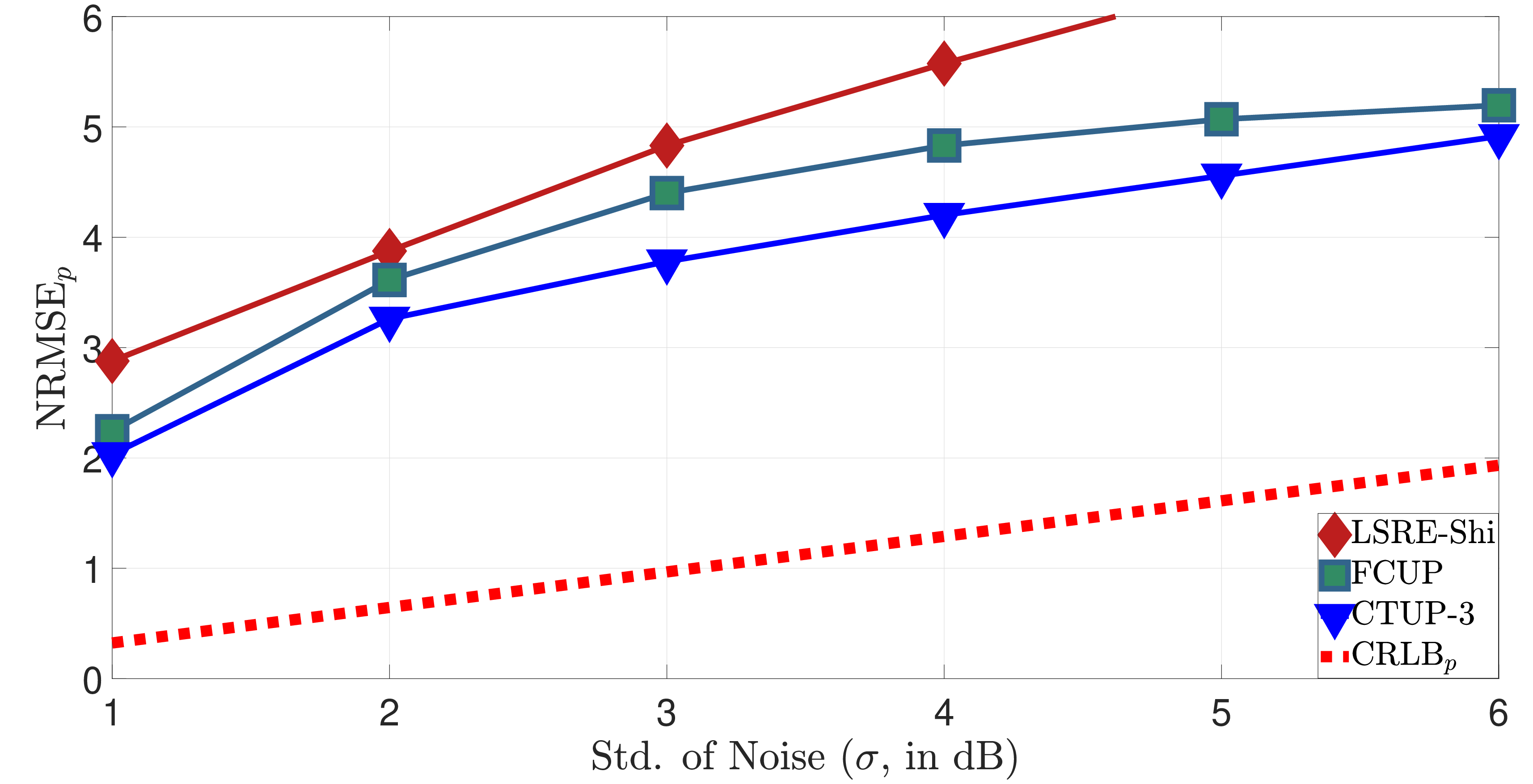}
    \label{Fig:S3_power_noise}
    \end{minipage}
    }
    \caption{Performance of the localization techniques for scenarios where transmit power is unknown ($\delta = 3$~m).}
    \label{Fig:S3}
\end{figure*}
\begin{table*}[!t]
\centering
\caption{Computational complexity of existing techniques.}
% \resizebox{\textwidth}{!}
{\begin{tabular}{|c|c|c|cc|c|}
\hline
\multirow{2}{*}{Algo.} & \multirow{2}{*}{Manner}      & \multirow{2}{*}{Computational complexity}                                                                                                                                                         & \multicolumn{2}{c|}{CPU runtime (in s)} & \multirow{2}{*}{Year} \\ \cline{4-5}
                       &                              &                                                                                                                                                                                                   & \multicolumn{1}{c|}{$\text{T}_1$}      & $\text{T}_2$       &                       \\ \hline
LSRE-Shi               & Non-cooperative              & $\mathcal{O}\big(N_t\sqrt{N_a+2}(N_a+4)^3\big)$                                                                                                                                                   & \multicolumn{1}{c|}{12.41}   & 13.17    & 2020                  \\ \hline
RWLS-AE                & Non-cooperative              & $\mathcal{O}\big(N_tN_a^{6.5}\sum_{j\in\mathcal{T}}i_j^{\text{itr}}\big)$                                                                                                                                   & \multicolumn{1}{c|}{92.03}   & 153.06   & 2021                  \\ \hline
SDP-Zou                & Non-cooperative              & $\mathcal{O}\big((N_t+N_a)N_tN_a^{6.5}\big)$                                                                                                                                                     & \multicolumn{1}{c|}{18.41}   & 94.71    & 2021                  \\ \hline
RLBM                   & Non-cooperative              & $\mathcal{O}\big(N_t(N_a^4+7N_a^2)\sum_{j\in\mathcal{T}}i_j^{\text{itr}}\big)$                                                                                                                             & \multicolumn{1}{c|}{11.02}   & 11.63    & 2021                  \\ \hline
MSL                    & Non-cooperative              & $\mathcal{O}\big(N_t\big)$                                                                                                                                                                        & \multicolumn{1}{c|}{0.001}   & 0.002    & 2022                  \\ \hline
IRGDL                  & Cooperative                  & $\mathcal{O}\big(\lvert\mathcal{H}\rvert k^{itr}\big)$                                                                                                                                             & \multicolumn{1}{c|}{0.07}    & 0.07     & 2022                  \\ \hline
SDP-$\ell_2$                 & Cooperative                  & $\mathcal{O}\big(\sqrt{2N_t(N_t+N_a)}(2N_t)^4(N_t+N_a)^4\big)$                                                                                                                                    & \multicolumn{1}{c|}{9.06}    & 17.75    & 2022                  \\ \hline
FCUP                   & Cooperative                  & $\mathcal{O}\Big(N_t^{0.5}\big(\lvert\mathcal{H}\rvert^2N_a(N_t+2)^2 + \lvert\mathcal{H}\rvert (N_t+2)^3\big)\Big)$                                                                                & \multicolumn{1}{c|}{2.29}    & 2.68     & 2023                  \\ \hline
CTUP-1                 & \multirow{4}{*}{Cooperative} & \multirow{4}{*}{$\mathcal{O}\Big(N_t^{0.5}\big(\lvert\mathcal{H}\rvert^4+\lvert\mathcal{H}\rvert^2\left(N_t+2\right)^2 + \lvert\mathcal{H}\rvert (N_t+2)^3 + \lvert\mathcal{H}\rvert^3\big)\Big)$} & \multicolumn{1}{c|}{2.74}    & 3.88     & \multirow{4}{*}{2023} \\ \cline{1-1} \cline{4-5}
CTUP-2                 &                              &                                                                                                                                                                                                   & \multicolumn{1}{c|}{2.81}    & 4.12     &                       \\ \cline{1-1} \cline{4-5}
CTUP-3                 &                              &                                                                                                                                                                                                   & \multicolumn{1}{c|}{2.79}    & 4.08     &                       \\ \cline{1-1} \cline{4-5}
CTUP-4                 &                              &                                                                                                                                                                                                   & \multicolumn{1}{c|}{2.87}    & 4.21     &                       \\ \hline
\end{tabular}}
\label{Tb:complexity}
\end{table*}
\begin{figure}[!t]
    \centering
    \subfigure[NRMSE of location estimate as a function of $\sigma$.]{
    \begin{minipage}[t]{.97\linewidth}
    \centering
    \includegraphics[width=\linewidth]{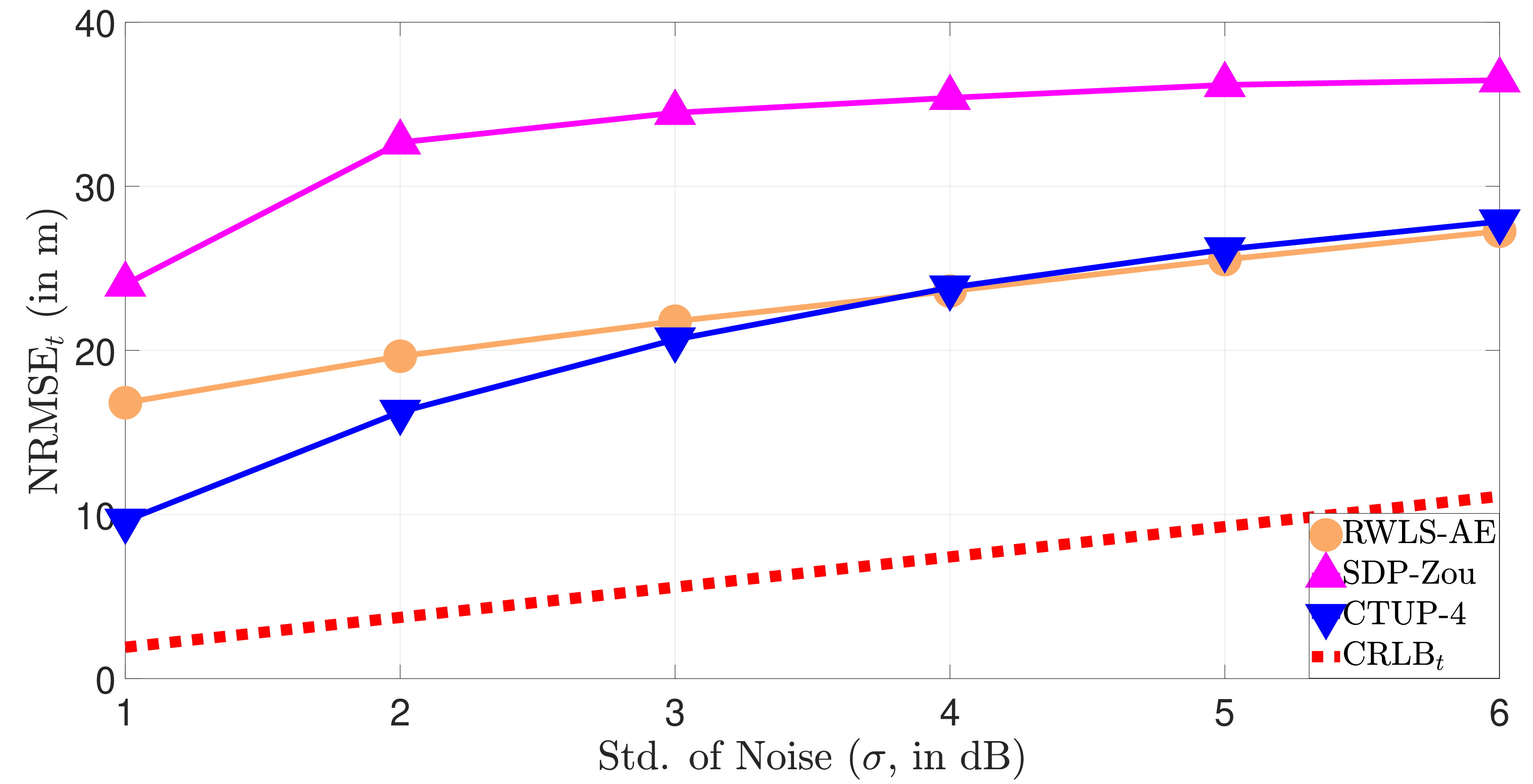}
    \label{Fig:S4_Loc_noise}
    \end{minipage}
    }
    \subfigure[NRMSE of transmit power estimate as a function of $\sigma$.]{
    \begin{minipage}[t]{.97\linewidth}
    \centering
    \includegraphics[width=\linewidth]{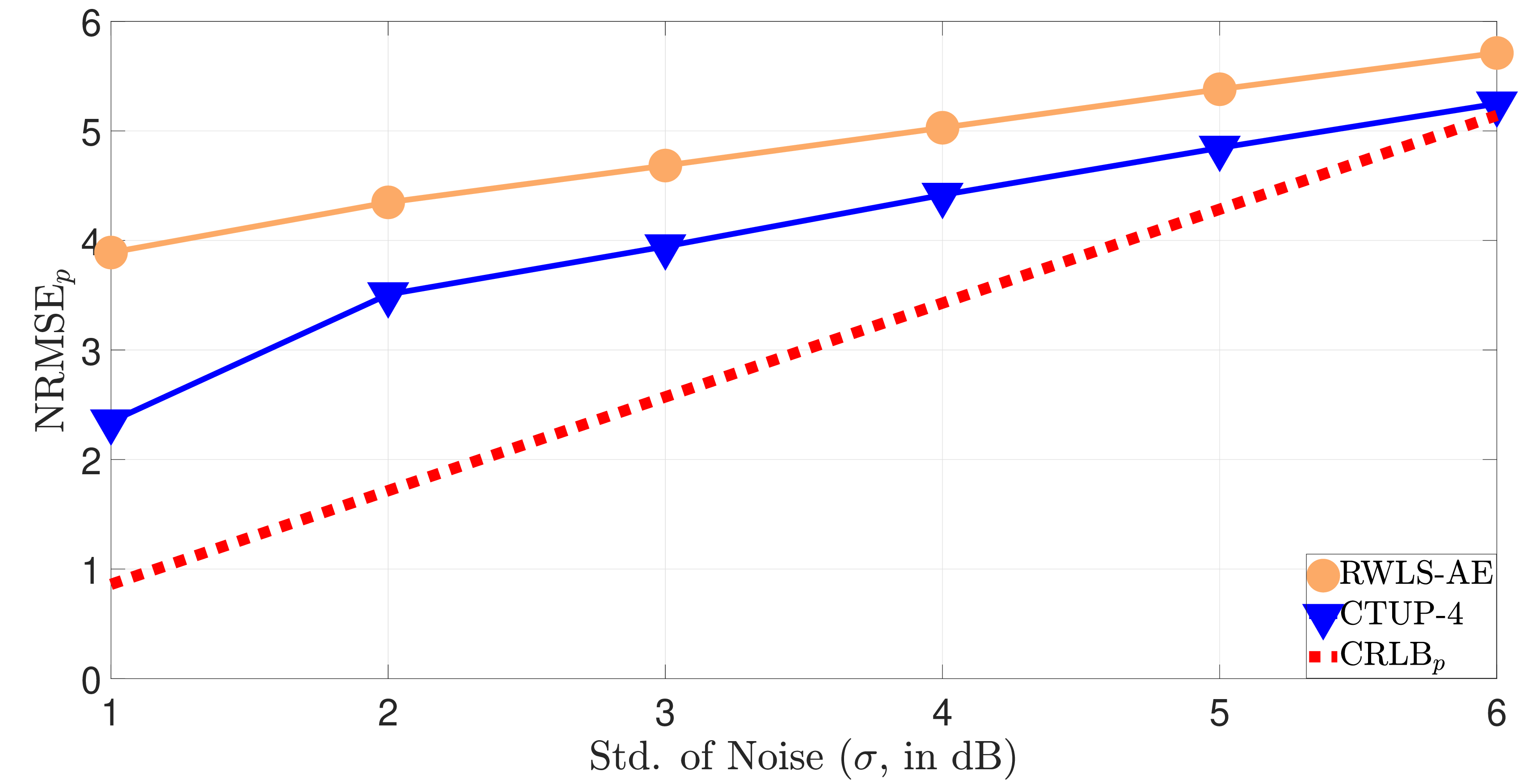}
    \label{Fig:S4_power_noise}
    \end{minipage}
    }
    \subfigure[NRMSE of PLE estimate as a function of $\sigma$.]{
    \begin{minipage}[t]{.97\linewidth}
    \centering
    \includegraphics[width=\linewidth]{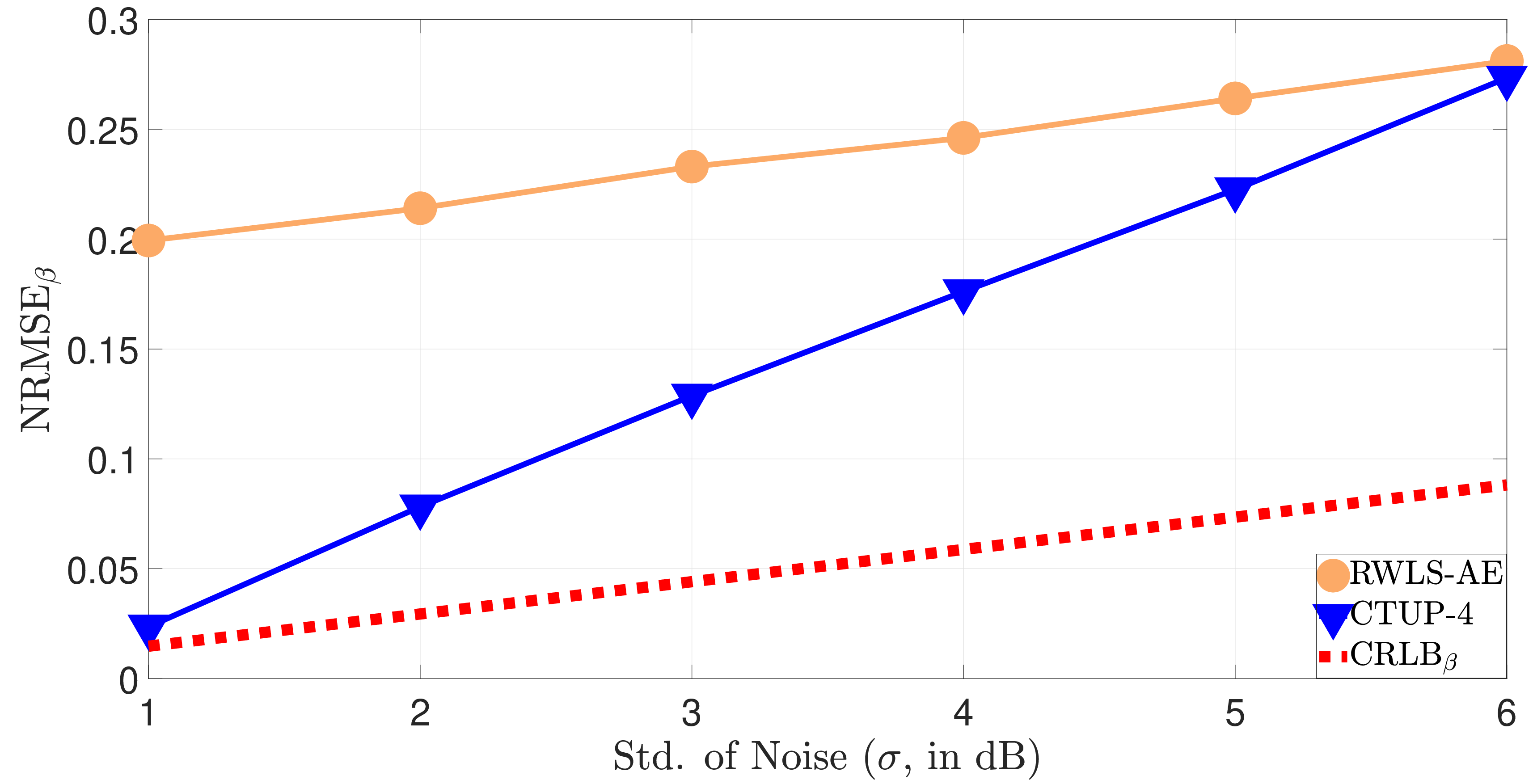}
    \label{Fig:S4_PLE_noise}
    \end{minipage}
    }
    \caption{Performance of the localization techniques for scenarios where transmit power and PLE are unknown ($\delta = 3$~m).}
    \label{Fig:S4}
\end{figure}
%
% Anchor uncertainty
\begin{figure*}[!t]
    \subfigure[NRMSE of location estimate as a function of $\delta$.]{
    \begin{minipage}[t]{0.47\linewidth}
    \centering
    \includegraphics[width=\linewidth]{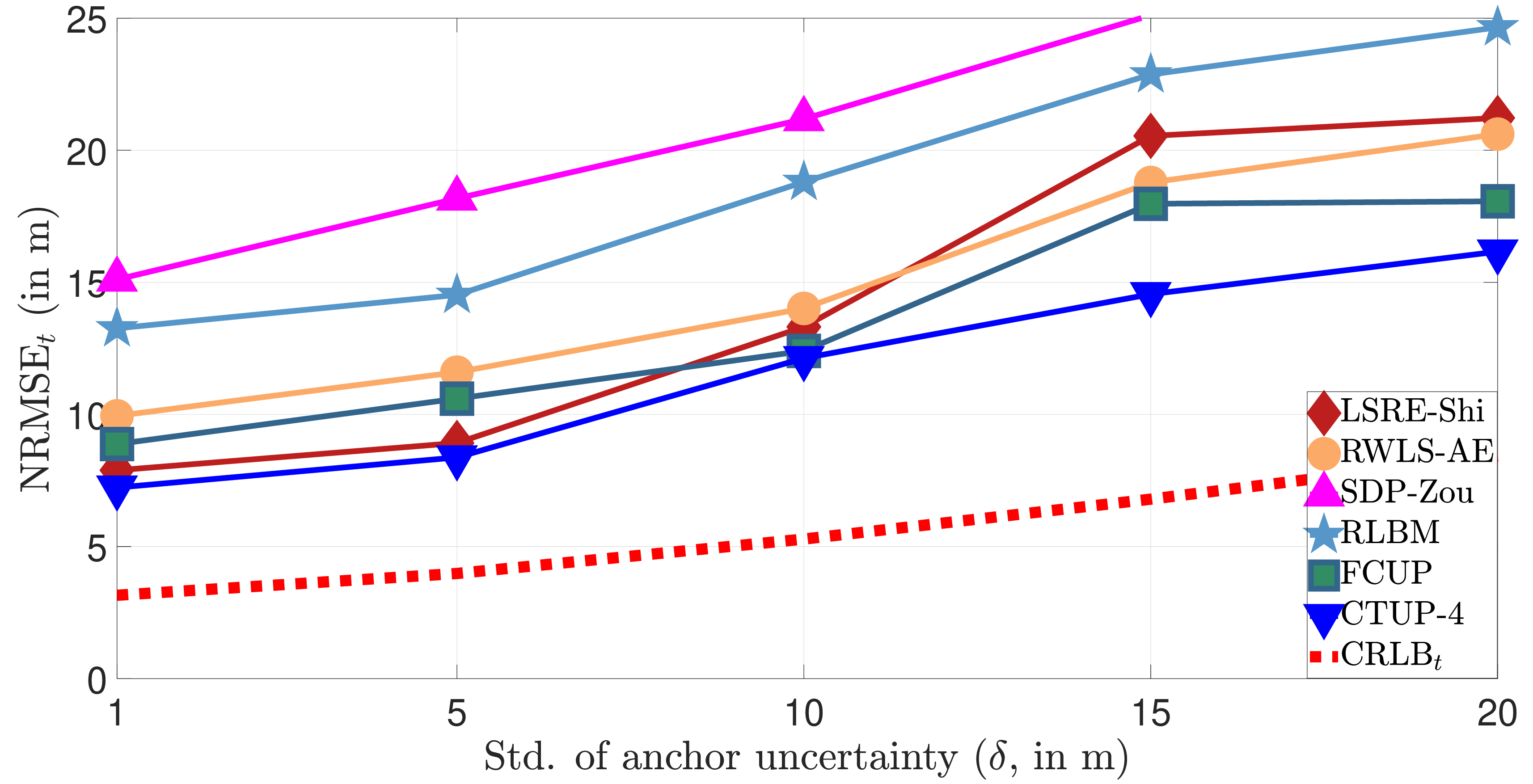}
    \label{Fig:AU_loc}
    \end{minipage}
    }
    \subfigure[NRMSE of transmit power estimate as a function of $\delta$.]{
    \begin{minipage}[t]{0.47\linewidth}
    \centering
    \includegraphics[width=\linewidth]{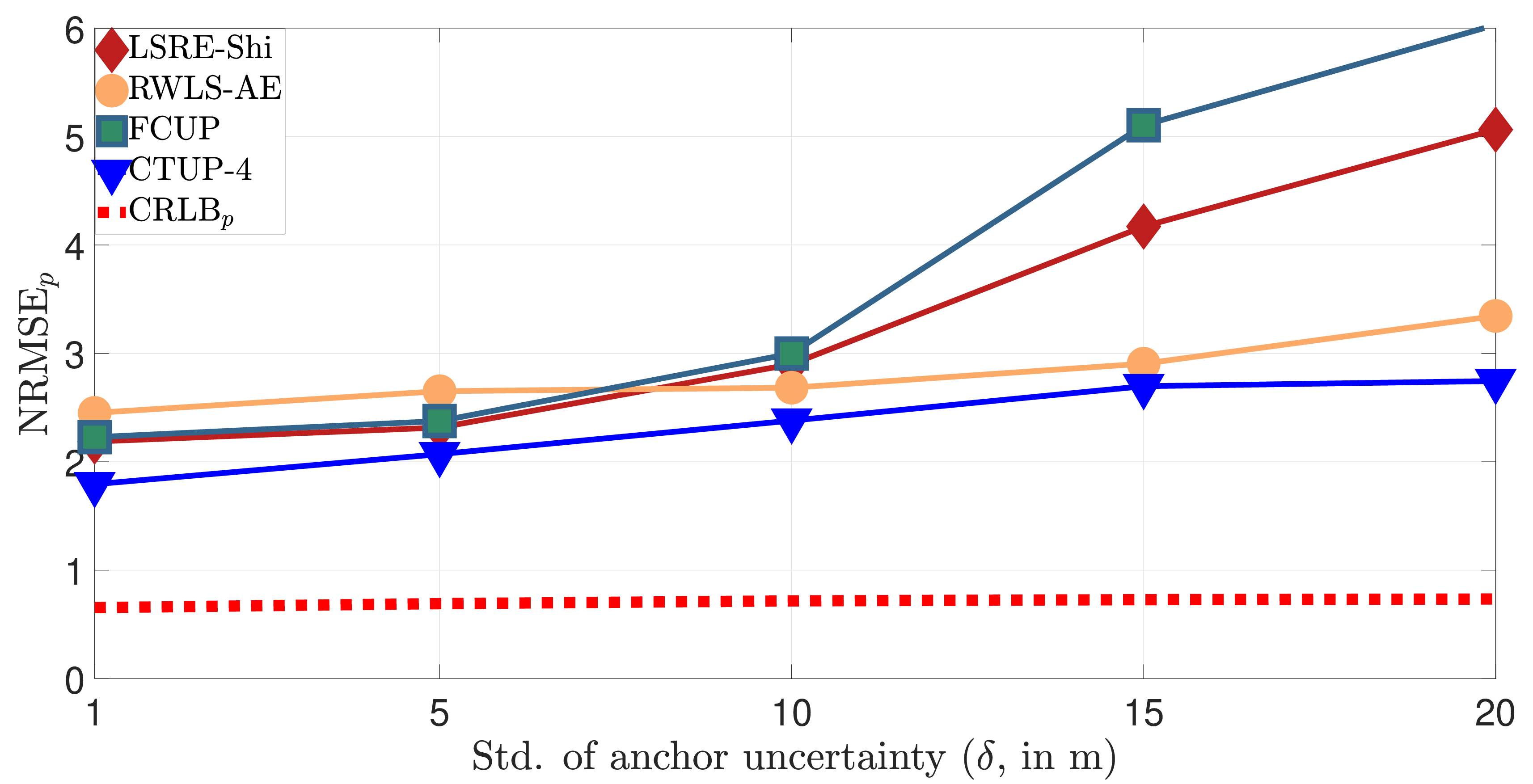}
    \label{Fig:AU_power}
    \end{minipage}
    }
    \caption{Performance of the localization techniques with respect to the anchor location uncertainty ($\sigma = 3$~dB).}
    \label{Fig:AU}
\end{figure*}
%
% Noncooperative localization
\begin{figure*}[!t]
    \subfigure[NRMSE of location estimate as a function of $\sigma$.]{
    \begin{minipage}[t]{0.47\linewidth}
    \centering
    \includegraphics[width=\linewidth]{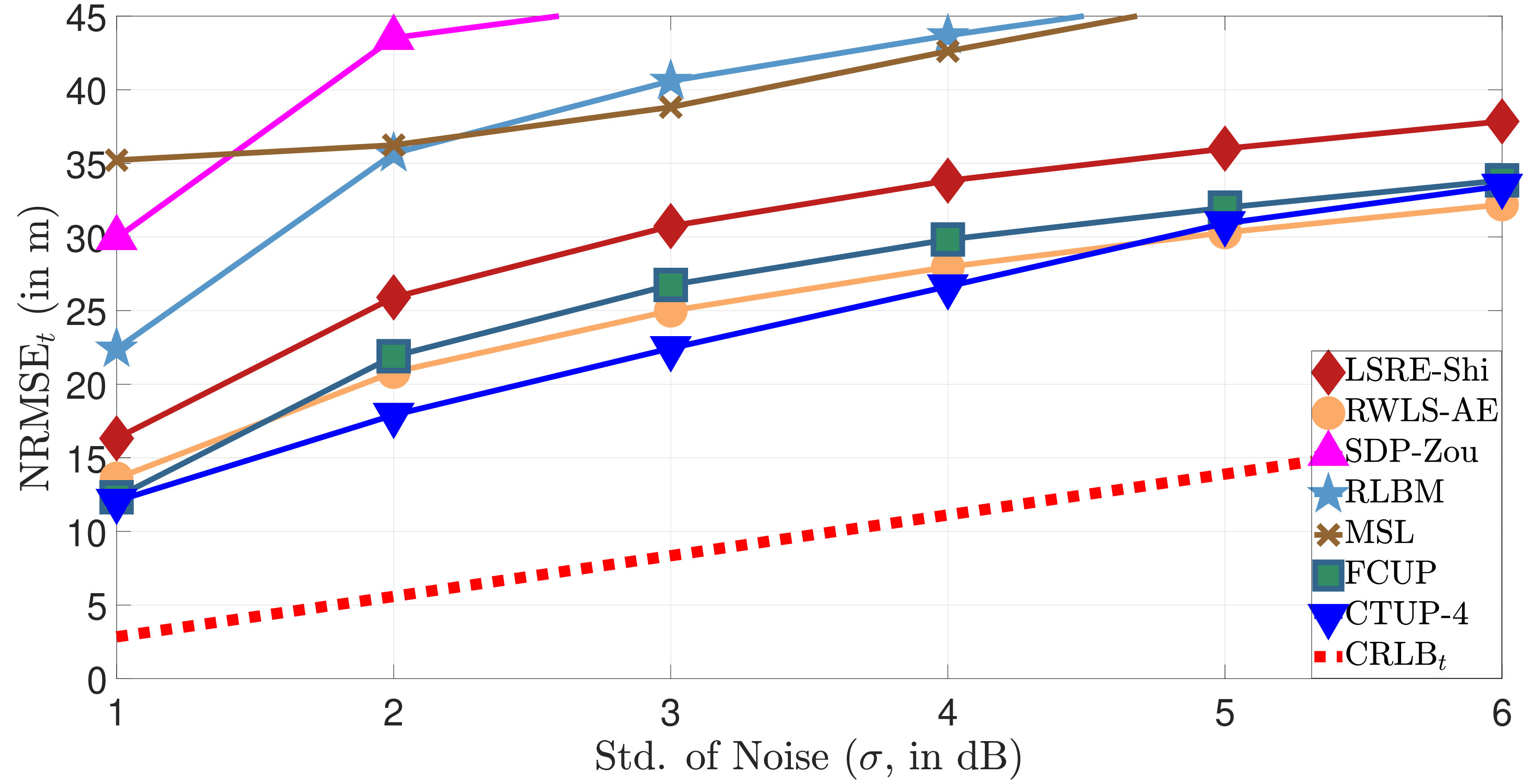}
    \label{Fig:Non_sigma}
    \end{minipage}
    }
    \subfigure[NRMSE of location estimate as a function of $\delta$.]{
    \begin{minipage}[t]{0.47\linewidth}
    \centering
    \includegraphics[width=\linewidth]{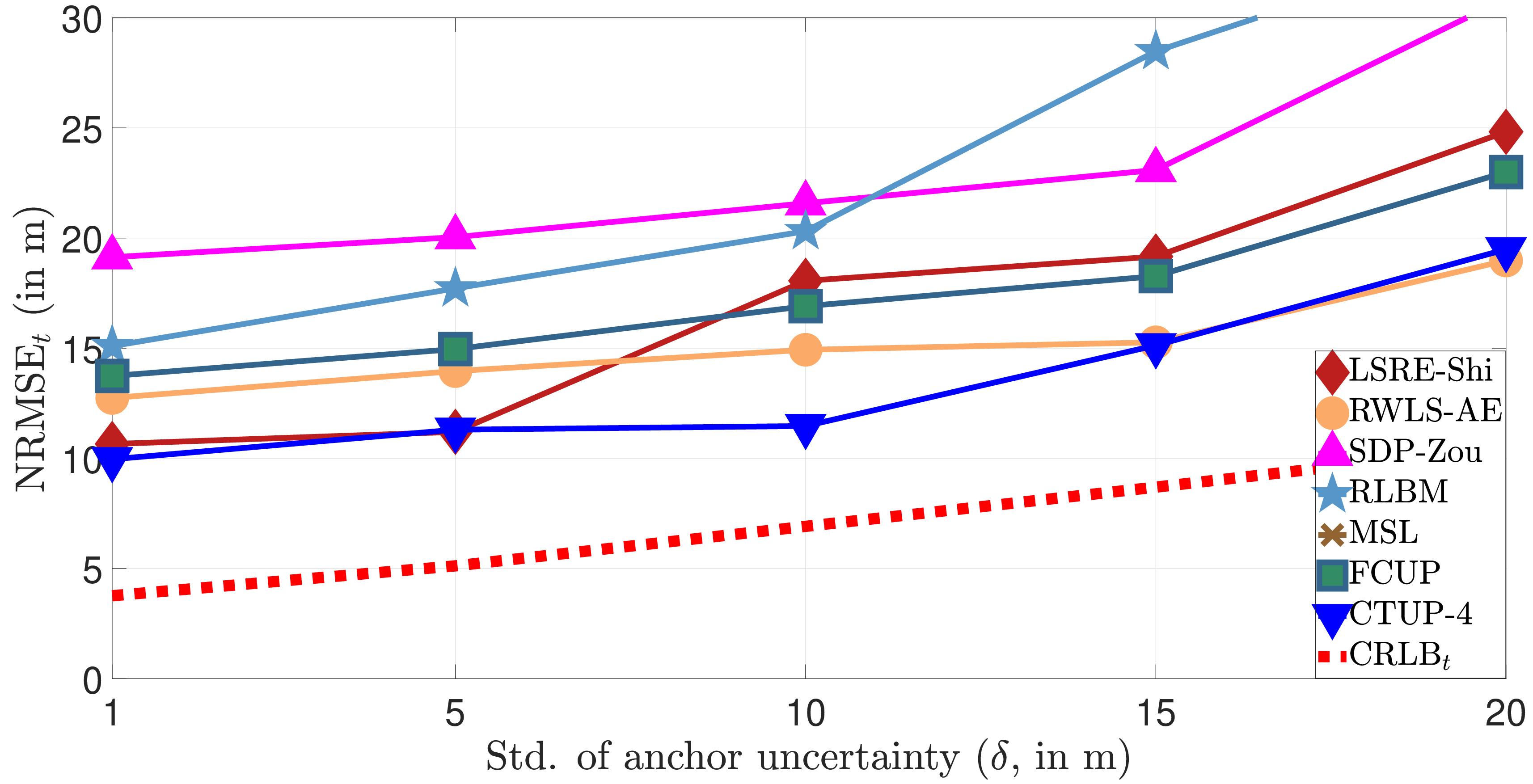}
    \label{Fig:Non_delta}
    \end{minipage}
    }
    \caption{Performance of the localization techniques in non-cooperative scenarios.}
    \label{Fig:Non}
\end{figure*}
\begin{figure*}
    \centering
    \includegraphics[width=\linewidth]{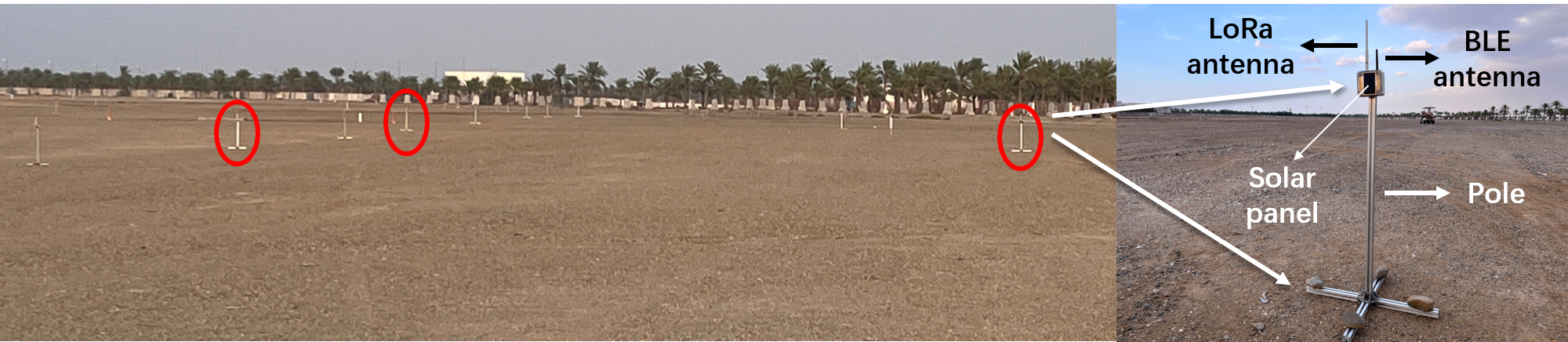}
    \caption{Photographs of the localization experimental site and setups.}
    \label{Fig:Experiment_site}
\end{figure*}
\begin{figure}[!b]
    \centering
    \includegraphics[width=\linewidth]{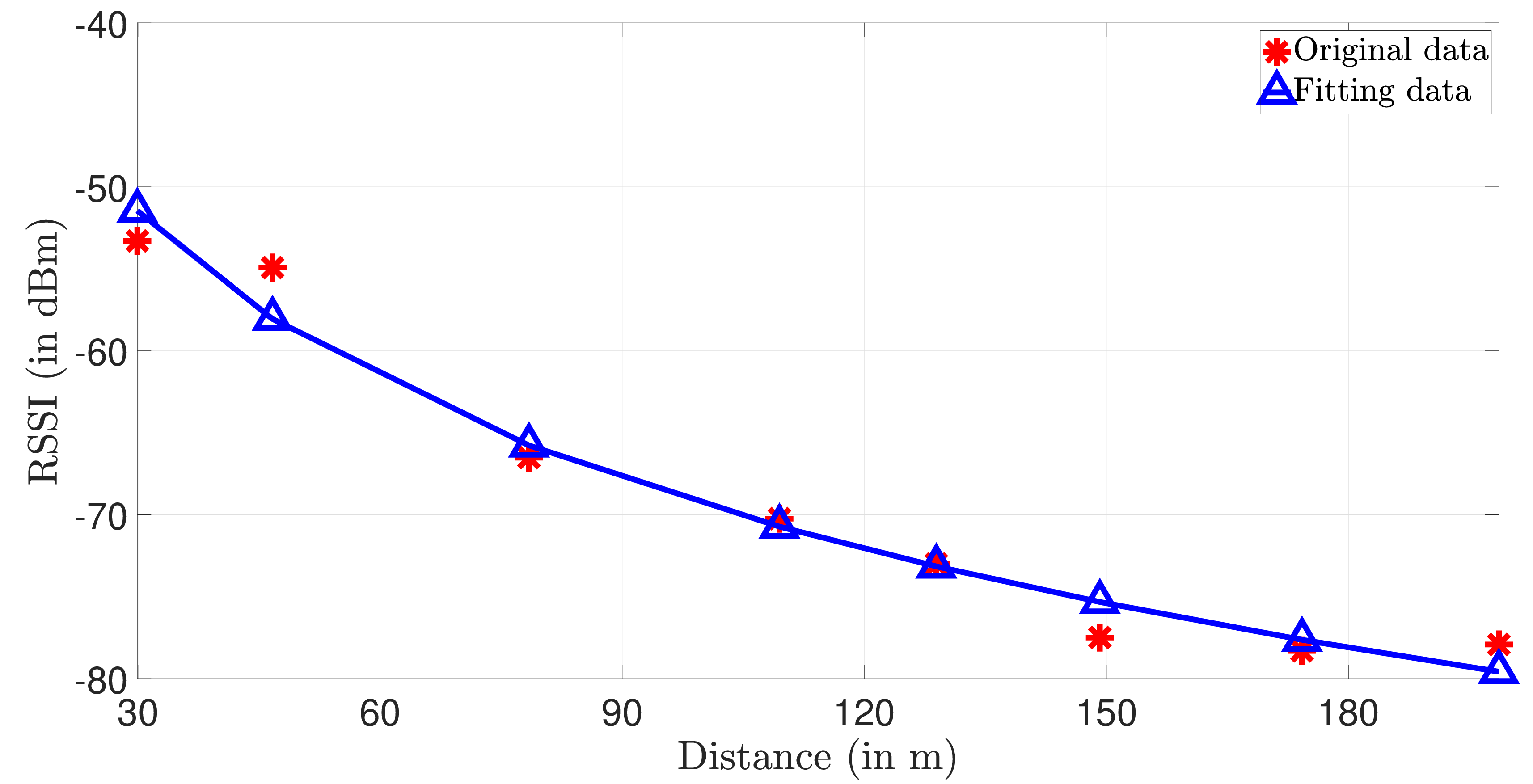}
    \caption{Fit the log-normal model to estimate the transmit power and PLE.}
    \label{Fig:FitLogNormal}
\end{figure}

\subsubsection{Scenario $\uppercase\expandafter{\romannumeral1}$}\label{Sim:S1}
We compare the performance of CTUP-1 with IRGDL~\cite{Invex} and SDP-$\ell_2$~\cite{SDPl2} when the transmit power of the nodes and PLE are known. Fig.~\ref{Fig:S1_Loc} demonstrates that CTUP-1 outperforms SDP-$\ell_2$ and IRGDL for all values of $\sigma$. This occurs because SDP-$\ell_2$ and IRGDL are affected by both $\sigma$ and $\delta$, while CTUP-1 mitigates the influence of anchor location uncertainty. For low noise scenarios, the performance of SDP-$\ell_2$ and CTUP-1 are similar; however, its effectiveness diminishes notably in scenarios where $\sigma > 3$ dB. The variations in $\sigma$ have less impact on the NRMSE of IRGDL, yet it exhibits high localization error. This is due to the loss of information caused by invexifying the ML objective function.

% Fig.~\ref{Fig:S1_Loc_AU} shows that the performance of localization techniques deteriorates with the increase in anchor location uncertainly ($\delta$). SDP-$\ell_2$ and IRGDL fail to mitigate the impact of anchor location uncertainty and have worse performance than CTUP-1. This is because CTUP-1 is derived from joint measurements including both RSS measurements and anchor location measurements, unlike SDP-$\ell_2$ and IRGDL, which rely simply on RSS measurements.

%shows the location estimation error increases with $\sigma$. CTUP-1 outperforms SDP-$l_2$ and IRGDL for the considered scenario. SDP-$\ell_2$ demonstrates comparable performance to CTUP-1 for small noise level scenarios. However, its performance degrades in scenarios that $\sigma > 3$~dB. 

%Fig.~\ref{Fig:S1_Loc_AU} shows that the performance of localization techniques deteriorates as the uncertainty in anchor location increases. This effect is apparent when a lot of anchor nodes are present. SDP-$l_2$ and IRGDL fail to mitigate the impact of anchor location uncertainty and have worse performance than CTUP-1.
%
\subsubsection{Scenario $\uppercase\expandafter{\romannumeral2}$}\label{Sim:S2}
In the scenario of unknown $\beta$, we compare the performance of CTUP-2 with MLS~\cite{MSL} in terms of the location estimate and PLE estimate. The performance of localization techniques is studied when PLE is unknown. The NRMSE of the location estimate as a function of $\sigma$ is presented in Fig.~\ref{Fig:S2_Loc_noise}. CTUP-2 demonstrates superior performance compared with MSL for considered scenarios. As the $\sigma$ increases, the performance gap between MSL and CTUP-2 widens, primarily due to the limitations of  LLS in managing high-noise scenarios. In Fig.~\ref{Fig:S2_PLE_noise}, it can observed that CTUP-2 outperforms MSL in the estimate of PLE. 
% Fig.~\ref{Fig:S2_Loc_AU} shows the superiority of CTUP-2 in location estimate when the anchor location uncertainty increases.
% Fig.~\ref{Fig:S2_PLE_AU} displays that the performance of MSL is significantly impacted by the anchor location uncertainty. However, CTUP-2 can consistently maintain satisfactory performance in considered instances. The increase in noise standard deviation or anchor location uncertainty has a minimal impact on $CRLB_\beta$.
%
\subsubsection{Scenario $\uppercase\expandafter{\romannumeral3}$}\label{Sim:S3}
In the case of unknown transmit power, we evaluate the performance of CTUP-3, LSRE-Shi~\cite{SDPShi}, RLBM~\cite{Lohrasbipeydeh_Gulliver_2021}, and FCUP~\cite{FCUP} in estimating the location and transmit power of target nodes. The performance of the localization techniques is shown in Fig.~\ref{Fig:S3_Loc_noise}.  CTUP-3 and FCUP exhibit comparable performances, surpassing other techniques. However, CTUP-3 showcases superior performance over FCUP due to FCUP's lack of consideration for anchor location uncertainty. CTUP-3 outperforms LSRE-Shi and FCUP in estimating the transmit power, as shown in Fig.~\ref{Fig:S3_power_noise}. RLBM and LSRE-Shi demonstrate inferior performance since they do not incorporate RSS readings of the target-target links.

% The superiority of CTUP-3 can also be observed in Fig.~\ref{Fig:S3_Loc_AU}. CTUP-3 achieves the best performance in considered cases. FCUP diverges from CTUP-3 in scenarios where considerable anchor location uncertainty exists. In addition, the anchor location uncertainty less impacts CTUP-3 in the transmit power estimate. Fig.~\ref{Fig:S3_power_AU} reveals that CTUP-3 is closest to $\text{CRLB}_{\mathbf{p}}$ and outperforms LSRE-Shi and FCUP. The advantage of CTUP-3 becomes increasingly evident as the anchor location uncertainty grows. 
%
\subsubsection{Scenario $\uppercase\expandafter{\romannumeral4}$}\label{Sim:S4}
In the fourth scenario where transmit power and PLE are unknown, we evaluate CTUP-4's performance in comparison to RWLS-AE~\cite{RWLSAE} and SDP-Zou~\cite{SDP_Zou}. In Fig.~\ref{Fig:S4_Loc_noise}, we present the localization accuracy of the techniques as a function of $\sigma$. CTUP-4 outperforms RWLS-AE and SDP-Zou at low noise levels. Although RWLS-AE performs slightly better than CTUP-4 at high noise levels, it has the highest computational complexity among the compared techniques (refer Table~\ref{Tb:complexity}). The accuracy of the transmit power estimate as a function of $\sigma$ is illustrated in Fig.~\ref{Fig:S4_power_noise}, where CTUP-4 outperforms RWLS-AE across all values of $\sigma$. In Fig.~\ref{Fig:S4_PLE_noise}, we present $\text{NRMSE}_\beta$ as a function of the noise standard deviation. CTUP-4 achieves the best performance, especially at low noise levels. RWLS-AE demonstrates comparable performance to CTUP-4 when $\sigma > 6$~dB. However, The noise standard deviation is typically less than 4~dB in practical scenarios (refer Table~\ref{Tb:ExperimentalParameters}), thereby indicating the superior performance of CTUP-4 in the realistic implementation.
% Fig.~\ref{Fig:S4_Loc_AU} shows that RWLS-AE and SDP-Zou are unable to mitigate the effects of anchor location uncertainty. CTUP-4 has superior performance to existing techniques. Fig.~\ref{Fig:S4_power_AU} illustrates that the accuracy of the transmit power estimate decreases with the anchor location uncertainty increasing. CTUP-4 surpasses RWLS-AE in performance and is closest to the CRLB. Fig.~\ref{Fig:S4_PLE_AU} demonstrates the superiority of CTUP-4 in estimating PLE as the anchor location uncertainty increases.
%
\subsubsection{Effect of the anchor location uncertainty}\label{Sim:AU}

In this section, we study the performance of localization techniques as a function of $\delta$ (refer Fig.~\ref{Fig:AU_loc}). Across all $\delta$ values, CTUP-4 outperforms other existing techniques. When $\delta$ is small, LSRE-Shi has a performance similar to CTUP-4; however, in scenarios where $\delta > 5$~dBm, its effectiveness decreases significantly. In Fig.~\ref{Fig:AU_power}, CTUP-4 shows superior performance in estimating the transmit power of target nodes. Although RWLS-AE and CTUP-4 are less influenced by anchor location uncertainty, CTUP-4's transmit power estimate is closest to $\text{CRLB}_{{p}}$, highlighting its increasing supriority with greater uncertainty in anchor locations.
% CTUP-4 and RWLS-AE are less affected by anchor location uncertainty; nevertheless, CTUP-4 is closest to $\text{CRLB}_{{p}}$ and surpasses RWLS-AE in \textcolor{blue}{accuracy} for estimating the transmit power of target nodes. 
In conclusion, numerical simulations have verified the effectiveness of proposed techniques in considered scenarios. CTUP shows superior performance in estimating location, transmit power, and PLE compared to existing techniques.
\subsubsection{Non-cooperative localization}\label{subsubsec:noncooperative}
\textcolor{blue}{In this section, we evaluate the performance of localization techniques in non-cooperative scenarios. We utilize two networks, denoted as NW-3 and NW-4, to examine the effects of noise standard deviation and anchor location uncertainty, respectively. NW-3 and NW-1 share identical configurations except for NW-3 containing only one randomly deployed target node. Similarly, NW-4 and NW-2 have identical configurations, with NW-4 featuring only one randomly deployed target node. Simulation results presented in Fig.~\ref{Fig:Non_sigma} and  Fig.~\ref{Fig:Non_delta} are obtained from NW-3 and NW-4, respectively. We study the performance of localization techniques as a function $\sigma$ in Fig.~\ref{Fig:Non_sigma}. CTUP-4 demonstrates superior performance across most scenarios. Despite RWLS-AE marginally outperforming CTUP-4 for $\sigma > 5$~dB, It has the highest complexity (refer to Table~\ref{Tb:complexity}). The performance order of localization techniques is similar to that observed in cooperative scenarios. Fig.~\ref{Fig:Non_delta} presents the localization performance as a function of $\delta$. Similar to the scenario depicted in Fig.~\ref{Fig:AU_loc}, LSRE-Shi and CTUP-4 exhibit comparable performance when $\delta < 5$~m. However, CTUP-4 significantly surpasses LSRE-Shi for $\delta > 5$~m. While RWLS-AE outperforms CTUP-4 for $\delta > 15$~m, it has extremely high complexity. Overall, CTUP-4 exhibits superior performance compared to existing techniques in non-cooperative scenarios.}

\begin{figure*}[!t]
    \centering
    \subfigure[The distribution of gathered latitude and longitude readings.]{
    \begin{minipage}[t]{0.47\linewidth}
    \centering
    \includegraphics[width=\linewidth]{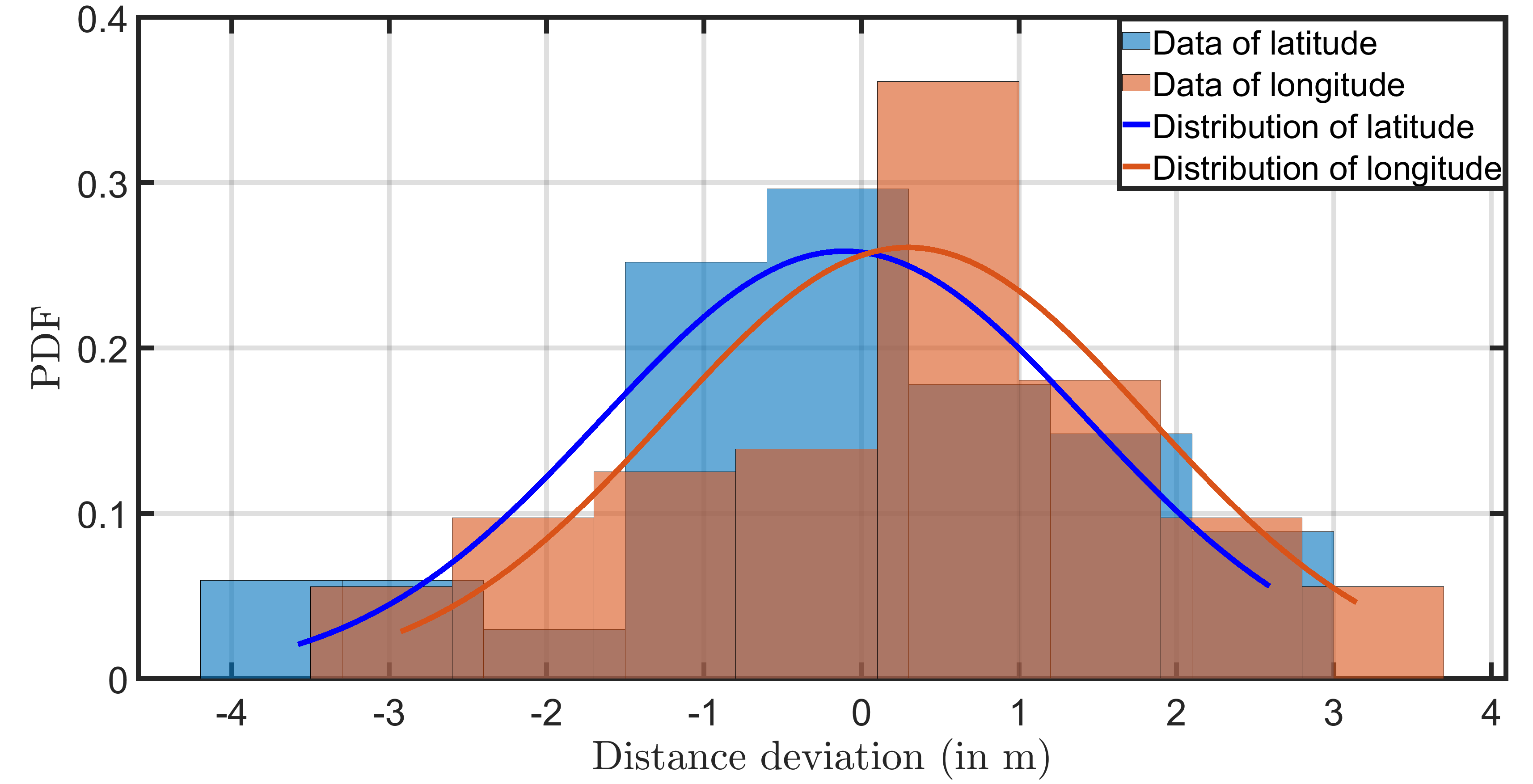}
    \label{Fig:DisAU}
    \end{minipage}
    }
    \subfigure[The distribution of RSS measurements.]{
    \begin{minipage}[t]{0.47\linewidth}
    \centering
    \includegraphics[width=\linewidth]{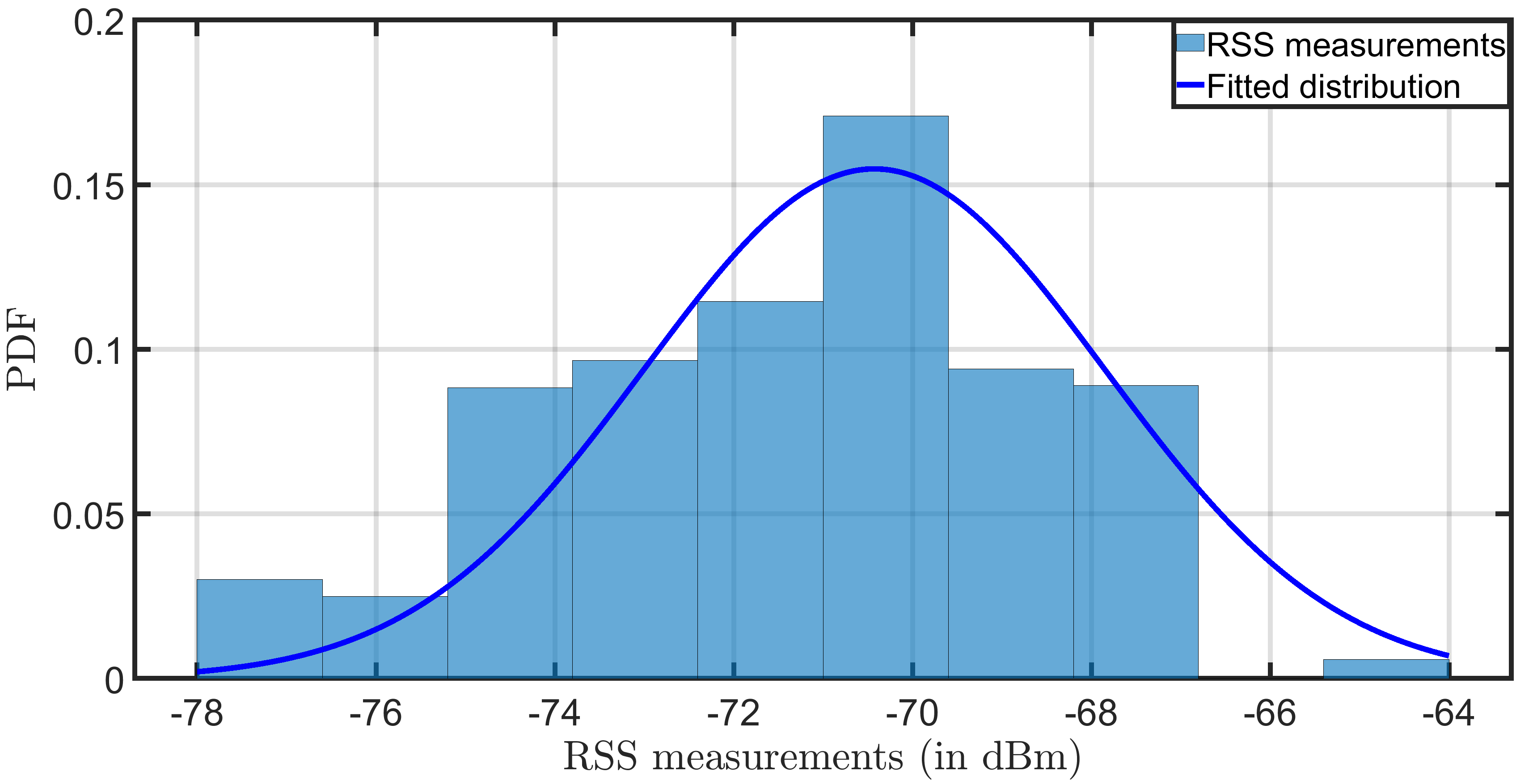}
    \label{Fig:DisRSS}
    \end{minipage}
    }
    \caption{The distribution of gathered experimental data: a) anchor location readings, and b) RSS measurements.}
    \label{Fig:Dis}
\end{figure*}
\subsection{Complexity comparison}\label{subsec:complexity}
The computational complexity of localization techniques is provided in Table~\ref{Tb:complexity}. 
\textcolor{blue}{We consider only the dominant elements to compute the computational complexity to ensure consistency with compared localization techniques~\cite{SDPShi,RWLSAE,SDP_Zou,Lohrasbipeydeh_Gulliver_2021,MSL, Invex,SDPl2,FCUP}. In Table~\ref{Tb:complexity}, we compute the CPU runtime of all techniques in MATLAB R2021a using an Intel(R) Xeon(R) W-2245 CPU @ 3.90GHz processor with 64GB RAM.} RLBM and RWLS-AE are non-cooperative iterative techniques, whereas IRGDL is a cooperative technique utilizing gradient descent. Therefore, in Table~\ref{Tb:complexity}, we employ two distinct notations to represent the number of iterations required for convergence. For RLBM and RWLS-AE, $i_j^{\text{itr}}$ denotes the iteration count necessary for the $j^{\text{th}}$ target node to converge. In IRGDL, $k^{\text{itr}}$ denotes the number of steps taken for gradient descent to achieve convergence. $\text{T}_1$ and $\text{T}_2$ represent the CPU running time of the localization techniques for NW-1 and NW-2, respectively. It can be observed in Table~\ref{Tb:complexity} that MSL has the least complexity, however, it exhibits significantly poor localization accuracy (refer Fig.~\ref{Fig:S2}). Non-cooperative techniques usually have high computational complexity since they are required to solve an SDP problem for each target node. IRGDL benefits from the simplicity of gradient descent, with its computational complexity increasing linearly in accordance with the network's size. Nevertheless, it demonstrates poor performance, as can be observed in Fig.~\ref{Fig:S1_Loc}. The computational complexity of SDP-$\ell_2$ scales proportionally to $N_t^{8.5}$, whereas CTUPs exhibit a computational complexity of $\sim N_t^{6.5}$, making them suitable for dense networks. However, CTUP-2 and CTUP-4 require additional time as they leverage links between anchor nodes prior to solving the mixed SDP-SOCP problem. CTUP-3 takes slightly longer than CTUP-1 due to the additional step of extracting the estimate of the transmit power. In comparison to FCUP, the CTUPs have marginally higher CPU runtime. However, this trade-off allows them to address the impact of anchor location uncertainty effectively and also consider scenarios where PLE is unknown (refer Section~\ref{S4}), leading to enhanced performance (refer Fig.~\ref{Fig:S3} and Fig.~\ref{Fig:AU}).

\subsection{Experimental results}\label{subsec:experiment}
\textcolor{blue}{Real-world network experiments are significant for the implementation and analysis of cooperative localization techniques~\cite{MoeWin1}.} We carried out extensive real-field experiments to further validate the effectiveness of localization techniques. We use the RAK 4631~\cite{nguyen2022development} to establish a Bluetooth low energy (BLE) mesh network employing an omnidirectional BLE antenna with a gain of 2~dBi~\cite{BLEAntenna}. The RSS measurements among the nodes are uploaded to the data center using a LoRa transceiver. The precise location of the nodes, accurate to the centimeter level, is obtained using real-time kinematic (RTK) GPS, serving as the ground truth. We also collected the location of the anchor nodes using standard GPS, providing meter-level accuracy. The imprecision associated with these readings is referred to as ``anchor location uncertainty''. Each node is equipped with a solar panel and a lithium battery to ensure long-term operation. The nodes are mounted on aluminium poles, each standing at a height of 1 meter. Across a vast outdoor expanse measuring $640\times180~m^2$, we deploy a total of 50 nodes. 
The photographs of the experiment site and the node are shown in Fig.~\ref{Fig:Experiment_site}. 
\begin{table}[!t]
\centering
\caption{Parameters of the experimental site.}
\resizebox{\linewidth}{!}{\begin{tabular}{|c|c|}
\hline
Area (in $m^2$)                            & $640\times180~m^2$ \\ \hline
Number of nodes                            & 50                 \\ \hline
Received RSS readings per link             & $\sim500$          \\ \hline
The nodes' height                          & 1~m           \\ \hline
Transmit power (in dBm)                    & -3.59            \\ \hline
PLE                                        & 3.27             \\ \hline
Average noise standard deviation (in dB)   & 2.47             \\ \hline
Average anchor location uncertainty (in m) & 1.53             \\ \hline
\end{tabular}}
\label{Tb:ExperimentalParameters}
\end{table}
Although the transmit power and PLE are not required for CTUP-4, these parameters are essential to implement other localization techniques. To obtain the transmit power of the nodes and PLE, we placed 9 nodes in a straight-line configuration. One node is configured as the BLE transmitter, and the remaining nodes are the BLE listener. Fig.~\ref{Fig:FitLogNormal} shows the variation of RSS as a function of distance. We estimated the transmit power of the node and PLE by applying the log-normal model (refer~\eqref{LogNormal}) to the dataset. The network details are shown in Table~\ref{Tb:ExperimentalParameters}. 

We gathered approximately 100 location readings of a node using a standard GPS, while the precise location was determined using RTK GPS. Fig.~\ref{Fig:DisAU} showcases the deviations in latitude and longitude of the GPS data compared to the precise location. We employed the MATLAB routine $histogram(data,N_b,\text{`Normalization'},\text{`pdf'})$ to normalize the counts within each bin, where $data$ represents the readings of interest, and $N_b$ stands for the number of bins. \textcolor{blue}{We select $N_b$ employing Sturges' rule~\cite{Sturges, Sturges2}: $N_b = \lceil 1+\log_2(N_d)\rceil$, where $N_d$ denotes the total sample size, and $\lceil\cdot\rceil$ represents the ceiling function. Our dataset comprises of 100 location readings from a single GPS node, hence, $N_b$ is set to be 8 bins. In Fig.~\ref{Fig:DisAU},} the mean deviation for latitude is -0.11~m, and for longitude, it is 0.29~m. Moreover, the standard deviation for longitude and latitude measurements is $\sim$1.53~m. Therefore, Fig.~\ref{Fig:DisAU} demonstrates that the anchor location uncertainties for both latitude and longitude follow a common zero-mean Gaussian distribution, affirming the efficacy of our considered model~\eqref{AnchorLoc}. \textcolor{blue}{Fig.~\ref{Fig:DisRSS} depicts the distribution of 500 RSS readings from a BLE communication link. The mean value and standard deviation are -70.43~dBm and 2.57~dB, respectively. Given the distance between the two nodes as 86.23~m, utilizing~\eqref{LogNormal} and parameters from Table~\ref{Tb:ExperimentalParameters}, the mean value and standard deviation of RSS measurements in this link are computed as -66.88~dBm and 2.47~dB, respectively. These values align with Fig.~\ref{Fig:DisRSS}, affirming the effectiveness of our approach in obtaining experimental parameters. In addition, Fig.~\ref{Fig:DisRSS} demonstrates that RSS measurements follow a Gaussian distribution. Moreover, as RSS measurements and anchor locations are mutually independent, the efficacy of~\eqref{PDF_theta} can be confirmed.}

\begin{figure}
    \centering
    \includegraphics[width=\linewidth]{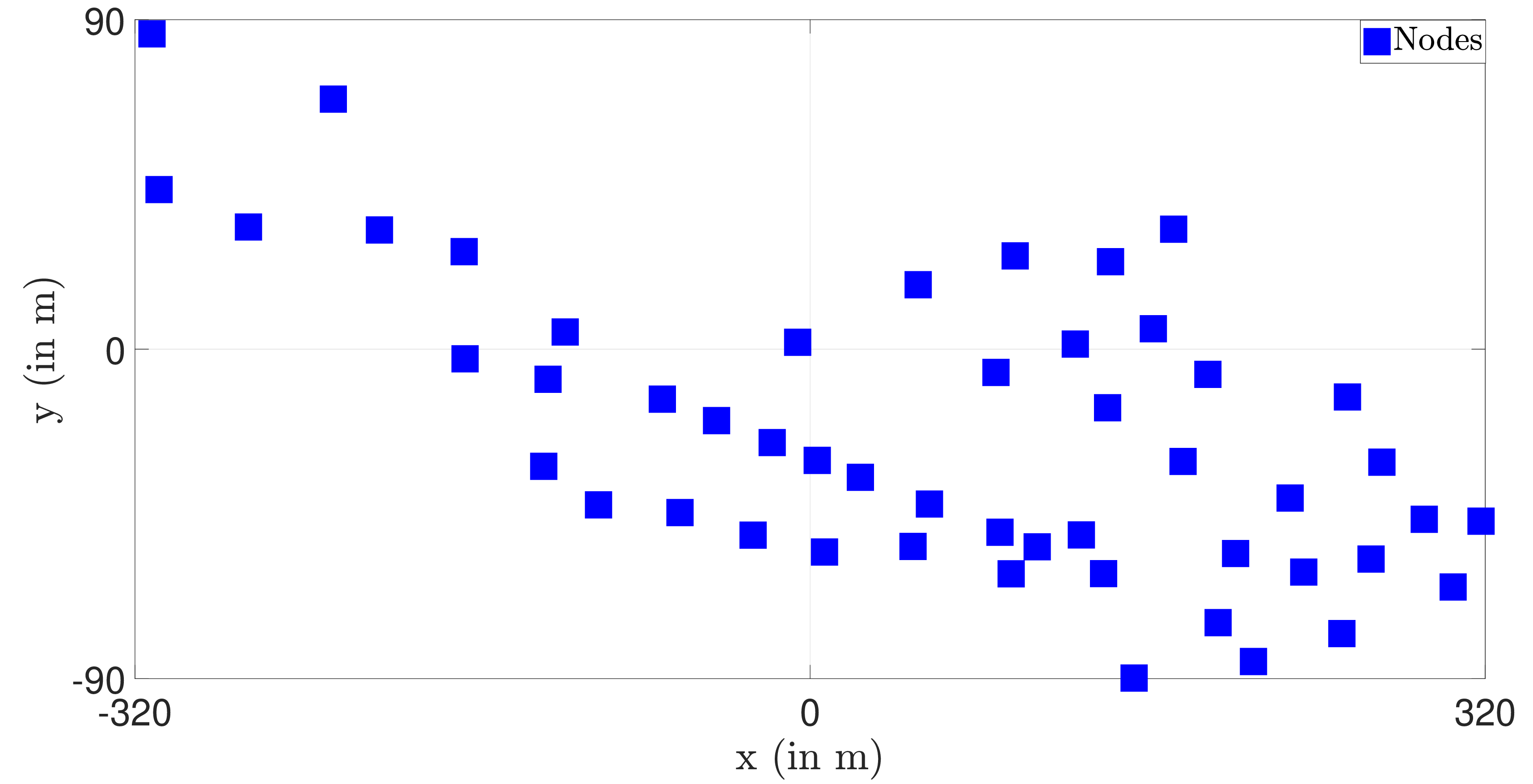}
    \caption{The network utilized in the experiment, comprising 50 nodes.}
    \label{Fig:Hardware_topo}
\end{figure}

The performance of localization techniques is evaluated using the experimental data, with each node receiving a minimum of 500 readings. Fig.~\ref{Fig:Hardware_topo} displays the topology of the network consisting of 50 nodes.
\begin{figure}[!t]
    \centering
    \subfigure[NRMSE of location estimate as a function of the number of anchor nodes.]{
    \begin{minipage}[t]{.97\linewidth}
    \centering
    \includegraphics[width=\linewidth]{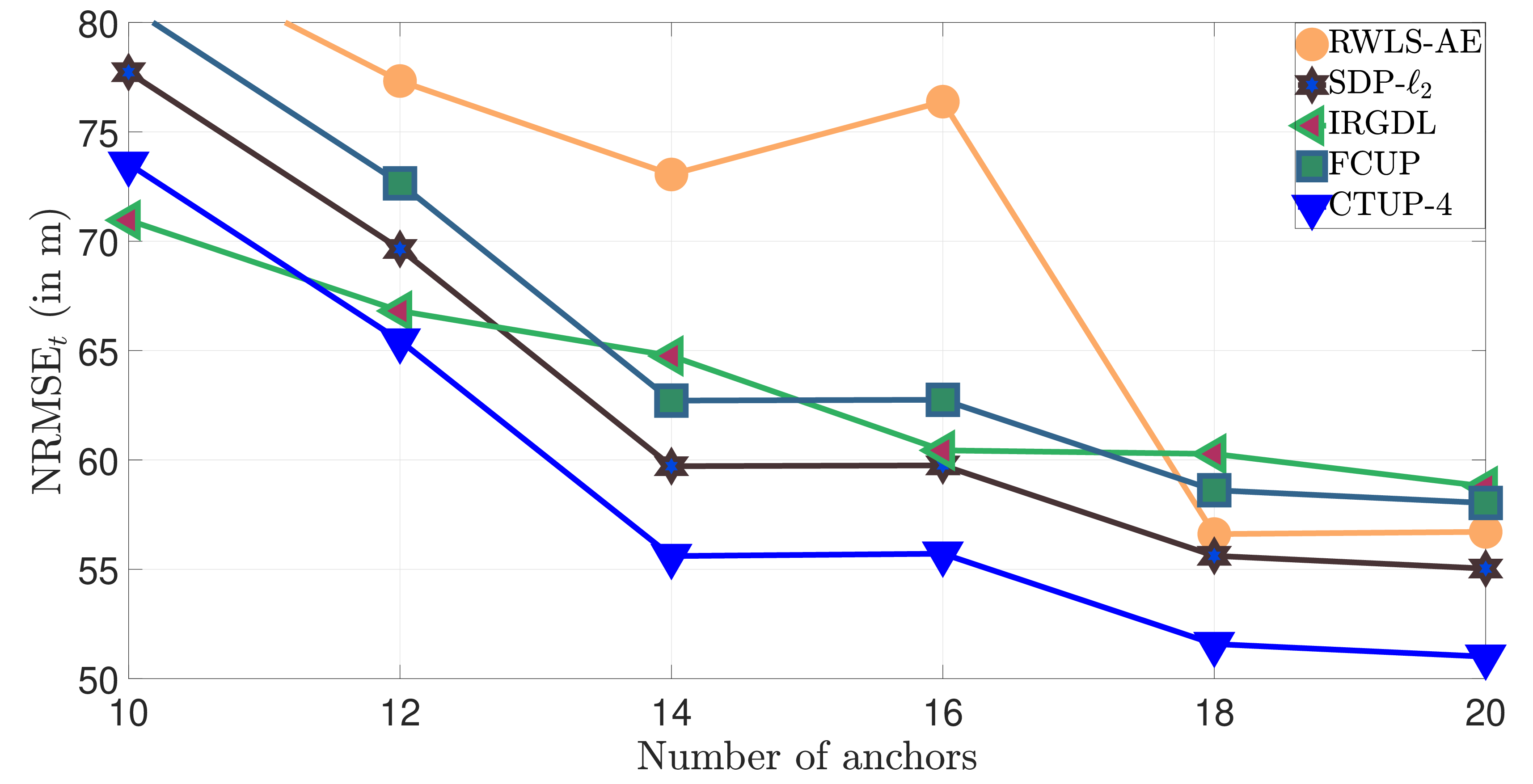}
    \label{Fig:Hardware_Loc_Na}
    \end{minipage}
    }
    \subfigure[\textcolor{blue}{NRMSE of transmit power estimate as a function of the number of anchor nodes.}]{
    \begin{minipage}[t]{.97\linewidth}
    \centering
    \includegraphics[width=\linewidth]{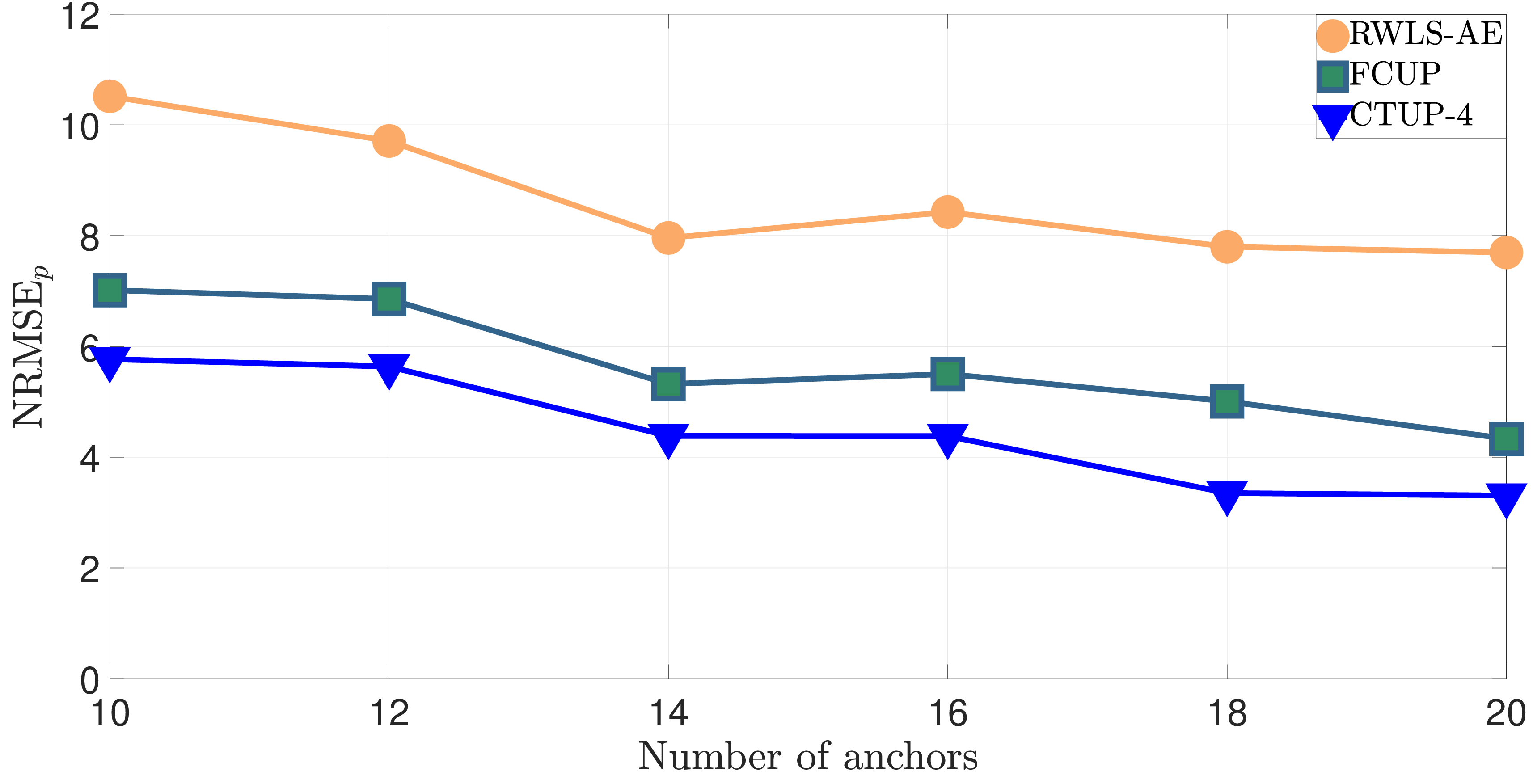}
    \label{Fig:Hardware_Power_Na}
    \end{minipage}
    }
    \subfigure[\textcolor{blue}{NRMSE of PLE estimate as a function of the number of anchor nodes.}]{
    \begin{minipage}[t]{.97\linewidth}
    \centering
    \includegraphics[width=\linewidth]{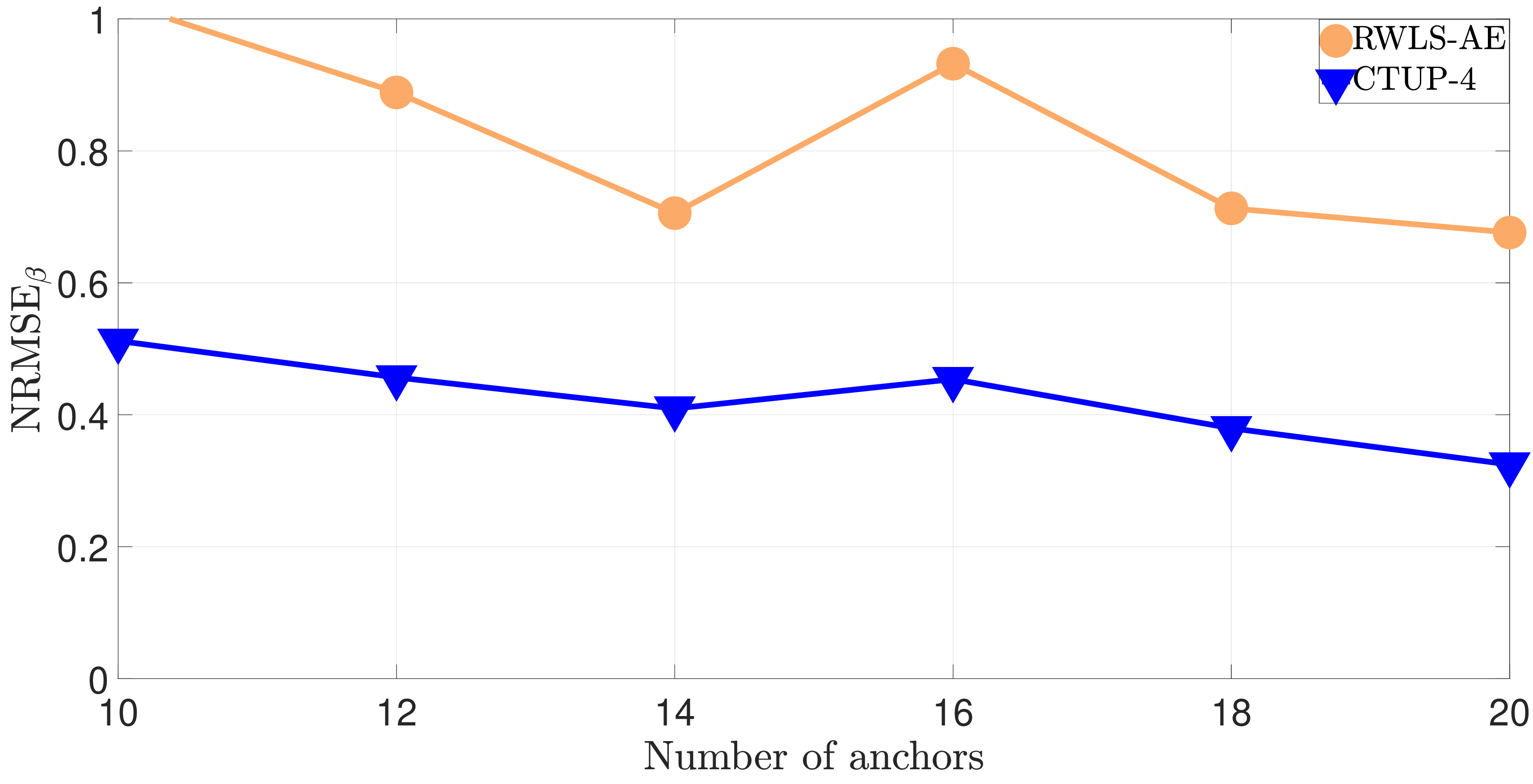}
    \label{Fig:Hardware_PLE_Na}
    \end{minipage}
    }
    \caption{\textcolor{blue}{Localization performance as a function of anchor node numbers utilizing experimental data.}}
    \label{Fig:Hardware_Na}
\end{figure}
\begin{figure}[!t]
    \centering
    \subfigure[NRMSE of location estimate as a function of the number of target nodes.]{
    \begin{minipage}[t]{.97\linewidth}
    \centering
    \includegraphics[width=\linewidth]{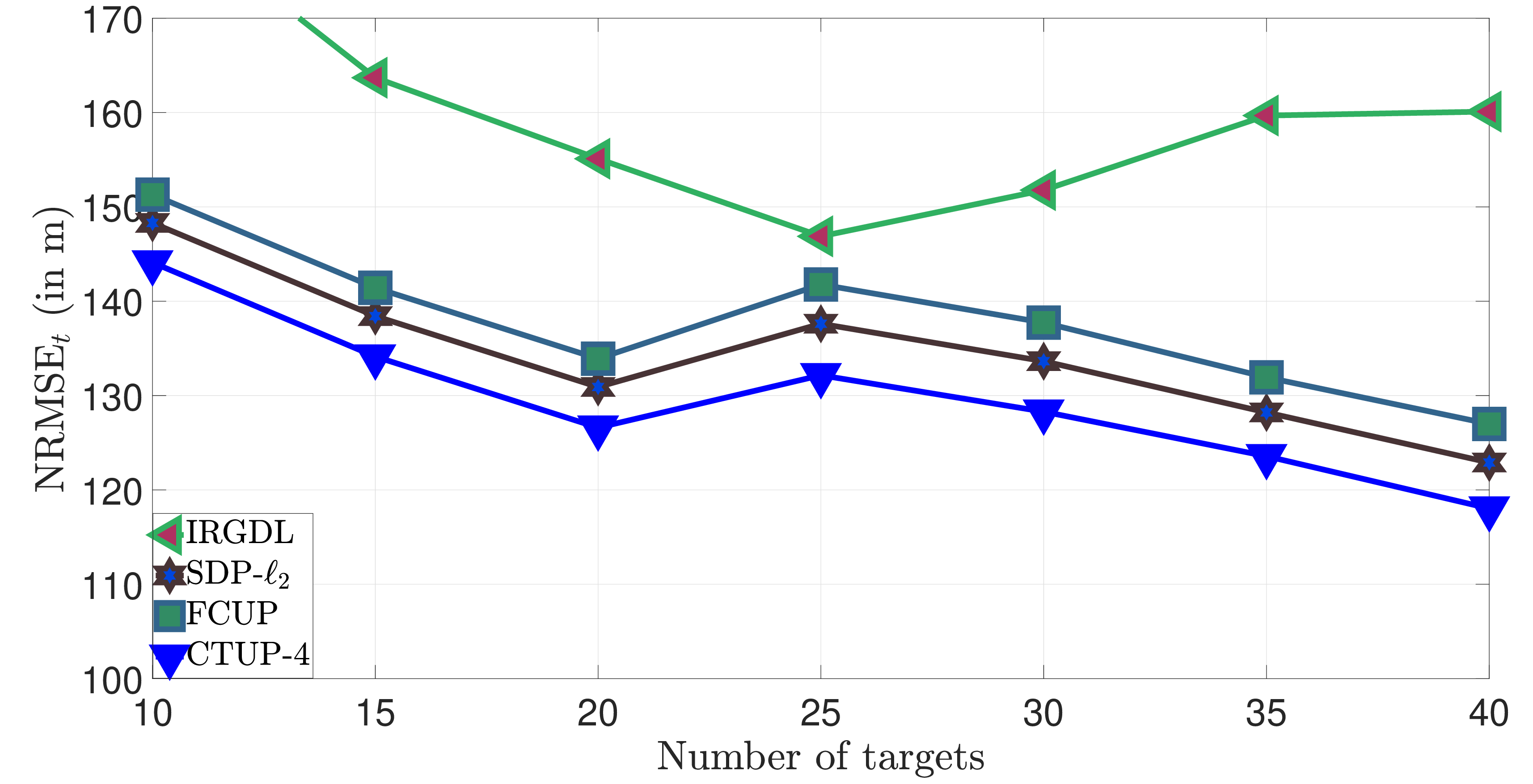}
    \label{Fig:Hardware_Loc_Nt}
    \end{minipage}
    }
    \subfigure[\textcolor{blue}{NRMSE of transmit power estimate as a function of the number of target nodes.}]{
    \begin{minipage}[t]{.97\linewidth}
    \centering
    \includegraphics[width=\linewidth]{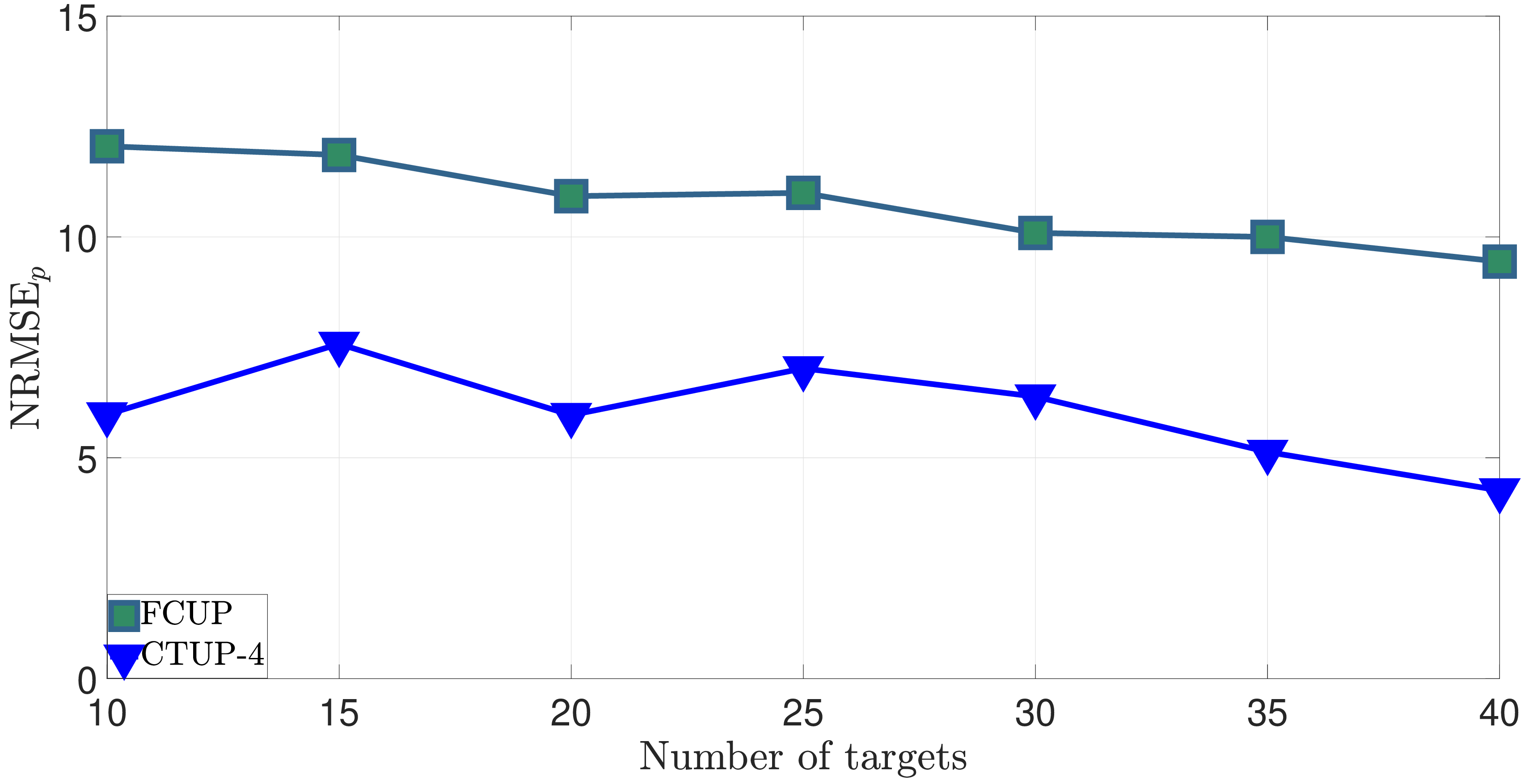}
    \label{Fig:Hardware_Power_Nt}
    \end{minipage}
    }
    \subfigure[\textcolor{blue}{NRMSE of PLE estimate as a function of the number of target nodes.}]{
    \begin{minipage}[t]{.97\linewidth}
    \centering
    \includegraphics[width=\linewidth]{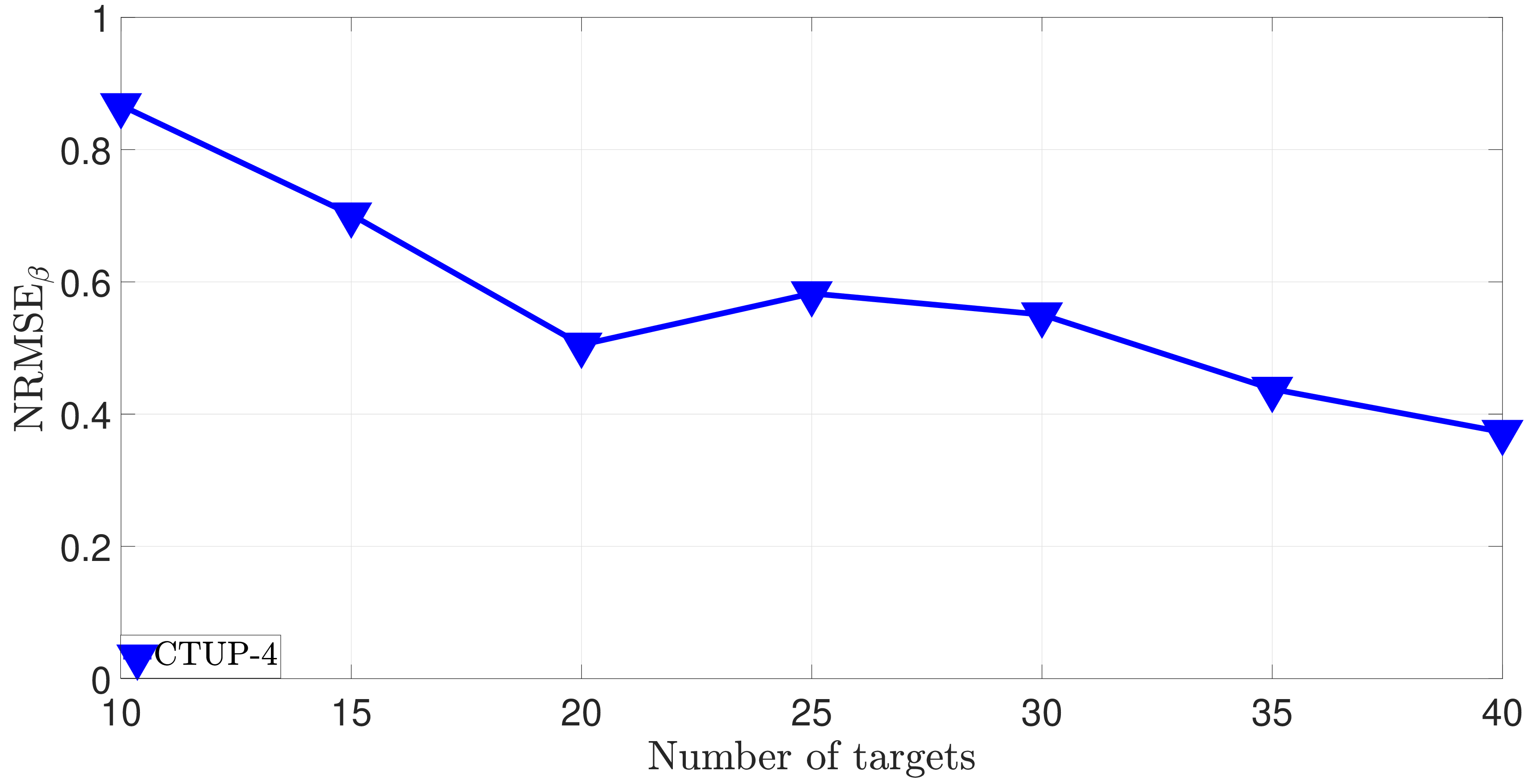}
    \label{Fig:Hardware_PLE_Nt}
    \end{minipage}
    }
    \caption{\textcolor{blue}{Localization performance as a function of target node numbers using experimental data.}}
    \label{Fig:Hardware_Nt}
\end{figure}

\subsubsection{Effect of the number of anchor nodes}

In this study, we randomly selected 10 target nodes to explore the impact of varying the number of anchor nodes on the location algorithms. In Fig.~\ref{Fig:Hardware_Loc_Na}, we present the performance of the localization technique as a function of the number of anchor nodes ($N_a$). LSRE-Shi, SDP-Zou, RLBM, and MSL are excluded from Fig.~\ref{Fig:Hardware_Loc_Na} due to their poor performance. Fig.~\ref{Fig:Hardware_Loc_Na} demonstrates that the increase in the number of anchor nodes tends to decrease the NRMSE of the location estimate. Non-cooperative techniques fail to achieve high accuracy since they cannot leverage the target-target links. CTUP-4 outperforms FCUP and SDP-$l_2$ by effectively mitigating the impact of anchor location uncertainty. IRGDL exhibits marginally superior performance compared to CTUP-4 at $N_a = 10$;  however, as  $N_a$ increases, IRGDL's performance deteriorates. This occurs since the error induced by the invex relaxation method escalates alongside the growing number of available links, resulting in a situation where IRGDL gains fewer benefits from the increased number of anchor nodes compared to SDP-based techniques. 

\textcolor{blue}{We utilize the transmit power and PLE presented in Table~\ref{Tb:ExperimentalParameters} as ground truths to evaluate the accuracy of localization techniques in estimating transmit power and PLE. The accuracy of the transmit power estimate with respect to the of number anchor node is depicted in Fig.~\ref{Fig:Hardware_Power_Na}. From Fig.~\ref{Fig:Hardware_Power_Na} it can be observed that CTUP-4 outperforms RWLS-AE and FCUP across all considered scenarios. In Fig.~\ref{Fig:Hardware_PLE_Na}, we compare the $\text{NRMSE}_\beta$ of CTUP-4 with that of RWLS-AE as a function of the number of anchor nodes, with CTUP-4 demonstrating superior performance. %Overall, Fig.~\ref{Fig:Hardware_Na} highlights CTUP-4's superior performance in practical implementations.
}

\subsubsection{Effect of the number of target nodes}

Fig.~\ref{Fig:Hardware_Loc_Nt} displays the performance of the location techniques as the number of target nodes is increased. Non-cooperative techniques are not included as they cannot leverage the target-target links. A tendency of performance enhancement is observed with the rising number of target nodes due to additional RSS measurements obtained from new links. CTUP-4 shows the best performance in comparison to other techniques. \textcolor{blue}{In Fig.~\ref{Fig:Hardware_Power_Nt}, CTUP-4 demonstrates superior performance compared to existing localization techniques. While CTUP-4 uniquely possesses the capability to estimate PLE among cooperative techniques (refer to Table~\ref{Tb:ExistingTech}), we present its performance in Fig.~\ref{Fig:Hardware_PLE_Nt}.} It's noteworthy that the performance enhancement resulting from an increased number of target nodes is less significant compared to the improvements obtained from expanding the number of anchor nodes. This disparity arises because, with an increase in target nodes, more decision variables (such as the location and transmit power of new nodes) are introduced to the optimization problem. 

\section{Conclusion}\label{sec:conclusions}
In this paper, we have addressed cooperative RSS-based localization in scenarios involving unknown transmit power, PLE, and anchor location uncertainty. We have proposed localization techniques (CTUPs) using mixed SDP-SOCP to jointly estimate target nodes' location, transmit power, and PLE. The reformulation of the ML estimator employs Taylor expansion, SDR, and the epigraph method, leveraging the accuracy of SDP and the computational efficiency of SOCP. We have carried out comprehensive simulations to demonstrate CTUPs' superior performance in terms of estimation accuracy and computational complexity compared to existing localization techniques. We have conducted extensive real-field experiments to curate an indigenous dataset encompassing RSS measurements among 50 nodes (with anchor location uncertainty) and further validate the effectiveness of CTUPs. Our future work will focus on developing localization techniques in the presence of malicious nodes, aiming to enhance robustness and security in cooperative localization.

\bibliography{IEEEabrv,ref}

\begin{thebibliography}{10}
\providecommand{\url}[1]{#1}
\csname url@samestyle\endcsname
\providecommand{\newblock}{\relax}
\providecommand{\bibinfo}[2]{#2}
\providecommand{\BIBentrySTDinterwordspacing}{\spaceskip=0pt\relax}
\providecommand{\BIBentryALTinterwordstretchfactor}{4}
\providecommand{\BIBentryALTinterwordspacing}{\spaceskip=\fontdimen2\font plus
\BIBentryALTinterwordstretchfactor\fontdimen3\font minus \fontdimen4\font\relax}
\providecommand{\BIBforeignlanguage}[2]{{%
\expandafter\ifx\csname l@#1\endcsname\relax
\typeout{** WARNING: IEEEtran.bst: No hyphenation pattern has been}%
\typeout{** loaded for the language `#1'. Using the pattern for}%
\typeout{** the default language instead.}%
\else
\language=\csname l@#1\endcsname
\fi
#2}}
\providecommand{\BIBdecl}{\relax}
\BIBdecl

\bibitem{Intro_Loc_Environ}
S.~Chen, B.~Liu, C.~Feng, C.~Vallespi-Gonzalez, and C.~Wellington, ``{3D} point cloud processing and learning for autonomous driving: Impacting map creation, localization, and perception,'' \emph{IEEE Signal Process. Mag.}, vol.~38, no.~1, pp. 68--86, Jan. 2021.

\bibitem{Intro_Loc_TOA}
Y.~Zou, H.~Liu, and Q.~Wan, ``Joint synchronization and localization in wireless sensor networks using semidefinite programming,'' \emph{IEEE Internet Things J.}, vol.~5, no.~1, pp. 199--205, Feb. 2018.

\bibitem{SDP_ToA_ma}
X.~Ma, B.~Hao, H.~Zhang, and P.~Wan, ``Semidefinite relaxation for source localization by {TOA} in unsynchronized networks,'' \emph{IEEE Signal Process. Lett.}, vol.~29, pp. 622--626, Feb. 2022.

\bibitem{Intro_Loc_IoT}
A.~Pandey, P.~Tiwary, S.~Kumar, and S.~K. Das, ``Fadeloc: Smart device localization for generalized ${\kappa {-} \mu}$ faded {IoT} environment,'' \emph{IEEE Trans. Signal Process.}, vol.~70, pp. 3206--3220, Jun. 2022.

\bibitem{MoeWin2}
M.~Z. Win, F.~Meyer, Z.~Liu, W.~Dai, S.~Bartoletti, and A.~Conti, ``\textcolor{blue} {Efficient Multisensor Localization for the Internet of Things: Exploring a New Class of Scalable Localization Algorithms},'' \emph{\textcolor{blue} {IEEE Signal Process. Mag.}}, vol. \textcolor{blue} {35}, no. \textcolor{blue} {5}, pp. \textcolor{blue} {153--167}, \textcolor{blue} {Sep.} \textcolor{blue} {2018}.

\bibitem{Intro_Gps}
T.~Stoyanova, F.~Kerasiotis, C.~Antonopoulos, and G.~Papadopoulos, ``{RSS}-based localization for wireless sensor networks in practice,'' in \emph{2014 9th International Symposium on Communication Systems, Networks \& Digital Sign (CSNDSP)}, Jul. 2014, pp. 134--139.

\bibitem{MoeWin4}
M.~Z. Win, A.~Conti, S.~Mazuelas, Y.~Shen, W.~M. Gifford, D.~Dardari, and M.~Chiani, ``\textcolor{blue} {Network localization and navigation via cooperation},'' \emph{\textcolor{blue} {IEEE Commun. Mag.}}, vol. \textcolor{blue} {49}, no. \textcolor{blue} {5}, pp. \textcolor{blue} {56--62}, \textcolor{blue} {May.} \textcolor{blue} {2011}.

\bibitem{MoeWin5}
Z.~Liu, W.~Dai, and M.~Z. Win, ``\textcolor{blue} {Mercury: An Infrastructure-Free System for Network Localization and Navigation},'' \emph{\textcolor{blue} {IEEE Trans. Mob. Comput.}}, vol. \textcolor{blue} {17}, no. \textcolor{blue} {5}, pp. \textcolor{blue} {1119--1133}, \textcolor{blue} {May.} \textcolor{blue} {2018}.

\bibitem{Intro_AOA}
Y.~Sun, K.~C. Ho, and Q.~Wan, ``Eigenspace solution for {AOA} localization in modified polar representation,'' \emph{IEEE Trans. Signal Process.}, vol.~68, pp. 2256--2271, Mar. 2020.

\bibitem{Intro_TOA}
X.~Ma, B.~Hao, H.~Zhang, and P.~Wan, ``Semidefinite relaxation for source localization by {TOA} in unsynchronized networks,'' \emph{IEEE Signal Process. Lett.}, vol.~29, pp. 622--626, Feb. 2022.

\bibitem{Intro_TDOA}
K.~C. Ho and T.-K. Le, ``Integrating {AOA} with {TDOA} for joint source and sensor localization,'' \emph{IEEE Trans. Signal Process.}, vol.~71, pp. 2087--2102, May. 2023.

\bibitem{DLOC_bodhi}
B.~Mukhopadhyay, S.~Srirangarajan, and S.~Kar, ``{RSS}-based cooperative localization and edge node detection,'' \emph{IEEE Trans. Veh. Technol.}, vol.~71, no.~5, pp. 5387--5403, Feb. 2022.

\bibitem{SignalStamp}
A.~Coluccia and A.~Fascista, ``On the hybrid {TOA/RSS} range estimation in wireless sensor networks,'' \emph{IEEE Trans. Wirel. Commun.}, vol.~17, no.~1, pp. 361--371, Jan. 2018.

\bibitem{Vaghefi_cooperative}
R.~M. Vaghefi, M.~R. Gholami, R.~M. Buehrer, and E.~G. Strom, ``Cooperative received signal strength-based sensor localization with unknown transmit powers,'' \emph{IEEE Trans. Signal Process.}, vol.~61, no.~6, pp. 1389--1403, Dec. 2013.

\bibitem{Intro_RWLS}
S.~Yang, G.~Wang, Y.~Hu, and H.~Chen, ``Robust differential received signal strength based localization with model parameter errors,'' \emph{IEEE Signal Process. Lett.}, vol.~25, no.~11, pp. 1740--1744, Nov. 2018.

\bibitem{SDPShi}
J.~Shi, G.~Wang, and L.~Jin, ``Least squared relative error estimator for {RSS} based localization with unknown transmit power,'' \emph{IEEE Signal Process. Lett.}, vol.~27, pp. 1165--1169, Jun. 2020.

\bibitem{RWLSAE}
Y.~Sun, S.~Yang, G.~Wang, and H.~Chen, ``Robust {RSS}-based source localization with unknown model parameters in mixed {LOS/NLOS} environments,'' \emph{IEEE Trans. Veh. Technol.}, vol.~70, no.~4, pp. 3926--3931, Mar. 2021.

\bibitem{SDP_Zou}
Y.~Zou and H.~Liu, ``{RSS}-based target localization with unknown model parameters and sensor position errors,'' \emph{IEEE Trans. Veh. Technol.}, vol.~70, no.~7, pp. 6969--6982, Jun. 2021.

\bibitem{Lohrasbipeydeh_Gulliver_2021}
H.~Lohrasbipeydeh and T.~A. Gulliver, ``{RSSD}-based {MSE}-{SDP} source localization with unknown position estimation bias,'' \emph{IEEE Trans. Commun.}, vol.~69, no.~12, pp. 8416--8428, Sep. 2021.

\bibitem{MSL}
X.~Mei, Y.~Chen, X.~Xu, and H.~Wu, ``{RSS} localization using multistep linearization in the presence of unknown path loss exponent,'' \emph{IEEE Sens. Lett.}, vol.~6, no.~8, pp. 1--4, Aug. 2022.

\bibitem{Invex}
B.~Mukhopadhyay, S.~Srirangarajan, and S.~Kar, ``Invex relaxation based cooperative localization using {RSS} measurements,'' \emph{IEEE Trans. Commun.}, vol.~70, no.~8, pp. 5482--5497, Jun. 2022.

\bibitem{SDPl2}
Q.~Wang, Z.~Duan, and F.~Li, ``Semidefinite programming for wireless cooperative localization using biased {RSS} measurements,'' \emph{IEEE Commun. Lett.}, vol.~26, no.~6, pp. 1278--1282, Apr. 2022.

\bibitem{FCUP}
Y.~Li, B.~Mukhopadhyay, and M.-S. Alouini, ``{RSS}-based cooperative localization and transmit power(s) estimation using mixed {SDP-SOCP},'' \emph{IEEE Trans. Veh. Technol.}, pp. 1--6, Jul. 2023.

\bibitem{LS_ref}
H.~C. So and L.~Lin, ``Linear least squares approach for accurate received signal strength based source localization,'' \emph{IEEE Trans. Signal Process.}, vol.~59, no.~8, pp. 4035--4040, Aug. 2011.

\bibitem{UTSDSOCP}
G.~Wang and K.~Yang, ``A new approach to sensor node localization using {RSS} measurements in wireless sensor networks,'' \emph{IEEE Trans. Wireless Commun.}, vol.~10, no.~5, pp. 1389--1395, Mar. 2011.

\bibitem{CRS}
C.~Soares, J.~Xavier, and J.~Gomes, ``Simple and fast convex relaxation method for cooperative localization in sensor networks using range measurements,'' \emph{IEEE Trans. Signal Process.}, vol.~63, no.~17, pp. 4532--4543, Jul. 2015.

\bibitem{ADMMSF}
N.~Piovesan and T.~Erseghe, ``Cooperative localization in {WSN}s: A hybrid convex/nonconvex solution,'' \emph{IEEE Trans. Signal Inf.}, vol.~4, no.~1, pp. 162--172, Dec. 2018.

\bibitem{SOCPU}
S.~Chang, Y.~Li, H.~Wang, W.~Hu, and Y.~Wu, ``{RSS}-based cooperative localization in wireless sensor networks via second-order cone relaxation,'' \emph{IEEE Access}, vol.~6, pp. 54\,097--54\,105, Sep. 2018.

\bibitem{TLLS}
K.~N. R. S.~V. Prasad and V.~K. Bhargava, ``Rss localization under gaussian distributed path loss exponent model,'' \emph{IEEE Wireless Communications Letters}, vol.~10, no.~1, pp. 111--115, 2021.

\bibitem{Intro_SDPLSREWang}
Z.~Wang, H.~Zhang, T.~Lu, and T.~A. Gulliver, ``Cooperative {RSS}-based localization in wireless sensor networks using relative error estimation and semidefinite programming,'' \emph{IEEE Trans. Veh. Technol.}, vol.~68, no.~1, pp. 483--497, Nov. 2019.

\bibitem{RSSTOA}
H.~Xiong, M.~Peng, S.~Gong, and Z.~Du, ``A novel hybrid {RSS} and {TOA} positioning algorithm for multi-objective cooperative wireless sensor networks,'' \emph{IEEE Sens. J.}, vol.~18, no.~22, pp. 9343--9351, Sep. 2018.

\bibitem{Battery_level}
N.~Saeed, A.~Celik, T.~Y. Al-Naffouri, and M.-S. Alouini, ``Localization of energy harvesting empowered underwater optical wireless sensor networks,'' \emph{IEEE Trans. Wireless Commun.}, vol.~18, no.~5, pp. 2652--2663, May. 2019.

\bibitem{Channel_state}
J.~Choi, ``Sensor-aided learning for {Wi-Fi} positioning with beacon channel state information,'' \emph{IEEE Trans. Wireless Commun.}, vol.~21, no.~7, pp. 5251--5264, Jul. 2022.

\bibitem{AntennaOrientation}
A.~Nagy, T.~Bigler, A.~Treytl, R.~Stenzl, S.~Wilker, T.~Sauter, and T.~Wien, ``{RSS}-based localization for directional antennas,'' in \emph{2020 25th IEEE International Conference on Emerging Technologies and Factory Automation ({ETFA})}, vol.~1, Sep. 2020, pp. 774--781.

\bibitem{simon2001digital}
M.~K. Simon and M.-S. Alouini, \emph{Digital Communication over Fading Channels}.\hskip 1em plus 0.5em minus 0.4em\relax New York: Wiley, 2001.

\bibitem{PLE_Radiofrequency}
J.~Miranda, R.~Abrishambaf, T.~Gomes, P.~Gonçalves, J.~Cabral, A.~Tavares, and J.~Monteiro, ``Path loss exponent analysis in wireless sensor networks: Experimental evaluation,'' in \emph{2013 11th IEEE International Conference on Industrial Informatics ({INDIN})}, Jul. 2013, pp. 54--58.

\bibitem{PLE_weather}
S.-H. Kim, S.-W. Moon, D.-G. Kim, M.~Ko, and Y.-H. Choi, ``A neural network-based path loss model for bluetooth transceivers,'' in \emph{2022 International Conference on Information Networking ({ICOIN})}, Jan. 2022, pp. 446--449.

\bibitem{PLE_temperature}
S.~Sun, T.~S. Rappaport, T.~A. Thomas, A.~Ghosh, H.~C. Nguyen, I.~Z. Kovács, I.~Rodriguez, O.~Koymen, and A.~Partyka, ``Investigation of prediction accuracy, sensitivity, and parameter stability of large-scale propagation path loss models for {5G} wireless communications,'' \emph{IEEE Trans. Veh. Technol.}, vol.~65, no.~5, pp. 2843--2860, May. 2016.

\bibitem{Beta0Ref}
M.~R. Gholami, R.~M. Vaghefi, and E.~G. Ström, ``{RSS}-based sensor localization in the presence of unknown channel parameters,'' \emph{IEEE Trans. Signal Process.}, vol.~61, no.~15, pp. 3752--3759, Aug. 2013.

\bibitem{PLE_AI}
L.~Wu, D.~He, B.~Ai, J.~Wang, H.~Qi, K.~Guan, and Z.~Zhong, ``Artificial neural network based path loss prediction for wireless communication network,'' \emph{IEEE Access}, vol.~8, pp. 199\,523--199\,538, Nov. 2020.

\bibitem{PLE_2}
Y.~Xu, J.~Zhou, and P.~Zhang, ``{RSS}-based source localization when path-loss model parameters are unknown,'' \emph{IEEE Commun. Lett.}, vol.~18, no.~6, pp. 1055--1058, Jun. 2014.

\bibitem{Anchor_GPS}
A.~G. Dempster and E.~Cetin, ``Interference localization for satellite navigation systems,'' \emph{Proc. IEEE}, vol. 104, no.~6, pp. 1318--1326, Jun. 2016.

\bibitem{Anchor_vary}
H.~Lohrasbipeydeh, T.~A. Gulliver, and H.~Amindavar, ``Unknown transmit power {RSSD} based source localization with sensor position uncertainty,'' \emph{IEEE Trans. Commun.}, vol.~63, no.~5, pp. 1784--1797, May. 2015.

\bibitem{A_Alink1}
P.~Barsocchi, S.~Lenzi, S.~Chessa, and G.~Giunta, ``A novel approach to indoor {RSSI} localization by automatic calibration of the wireless propagation model,'' in \emph{{VTC} Spring 2009 - IEEE 69th Vehicular Technology Conference}, Apr. 2009, pp. 1--5.

\bibitem{A_Alink2}
H.~Lim, L.-C. Kung, J.~C. Hou, and H.~Luo, ``Zero-configuration, robust indoor localization: Theory and experimentation,'' in \emph{Proceedings IEEE INFOCOM 2006. 25TH IEEE International Conference on Computer Communications}, Apr. 2006, pp. 1--12.

\bibitem{A_Alink3}
A.~Coluccia and F.~Ricciato, ``On {ML} estimation for automatic {RSS}-based indoor localization,'' in \emph{IEEE 5th International Symposium on Wireless Pervasive Computing 2010}, May. 2010, pp. 495--502.

\bibitem{kay1993fundamentals}
S.~M. Kay, \emph{Fundamentals of Statistical Signal Processing}.\hskip 1em plus 0.5em minus 0.4em\relax Prentice Hall PTR, 1993.

\bibitem{vandenberghe1996semidefinite}
L.~Vandenberghe and S.~Boyd, ``Semidefinite programming,'' \emph{SIAM review}, vol.~38, no.~1, pp. 49--95, 1996.

\bibitem{cvx}
M.~Grant and S.~Boyd, ``{CVX}: Matlab software for disciplined convex programming, version 2.1,'' \url{http://cvxr.com/cvx}, Mar. 2014.

\bibitem{MoeWin3}
N.~Patwari, J.~Ash, S.~Kyperountas, A.~Hero, R.~Moses, and N.~Correal, ``\textcolor{blue} {Locating the nodes: cooperative localization in wireless sensor networks},'' \emph{\textcolor{blue} {IEEE Signal Process. Mag.}}, vol. \textcolor{blue} {22}, no. \textcolor{blue} {4}, pp. \textcolor{blue} {54--69}, \textcolor{blue} {Jul.} \textcolor{blue} {2005}.

\bibitem{SDP_Tomic}
S.~{Tomic \textit{et al.}}, ``{RSS}-based localization in wireless sensor networks using convex relaxation: Noncooperative and cooperative schemes,'' \emph{IEEE Trans. Veh. Technol.}, vol.~64, no.~5, pp. 2037--2050, Jul. 2015.

\bibitem{DeltaSame}
Y.~Zou and H.~Liu, ``Semidefinite programming methods for alleviating clock synchronization bias and sensor position errors in {TDOA} localization,'' \emph{IEEE Signal Process. Lett.}, vol.~27, pp. 241--245, Jan. 2020.

\bibitem{MoeWin1}
A.~Conti, M.~Guerra, D.~Dardari, N.~Decarli, and M.~Z. Win, ``\textcolor{blue} {Network Experimentation for Cooperative Localization},'' \emph{\textcolor{blue} {IEEE J. Sel. Areas Commun.}}, vol. \textcolor{blue} {30}, no. \textcolor{blue} {2}, pp. \textcolor{blue} {467--475}, \textcolor{blue} {Feb.} \textcolor{blue} {2012}.

\bibitem{nguyen2022development}
T.~N. Nguyen, ``Development of wireless sensor network to detect lameness in dairy cows,'' Ph.D. dissertation, Massachusetts Institute of Technology, 2022.

\bibitem{BLEAntenna}
\BIBentryALTinterwordspacing
T.~G.~H. Ltd. (2018, May.) Specification of {GW.11.A153}. [Online]. Available: \url{https://cdn.taoglas.com/datasheets/{GW.11.A153}.pdf}
\BIBentrySTDinterwordspacing

\bibitem{Sturges}
H.~A. Sturges, ``The choice of a class interval,'' \emph{Journal of the american statistical association}, vol.~21, no. 153, pp. 65--66, 1926.

\bibitem{Sturges2}
D.~W. Scott, ``Sturges' rule,'' \emph{Wiley Interdisciplinary Reviews: Computational Statistics}, vol.~1, no.~3, pp. 303--306, 2009.

\end{thebibliography}
\bibliographystyle{IEEEtran}

\end{document}